\documentclass[a4paper,fleqn,usenatbib]{mnras}

\usepackage{newtxtext,newtxmath}

\usepackage[T1]{fontenc}
\usepackage{ae,aecompl}

\usepackage{epsf}
\usepackage{graphicx}	
\usepackage{amsmath}	

\newcommand{\xhiv}{X_\mathrm{H\,I,V}}
\newcommand{\xhim}{X_\mathrm{H\,I,M}}

\newcommand{\hi}{\mathrm{H\,I}}
\newcommand{\hei}{\mathrm{He\,I}}
\newcommand{\heii}{\mathrm{He\,II}}
\newcommand{\heiii}{\mathrm{He\,III}}
\newcommand{\oi}{\mathrm{O\,I}}
\newcommand{\hii}{\mathrm{H\,II}}
\newcommand{\cii}{\mathrm{C\,II}}

\newcommand{\civ}{\mathrm{C\,IV}}
\newcommand{\siii}{\mathrm{Si\,II}}
\newcommand{\siiv}{\mathrm{Si\,IV}}

\newcommand{\cm}{{\rm cm}}

\newcommand{\kpc}{{\rm kpc}}
\newcommand{\Mpc}{{\rm Mpc}}
\newcommand{\yr}{{\rm yr}}

\newcommand{\hkpc}{h^{-1}{\rm kpc}}
\newcommand{\hmpc}{h^{-1}{\rm Mpc}}

\newcommand{\kms}{\;{\rm km}\,{\rm s}^{-1}}
\newcommand{\msun}{M_{\odot}}
\newcommand{\mstar}{M_*}
\newcommand{\MUV}{M_{\rm UV}}

\newcommand{\fesc}{f_{\rm esc}}
\newcommand{\taues}{\tau_{\rm es}}

\newcommand{\scnot}[2]{#1\times10^{#2}}

\newcommand{\fescg}{f_\mathrm{esc,gal}}
\newcommand{\fescq}{f_\mathrm{esc,QSO}}
\newcommand{\fescmax}{f_\mathrm{esc,max}}
\newcommand{\td}{\textit{Technicolor Dawn}}
\newcommand{\etai}{\eta_\mathrm{i}}
\newcommand{\etaw}{\eta_\mathrm{W}}

\title[Reionization in Technicolor]{Reionization in Technicolor} 

\author[K.\ Finlator et al.]{
\parbox[t]{\textwidth}{\vspace{-1cm}
Kristian Finlator$^{1,8,11}$, 
Laura Keating$^{2}$,
Benjamin D.\ Oppenheimer$^{4}$,
Romeel Dav\'e$^{5,6,7,9,10}$,
Erik Zackrisson$^3$
}
\\\\$^1$ New Mexico State University, Las Cruces, NM, USA
\\$^2$ Canadian Institute for Theoretical Astrophysics, 60 St.\ George Street, University of Toronto, ON M5S 3H8, Canada
\\$^3$ Department of Physics and Astronomy, Uppsala University, 751 20 Uppsala, Sweden
\\$^4$ CASA, Department of Astrophysical and Planetary Sciences, University of Colorado, 389-UCB, Boulder, CO 80309, USA
\\$^5$ University of the Western Cape, Bellville, Cape Town 7535, South Africa
\\$^6$ South African Astronomical Observatory, Observatory, Cape Town 7925, South Africa
\\$^7$ African Institute for Mathematical Sciences, Muizenberg, Cape Town 7945, South Africa
\\$^8$ Cosmic Dawn Center (DAWN), Niels Bohr Institute, University of Copenhagen / DTU-Space, Technical University of Denmark
\\$^{9}$ Center for Computational Astrophysics, Simons Foundation, New York, NY 10010, USA
\\$^{10}$ Institute for Astronomy, Royal Observatory, Edinburgh EH9 3HJ, UK
\\$^{11}$ finlator@nmsu.edu
\author[The Technicolor Dawn Simulations]{
K.\ Finlator,
C.\ Doughty,
B.\ D.\ Oppenheimer,
R.\ Dav\'e,
E.\ Zackrisson,
\& S.\ Huang
}
}

\date{Accepted XXX. Received YYY; in original form ZZZ}

\pubyear{2018}

\begin{document}
\label{firstpage}
\pagerange{\pageref{firstpage}--\pageref{lastpage}}
\maketitle

\begin{abstract}
We present the \td~simulations, a suite of cosmological
radiation-hydrodynamic simulations of the first 1.2 billion years.  By modeling a
spatially-inhomogeneous UVB on-the-fly with 24 frequencies and resolving
dark matter halos down to $10^8\msun$ within 12$\hmpc$
volumes, our simulations unify observations of
the intergalactic and circumgalactic media, galaxies, and reionization into a common
framework.  The only empirically-tuned parameter, the fraction $\fescg(z)$ of
ionizing photons that escape the interstellar medium, is adjusted to match
observations of the Lyman-$\alpha$ forest and the cosmic microwave
background.  With this single calibration, our simulations reproduce the
history of reionization; the stellar mass-star formation rate relation of 
galaxies; the number density and metallicity of damped Lyman-$\alpha$ absorbers 
(DLAs) at $z\sim5$; the abundance of weak metal absorbers; the ultraviolet 
background (UVB) amplitude; and the Lyman-$\alpha$ flux power 
spectrum at $z=5.4$.  The galaxy stellar mass and UV luminosity functions are 
underproduced by $\leq2\times$, suggesting an overly vigorous feedback 
model.  The mean transmission in the Lyman-$\alpha$ forest is
underproduced at $z<6$, indicating tension between measurements of the UVB
amplitude and Lyman-$\alpha$ transmission.  The observed $\siiv$ column density
distribution is reasonably well-reproduced ($\sim1\sigma$ low).  By 
contrast, $\civ$ remains significantly underproduced despite being boosted by 
an intense $>4$ Ryd UVB.  Solving this problem by increasing metal 
yields would overproduce both weak absorbers and DLA metallicities.  Instead, 
the observed strength of high-ionization
emission from high-redshift galaxies and absorption from their
environments suggest that the ionizing flux from conventional stellar
population models is too soft.
\end{abstract}




\section{Introduction} \label{sec:intro}
 
The first billion years of cosmological time witnessed a number of physical transformations 
that will constitute the subject of detailed scrutiny over the next decade: The formation of 
the first stars, galaxies, and quasars; the enrichment and heating of the circumgalactic 
medium (CGM) owing to feedback processes occurring deep within galaxies; and 
the transformation of the intergalactic 
medium (IGM) from a cold, neutral medium into a hot, ionized one.  Dedicated observational 
campaigns are now probing separately the CGM around early 
galaxies~\citep{dodo13,bosm17,chensf17}; the nature of their stellar 
populations~\citep{stei16,smit17,berg18}; and the IGM's temperature~\citep{viel13a,garz17} 
and ionization state~\citep{calv11,wyitbolt11,davi18a,dalo18}.  

Physical insight into what these observations imply is being pursued through a 
variety of theoretical approaches, which are usually optimized separately to address 
a small subset of available observations~\citep{trac11}.
These dedicated efforts have unearthed several problems that seem resistant 
to solution.  For example, models that match the reported amplitude of the extragalactic
ultraviolet background (UVB) at $z=5.4$, often quantified by the hydrogen photoionization 
rate $\Gamma_\hi$, underproduce the mean transmission in the Lyman-$\alpha$ 
forest~\citep{bosm18}.  Moreover, assuming a model for the UVB that matches the 
observed $\Gamma_\hi$ nearly invariably leads to underproducing the CGM's $\civ$ 
abundance at $z>3$~\citep{rahm16,keat16,turn16}.  Both of these observations hint
at the need for a more intense UVB or a hotter IGM, perhaps the result of a harder
ionizing continuum~\citep{dalo15,keat18}.  Such a revision would also modulate the 
predicted abundance of low-ionization metal absorbers and damped Lyman-$\alpha$ 
absorbers~\citep[DLAs;][]{rafe14,bird17}.  Can a single model satisfy all of these
constraints, or are the observations fundamentally in tension with one another? 
If indeed it turns out that a more intense UVB yields generally improved agreement 
with observations, can it be attributed to the observed level of early star formation 
given standard stellar population synthesis models, or are updates required for the 
physics of young, low-metallicity stars~\citep{stan16,rosd18}? 

In the current, data-rich era, it is worth considering whether combining observations 
that are traditionally considered separately can yield new insights when subsumed 
within a single theoretical framework.  This work takes the view that the critical 
link is the UVB, which couples stars, the CGM, and the IGM.  A generation of 
theoretical studies has modeled the IGM and CGM during the hydrogen reionization 
epoch under the assumption of a spatially-homogeneous, externally-imposed UVB, usually 
coupled with the assumption that gas is in ionization 
equilibrium~\citep{oppe09,luki15,keat16}.  Both assumptions are inaccurate: 
reionization is directly observed to be a spatially-inhomogeneous 
process~\citep{beck15a,bosm18} that is generally believed to have been driven by 
ionizing flux from small 
galaxies~\citep{finl16,dalo17,pars18,qin17,mcgr18,hass18}, although a significant 
contribution from rare, bright sources remains possible~\citep{char15,mada15}.  Treating
reionization as a spatially-homogeneous process severs its links to the sources
that drove it.  Moreover, it limits the ability of models to 
address small-scale tracers of reionization such as metal absorbers or small-scale 
fluctuations in the Lyman-$\alpha$ flux power spectrum.  This is even more true when
high-ionization absorbers such as $\civ$ are considered because they are sensitive to
$\heii$-ionizing photons, which were not homogeneously distributed until the close of
$\heii$ reionization around $z=2.7$~\citep{wors16}.  Meanwhile, forcing the diffuse 
IGM into ionization equilibrium boosts the $\hi$ fraction and the collisional 
excitation cooling rate, leading to an artificially-cold IGM~\citep{puch15}. 

Reionization affects overdense and underdense regions in different ways and at
different times~\citep{finl09}, hence it is important to consider how measurements
that probe the diffuse and condensed gas phases complement one another.
The physical state of the IGM is most readily probed through 
observations of the Lyman-$\alpha$ forest (LAF) while material that is associated with 
galaxy formation (that is, the CGM)  is probed through observations 
of metal absorbers and DLAs.  Modeling these media simultaneously is useful for
two reasons.  First, at $z>5$, the LAF probes primarily underdense
IGM gas while the CGM probes overdense gas; the two phases therefore provide 
complementary probes that can be combined to gain insight into the topology of 
reionization.  Second, whereas the LAF is an exquisite probe of 
the post-reionization IGM~\citep{beck15b,mcqu16}, it saturates at $z>6$, leaving 
metal absorbers as the only viable probe of the UVB's growth at earlier 
times~\citep{oh02,keat14,keat16}.  This indicates the need for theoretical models 
that can make the connection between observables that track the progress of 
reionization before and after $z=6$. 

Metal absorbers are more 
than just a complicated substitute for the LAF, however: given their proximity to 
galaxies, they are the \textit{most direct} observational probe of the galaxy-driven 
reionization hypothesis: they probably trace activity in the faint---currently 
unobservably so---systems that dominated reionization~\citep{diaz14,cai17}, and 
their physical state bears the chemical and radiative signatures of outflows 
and ionizing flux from the young stars in those galaxies.

Combining these arguments leads to the need for radiation-hydrodynamic
simulations of the reionization epoch with realistic star formation histories and
metal yields as well as a spatially-inhomogeneous, multifrequency UVB.  In this way,
observations of all baryon phases can be treated as complementary tracers of the 
progress of reionization.  As a step in this direction, we present the
\td~simulations.  At small scales, these simulations adopt
subgrid prescriptions for star formation and feedback that, while heavily updated 
to incorporate insight from observations and complementary modeling efforts, are
directly descended from the pathbreaking framework presented in~\citet{spri03}.
This means that gas inflows into galaxies, star formation, enrichment, 
and feedback are modeled in a way that allows predictions of bulk galaxy properties 
such as stellar mass, star formation rate, color, metallicity, and environment.
The UVB, however, is modeled using an on-the-fly radiation transport solver using
24 independent frequency bins. The physical conditions of the CGM and---crucially 
for reionization---the IGM are treated self-consistently, with the result that the
detailed way in which the IGM is heated, pressurized, and enriched connects
directly to the underlying model for galaxy growth.  In the process, it allows us 
to address simultaneously observations of galaxies, the CGM, and the IGM and 
evaluate whether their implications are mutually consistent.

In Section~\ref{sec:sims}, we present an overview of the \td~simulations.  In 
Section~\ref{sec:results}, we present initial results.  In Section~\ref{sec:sumDisc},
we summarize and discuss our results.

\section{Simulation}\label{sec:sims}
Our simulations assume a \textit{Planck} cosmology~\citep{plan16a} in which 
$(\Omega_M, \Omega_\Lambda, \Omega_b, H_0, X_H) = (0.3089, 0.6911, 0.0486, 67.74, 0.751)$.  
Initial conditions are generated at $z=199$ using {\sc music}~\citep{hahn11}.  
They incorporate an extensive suite of updates with respect our previous work, 
hence we review our physical model in detail. 

\subsection{Hydrodynamics and Star Formation}
We run our simulations using a custom version of {\sc Gadget-3} (last 
described in~\citealt{spri05}).  Hydrodynamics are modelled using a density-independent 
formulation of smoothed particle hydrodynamics (SPH) that treats fluid 
instabilities accurately~\citep{hopk13}.  We compute the physical properties of 
each gas particle using a 5th-order B-spline kernel that incorporates information
from up to 32 neighbours.  Increasing the minimum number of neighboring particles
leads to a more accurate hydrodynamic calculation, but it also suppresses star 
formation in low-mass systems at early times, hence we retain the coarser kernel
for the present.

Gas particles cool radiatively owing to collisional 
excitation of hydrogen and helium using the processes and rates in Table 1 
of~\citet{katz96}.  In the case of our radiation transfer simulations, however,
we relax the assumption that hydrogen and helium 
are in ionization equilibrium.  Gas particles that have nonzero metal mass fraction
are further allowed to cool via collisional excitation of metals cooling using the 
collisional ionization equilibrium tables of~\citet{suth93}.

Gas whose proper hydrogen number density exceeds 0.13 cm$^{-3}$ acquires
a subgrid multiphase structure~\citep{spri03} and forms stars at a rate 
that is calibrated to match the observed Kennicutt-Schmidt law.  We
assume that stars form with a universal~\citet{krou01} initial mass
function (IMF) from 0.1--100$\msun$ (although this may underpredict the
mass fraction in even more massive stars;~\citealt{crow16,schn18}).  
We assume that 
all stars more massive than $10\msun$ end their lives instantaneously
as supernovae (SNe), implying a SN mass fraction of 0.187 
($=\beta$ in~\citealt{spri03}).  Following~\citet{nomo06}, we further 
assume that half of SNe are hypernovae and that SNe from stars more massive 
than 50$\msun$ do not release metals into the ISM.  With these assumptions,
we find that SNe release $\scnot{3.75}{49}$ergs of energy per 
$\msun$ of new stars.  This is an order of magnitude more feedback energy 
than is assumed in~\citet{spri03}, but we leave the local heating parameter 
$u_{\mathrm{SN}}$ unchanged.  This corresponds to the assumption that 
most of the feedback energy is radiated away or converted efficiently to 
kinetic energy associated with galactic outflows.

\subsection{Metal Enrichment} 
Our simulations independently model metal enrichment of C, O, Si, Fe, N, Ne, Mg, S, 
Ca, and Ti.  We compute the Type II SNe metal yields by weighting the 
supernovae yields of~\citet{nomo06} by our assumed IMF and assuming a 
50\% hypernovae (HNe) fraction.  Each star-forming particle's Type II 
self-enrichment rate is then obtained by interpolating to its 
metallicity.  The resulting assumed metal yields are in 
Table~\ref{table:yields}.

As is well-known, the metal yields are a source of 
uncertainty in cosmological simulations, with $\sim$factor of two variations
among yields that are commonly assumed~\citep{wier09}.  In order to estimate
how the unknown HNe fraction contributes to our metal production, we recompute
our metal yields for a $Z=0.001$ stellar population with assumed HNe fractions
of 0, 0.5, and 1.0 in Table~\ref{table:yieldDep} (note that typical star-forming
gas at $z=6$ has a metallicity of $Z\sim0.1 Z_\odot$).  Broadly, HNe in 
low-metallicity stars boost the yields of C, O, Si, Fe, N, Ne, Mg, and Ti by
factors of up to 65\% (in the case of Ti).  By contrast, they suppress the 
yields of S and Ca by up to 10--15\%.  Evidence that the boosted yields 
may be realistic comes from the fact that simulations without
HNe often underproduce the normalization of the mass-metallicity 
relation, which is generally calibrated to the oxygen mass 
fraction~\citep{somd15,dave16}.  Hence while the unknown HNe 
fraction contributes an overall uncertainty of 2--50\% to our enrichment 
model (depending on the species), the addition of HNe may alleviate some
tension between simulations and observations.

We model enrichment from Type Ia SNe assuming a continuous~\citet{heri17}
delay time distribution: A stellar population of age $t$ has a Type Ia 
rate $\dot{N}_{\mathrm{Ia}}$ of
\begin{equation}\label{eqn:typeIaRate}
\dot{N}_{\mathrm{IA}} = \scnot{3.2}{-13} \left(\frac{t}{\tau}\right)^{-1.5} \msun^{-1}\yr^{-1}.
\end{equation}
Metal yields from Type Ia SNe are adopted from the W7 model in Table 2
of~\citet{maed10}.

Star particles that are old enough to form asymptotic giant branch (AGB) stars
lose mass and metals to the nearest gas particle.  The metal content in C, O, and
N is enhanced with respect to the star particle's original metallicity whereas
the metal mass fraction in all other species is unchanged; see~\citet{oppe08} 
for details.

\subsection{Feedback} 
Galactic outflows form via a Monte Carlo model in which star-forming 
gas particles receive ``kicks" in momentum space and are thereafter 
temporarily decoupled hydrodynamically.  The probability that a star-forming
gas particle is ejected at any timestep depends on its star formation rate and
its host galaxy's stellar mass in a way that is tuned to match the ratio $\etaw$ 
of the mass outflow rate to the star formation rate predicted from high-resolution 
simulations~\citep[][their Equation 8]{mura15}\footnote{
We have also experimented with the dependence on halo mass and redshift that
they give in their Equations 4 and 5 but found that this drastically 
over-suppresses star formation at high redshift.}:
\begin{equation}\label{eqn:mlf}
\etaw(\mstar) = 3.6 (\mstar/10^{10}\msun)^{-0.35}.
\end{equation}

While the~\citet{mura15} mass-loading factor seems to have substantial
predictive power~\citep{dave16,font17}, we have found that directly adopting
their outflow velocities (their Equation 9) leads to very little gas escaping
the host galaxy.  Accordingly, we follow~\citet{dave16} and add two boost
factors.  The first accounts for the fact that stars characteristically form
at a halocentric radius of 0.034 times the virial radius 
$R_\mathrm{vir}$~\citep{huan17} whereas the reported outflow scalings 
correspond to 0.25$R_\mathrm{vir}$.  We account for this energy loss $\Delta E$
by assuming the host halo, whose mass is computed by an on-the-fly group
finder, has a~\citet{nfw96} density profile with a concentration
given by the concentration-mass relation of~\citet{ange16}.  The second is
a multiplicative factor that insures that some of the gas is boosted up 
to velocities approaching 3 times the circular velocity, as 
observed~\citep{chis15}.  The final adopted velocity scaling is
\begin{equation}\label{eqn:vwind}
v = 2.3485 v_c^{1.12} + \Delta E
\end{equation}
where $v_c$ is the circular velocity in $\kms$.  We multiply each outflowing 
particle's velocity by a random number drawn uniformly from the range $[0.75,1.25]$.  

Equation~\ref{eqn:vwind} yields a median velocity that exceeds 
predictions from high-resolution simulations (Equation 10 of~\citealt{mura15}).  
However, the energy requirements for powering these outflows do not greatly exceed 
the energy available from Type II SNe.  For a constant stellar baryon
fraction $\mstar/M_b = 0.1$, the ratio of the wind kinetic energy to the Type II 
SNe energy scales with the halo mass as $M_h^{0.4}$ and exceeds unity for halos 
more massive than $\log(M_h/M_\odot)=(11.4,10.9)$ at $z=(6,10)$.  It does not 
exceed 3 for any halo less massive than $10^{12}\msun$
during the same epoch.  Given the order-of-magnitude uncertainty in the available
energy from SNe, the fact that halos more massive than $10^{11}\msun$ do not form
by $z=6$ in our modest simulation volumes, and the possible contribution of other 
processes such as stellar winds, we do not believe that this energy requirement 
is excessive.  Nonetheless, we will show in 
Figures~\ref{fig:dNdEWdX_CII}--\ref{fig:dNdEWdX_OI} that the abundance
of weak low-ionization absorbers may be overproduced in our model, which
may indicate that more moderate outflow velocities are indeed 
preferred~\citep[see also][]{keat16}.

As our simulations do not resolve the ISM with sufficient detail to treat
the emergence of outflows self-consistently, we retain the convention of
hydrodynamically decoupling gas that is kicked into 
outflows~\citep{spri03}.  Gas recouples once its density falls to 10\% 
of the minimum density for star formation, or after a time delay equal 
to the host halo's virial radius divided by the gas particle's initial 
kick velocity.  The virial radius is derived from the halo mass
combined with the~\citet{brya98} fitting-formula for virial overdensity 
as a function of redshift.

\subsection{Radiation Transport}\label{ssec:rt}
A key ingredient in our effort to model realistically the photoionization 
feedback that occurs at $z>5$ is the inclusion of a self-consistent, 
inhomogeneous, multifrequency UVB.   We model the evolving UVB on-the-fly by 
solving the moments of the radiation transport (RT) equation on a uniform 
Cartesian grid that is superposed on our simulation volume~\citep{finl09}.  
We discretize the radiation 
field owing to galaxies spatially on a regular grid and spectrally into 24 
frequency groups spaced evenly between 1--10 Ryd.  The quasar field is spatially
uniform except for the subgrid self-shielding prescription (see below) but 
evolved in the same frequency bins using the volume-averaged opacity field.

\subsubsection{Limiting Cosmological Redshifting}
A multifrequency radiation transport solver allows us to account for cosmological
redshifting, but our numerical implementation requires us to limit this effect
in order to avoid introducing numerical noise when computing the 
derivative of the intensity.  To see this, we integrate over the cosmological 
radiation transport equation~\citep[for example, ][]{gned97,finl09} in
frequency to yield a multigroup method:
\begin{equation}\label{eqn:multigroup}
\frac{\partial I}{\partial t} = \frac{-cn^i}{a} \frac{\partial I}{\partial x^i} + c\etai
- c\left[ \chi + \frac{H}{c}\left(2-\frac{1}{I}\int\nu\frac{\partial I_\nu}{\partial \nu}\mathrm{d}\nu\right)\right]I
\end{equation}
Here, $I_\nu$ is the proper photon number density per unit frequency; 
$I = \int I_\nu \mathrm{d}\nu$ is the density in a particular frequency bin; 
$n^i$ is the $i$th component of a unit vector; $a$ is the cosmological 
expansion factor; $c$ is the speed of light; $\etai$ 
is the ionizing emissivity; $H=\dot{a}/a$ is the cosmogical expansion rate; 
and $\chi$ is the photoelectric opacity.  The term with the prefactor $(H/c)$
is the cosmological term.  This functions as an opacity $\chi_H$ that accounts
for both dilution and redshifting.  Carrying out the integral over a frequency 
bin from $\nu_1$ to $\nu_2$ yields
\begin{equation}\label{eqn:chi_H1}
\chi_H \equiv \frac{H}{c} \left[3 - \frac{1}{I}(\nu_2 I_{\nu}(\nu_2) - \nu_1 I_\nu(\nu_1))\right].
\end{equation}
If the photon number density can be approximated as a power-law 
$I_\nu \propto \nu^{-\alpha}$, then this reduces to
\begin{equation}\label{eqn:chi_H2}
\chi_H \equiv \frac{H}{c} [3 - (1-\alpha)]
\end{equation}
To understand Equation~\ref{eqn:chi_H2}, consider the case of a uniform,
flat-spectrum field with no sources and sinks; that is, $\alpha = \etai = \chi = 0$.
In this case, the number density $I$ of photons in a fixed energy bin evolves at a rate
$\partial I / \partial t = -2 H$.  $I$ evolves more slowly than the
proper hydrogen number density ($\partial n_H / \partial t = -3H$) because photons
gradually accumulate into bins of fixed width in energy.  If we allow $\alpha>0$, then 
there are fewer high-energy than low-energy photons and cosmological redshifting moves 
more energy out of the bin than into it, leading to more rapid dimming than in the 
flat-spectrum case.  By contrast, if $\alpha<0$, then the UVB's spectral slope is 
locally-hard and redshifting moves more energy into the bin than in the 
flat-spectrum case.

Trouble arises because we compute $\alpha$ numerically using the difference 
between the photon densities in successive frequency bins.  When $I_\nu$ 
is small enough, numerical noise can lead to spuriously large values of $\lvert\alpha\rvert$.  
We find that enforcing $\alpha\geq-10$ suppresses the problem 
adequately.\footnote{In principle, this approximation
smooths over sharp spectral features such as the sawtooth absorption spectrum
from the $\heii$ Lyman series.  However, this does not affect our current work
because our model for the photoelectric opacity does not yet account for
bound-bound transitions.}

\subsubsection{The Galaxy UVB}
The galaxy emissivity field is computed from the star formation rate of star-forming 
gas particles, which improves spatial resolution with respect to models that derive
the emissivity directly from star particles~\citep{rahm13}.  Each gas particle's 
emissivity is proportional to its instantaneous star formation rate, which is assumed
to have been constant for the past 100 Myr.  The emissivity's metallicity 
dependence is computed from a modified version of {\sc Yggdrasil}~\citep{zack11}.  
The emissivity is tabulated at 7 distinct metallicities between $Z=0$--$0.04$, and 
each gas particle's actual emissivity is computed by interpolating to its $Z$.  
The $Z=0$ emissivity comes from~\citet{scha02};
the $Z=10^{-5}$ and $Z=10^{-7}$ emissivities come from~\citet{rait10}; and 
the $Z=0.001$--$0.040$ emissivities come from Starburst 99~\citep{leit99}, 
running with the Geneva tracks (high mass-loss version, without rotation) 
and Pauldrach/Hillier atmospheres.  Each model is adjusted to a~\citet{krou01}
IMF from 0.1--100$\msun$, consistent with our simulation's star formation model.

In order to close the moment hierarchy, we use a time-independent ray-casting
calculation to compute the Eddington tensor field separately for each frequency bin.  
After each timestep, any ``target" cell in which the radiation field in at least one 
frequency bin changes by a factor greater than 100\% is flagged for update.  A 
ray-tracing code then computes the total optical depth from the target cell to each 
source cell.  Periodic boundary conditions are accounted for by replicating the 
simulated emissivity and opacity fields in 26 replicas that surround the actual 
computation volume.  In other words, each source gives rise to 27 ray-casting 
calculations per target cell.  To save computation time, we halt ray-casting
calculations whenever the optical depth from the target cell to the source cell 
exceeds $6$ in all frequency bins.  The local Eddington tensor field is then 
computed based on the contribution from those sources that were unobscured.  
Note that this calculation is not used to compute the actual radiation field, 
which results from solving the moments of the RT equation.

Even with these optimizations, the computation time for Eddington tensor updates 
scales too severely with the problem's dynamic range.  To tame this problem, we 
compute Eddington tensor updates on a coarse grid.  For example, for simulations 
that include $64^3$ RT voxels, we smooth the simulated
opacity and emissivity fields onto an $8^3$ grid (where each coarse grid cell
contains 512 RT voxels), compute the updated Eddington tensors on the coarse grid, 
and then assume that the Eddington tensor field is uniform throughout each coarse
grid cell.  The price for such approximations is the introduction of small errors 
in the shapes of HII regions at early times~\citep{gned01}, but the impact on the
overall history of reionization is expected to be weak.

At each timestep, the UVB is used to evolve each particle's H and He ionization 
states forward, and the resulting opacity field is used to evolve the UVB.  
The opacity field accounts only for bound-free transitions in $\hi,\hei,$ and 
$\heii$.  An iterative solver repeats the ionization and moment updates throughout
the simulation volume if the UVB changes over the course of a timestep in at least 
one voxel and frequency by more than 1\%.  If the iteration does 
not yield a converged UVB within 1\% at all frequences and locations within 10 
iterations, then a substepper collapses the timestep and evolves the UVB and 
ionization fields forwards at the reduced timestep, iterating to 1\% convergence 
at each substep as before.

A chemical courant condition attempts to capture the progress of ionization
fronts by limiting a particle's timestep to $\tau_e = 0.5 n_e/\dot{n_e}$ (where $n_e$
is the local electron density and its derivative includes all contributions from
ionizations and recombinations of H and He).  Unfortunately, $n_e/\dot{n_e}$ drops
to extremely short values at ionization fronts, which can grind the simulation to 
a halt.  We therefore impose a lower limit $\tau_e \geq $ 100,000 years; in other
words, an individual particle's timestep may not be cut shorter than this owing to the
chemical courant condition.  This smooths ionization fronts over a spatial scale of 
order $c\tau_e \sim 30 \kpc$, which is comparable to the spatial resolution of our 
radiation transport solver.  The fully-implicit ionization solver can treat evolution 
on timesteps larger than $n_e/\dot{n_e}$ robustly.

For our highest-resolution simulation ($12\hmpc$ volume with $64^3$ RT voxels), 
each voxel has a side length equal to 1/64th of the total simulation volume,
or 187.5 comoving $\hkpc$.  This means that, throughout the reionization epoch ($z>6$),
the radiation field is directly resolved down to scales of $<50$ physical kpc.
This remains too coarse to resolve the escape of ionizing radation from galaxies'
interstellar media and somewhat too large to resolve self-shielding within Lyman-limit
systems, hence we employ two further prescriptions to account for subgrid effects: one
accounts for UVB attenuation as it penetrates into dense regions from the outside, and 
one accounts for absorption of locally-produced light as it exits dense star-forming 
regions.

The first, a self-shielding prescription that generalizes ideas originally
presented in~\citet{scha01} and~\citet{haeh98}, attenuates the UVB in dense regions 
based on the assumption that dense gas is in hydrostatic and photoionization 
equilibrium (see section 2.2 of~\citealt{finl15} for details).  The assumption of
photoionization equilibrium is invoked only to compute the self-shielding.  Once the
locally-attenuated UVB is computed, it is used to evolve the gas' ionization state
via a nonequilibrium solver.  The contribution of self-shielded gas particles to
the local opacity is reduced by the same factor as the local UVB; that is, if the
UVB in a particular frequency is suppressed by $90\%$ owing to self-shielding, then
that particle's contribution to the local opacity in that frequency is likewise 
reduced by 90\% with respect to the optically-thin case.  This treatment has no 
free physical parameters and yields excellent agreement with high-resolution
calculations that resolve ionization fronts at IGM filaments~\citep{finl15}.

Our second treatment for subgrid radiation transport effects involves the 
fraction $\fescg$ of ionizing photons that escape from galaxies, which is much more 
difficult to model.  Good agreement with observations of the Thomson scattering 
optical depth and the observed ionizing emissivity at $z\leq6$ are generally 
obtained by arguing that $\fescg$ could increase either to low 
masses~\citep{alva12,wise14} or luminosities~\citep{paar15,shar16}, or---perhaps 
equivalently---to high redshift~\citep{haar12,kuhl12,khai16,dous17}.  The most 
general model for $\fescg$ would permit dependence on the galaxy's mass and 
luminosity as well as its recent star formation history and ISM 
structure~\citep{ma15,paar15,katz18}. Unfortunately, despite over a decade of 
effort, the way in which $\fescg$ varies with these factors is not yet robustly 
constrained.  Moreover, our simulations do not resolve the ISM-scale processes 
that determine $\fescg$~\citep{ma15}. Consequently, in this work we 
follow~\citet{haar12} and consider a pure redshift dependence 
$\fescg(z)$:
\begin{equation}\label{eqn:fesc}
\fescg(z) = 0.176 \left(\frac{1+z}{6}\right)^{A}
\end{equation}
We also cap $\fescg$ at a value $\fescmax$, leading to a constant $\fescg$ at 
sufficiently high redshifts.

This model for $\fescg$ contains three parameters: a normalization, a slope, and
a maximum.  These parameters are adjusted in three steps.  First, the normalization 
is fixed to a constant value in order to match roughly the 
observationally-inferred ionizing emissivity at 
$z=5$~\citep[for example, Table 2 of][]{kuhl12}.  Next, we run a simulation
with a trial model for $\fescg$ in order to establish how many photons per
hydrogen atom are required to bring the volume-averaged neutral fraction 
$\xhiv$ below 1\%.  This number depends mostly on the simulation's resolution 
rather than on the model for $\fescg(z)$.  Finally, we evaluate a broad range 
of combinations of $A$ and $\fescmax$ in post-processing to determine which 
ones emit enough ionizing photons to complete reionization by $z=6$.  This
calibration step involves integrating the product of the comoving star formation 
rate density $\dot{\rho}_*$ from the trial simulation with a representative 
ionizing efficiency and each combination of $A$ and $\fescmax$, and then 
integrating in time:    
\begin{equation}\label{eqn:calib_fesc}
n_\gamma(t) = \int_0^t \dot{\rho}_*(t') Q \fescg(z(t')) dt'
\end{equation}
Here, $n_\gamma(t)$ is the cumulative number of ionizing photons emitted into
the IGM per unit volume by time $t$, and $Q$ is the
ionizing efficiency in photons sec$^{-1} \msun^{-1}$ yr.  While $Q$ varies
self-consistently with the local metallicity in ``live" simulations, we have 
found that $\log(Q) = 53.51$ approximates the predicted volume-averaged 
emissivity reasonably well.  

The final choice of $A$ and 
$\fescmax$ involves fitting two parameters to one measurement and thus 
introduces a measure of freedom.  Without further observational guidance, 
we favor combinations that do not involve either large $\fescmax$ or the 
rapid evolution in $\fescg$ that occurs when $A > 3.5$.  In principle,
measurements of the duration of hydrogen reionization can reduce this
freedom~\citep{dous17}; however, we have not considered this.

The number of ionizing photons per hydrogen atom that must be emitted in
order for reionization to complete depends on the resolution of our 
radiation transport solver: For a $12\hmpc$ box with $2\times256^3$ particles, 
simulations in which the RT voxels have a comoving side length of 
(375, 187.5)$\hkpc$ emit (2.44, 3.02) ionizing photons per hydrogen atom before 
$\xhiv$ drops below 1\%.  The slight lack of 
convergence is expected: If the resolution is higher, then fewer absorptions 
occur at subgrid scales, and $\fescg$ must be higher.  Likewise, the total 
number of photons per baryon that are absorbed at scales larger than the grid 
scale increases.  In order for our simulations to yield consistent reionization 
histories, we therefore allow $\fescg(z)$ to vary with resolution so that 
$A=(1.95, 3.4)$ and $\fescmax=(0.36,0.5)$ for the same resolutions 
(see Table~\ref{table:sims}).  We expect that the preferred redshift dependence
will weaken as we scale up to larger volumes. This is because because periodic 
cosmological volumes that are smaller than $\sim100 \hmpc$~experience 
artificially sudden reionization histories owing to their lack of large-scale 
density fluctuations~\citep{bark04,ilie14,dous17}, which we effectively 
compensate for via a large $\fescg$ at earlier times.

\begin{figure}
\centerline{
\setlength{\epsfxsize}{0.5\textwidth}
\centerline{\epsfbox{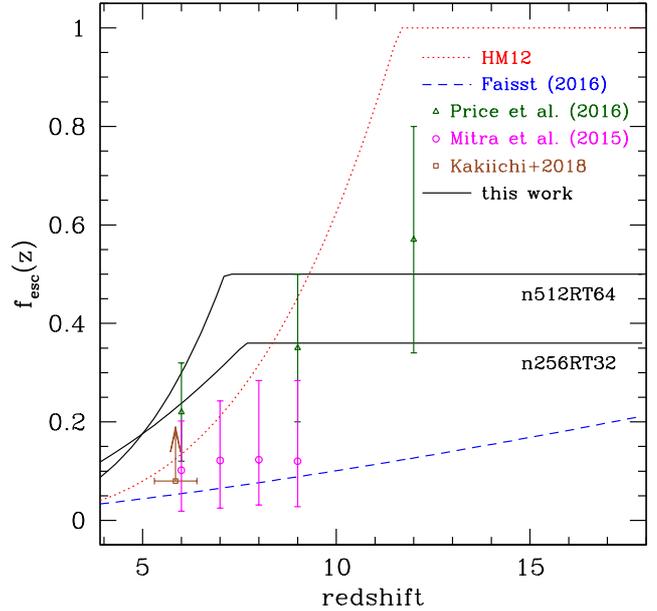}}
}
\caption{Our adopted models for $\fescg(z)$ versus several other models from
the literature.  Our simulations adopt values of $\fescg$ that are consistent
with inferences from semi-analytical models but slightly higher than more direct 
observational inferences.  They also 
require higher $\fescg$ if the resolution of the radiation transport solver is 
higher, as indicated by the numbers beneath the solid curves.
}
\label{fig:fesc}
\end{figure}

In Figure~\ref{fig:fesc}, we compare our model for $\fescg(z)$ versus other recent
determinations.  Our simulations assume a volume-averaged $\fescg$ that exceeds what
is assumed by the~\citet{haar12} model for $5<z<9$, but is lower at earlier times.  
At all redshifts, the adopted dependence is consistent with what~\citet{pric16} infer 
when using a non-parametric analysis to match the \emph{Planck} 2015 TT,TE, EE+lowP 
measurements using the~\citet{bouw15} UV luminosity function.  The disagreement with 
respect to the complementary semi-analytical modeling by~\citet{mitr15} is more
pronounced, likely reflecting an overall lower $\dot{\rho}_*$ in our model 
(see Figure~\ref{fig:lfs}).  Perhaps more troublingly, 
the purely-observational inference on the mean escape fraction of galaxies with stellar 
mass $\mstar = 10^9\msun$ based on H$\alpha$ observations
by~\citet{fais16} is lower than any of the models that are tuned to reproduce 
reionization.  This includes inferences from updated analytical models, which tend 
to require $\fesc>0.1$~\citep{mada17}.  While galaxies with $\fesc\geq0.4$ have been
reported~\citep{izot18,vanz18}, they do not seem to be the norm.  On the other hand,
they are consistent with the lower limit of 0.08 recently inferred from observations 
of transmissive regions along a sightline to a $z=6.42$ QSO~\citep{kaki18}, which
in turn is slightly inconsistent with~\citet{fais16}.  Our work reinforces the 
already well-known tension between direct observations of ionizing photon escape 
and the requirements of reionization (see also~\citealt{rutk17}).  We return to
this point in Section~\ref{ssec:disc}.

\subsubsection{The Quasar UVB}\label{ssec:qsoModel}
The space density of bright quasars (QSOs) is too low for a representative number of 
them to occur in our limited simulation volumes (see also~\citealt{gned14}).  Nonetheless, their
contribution must be included as they account for an increasing fraction of the total
flux at lower redshifts ($z<5$) and high energies ($h\nu > 4$ Ryd).  We model the
QSO UVB as a spatially-homogeneous field, appropriate for the case in which they
are predominantly distant, using a volume-averaged radiation transport calculation.

We take the emissivity at 1 Ryd as a function of redshift from Equation 9 
of~\citet{mant17} and assume that quasars make no contribution above $z=8$.  We use 
the~\citet{luss15} continuum slope to obtain the
emissivity at higher energies.  This leaves only the question of what fraction 
$\fescq$ of ionizing photons escape into the IGM.  While it is traditionally assumed 
that $\fescq=1$~\citep[for example,][]{haar12}, in principle it is possible that 
$\fescq$ varies with luminosity and redshift for QSOs just as in the case of galaxies.
Indeed, recent observations indicate that $\fescq\approx0.7$ in the case of bright 
QSOs at $z\sim3$~\citep{cris16}.  Furthermore,~\citet{mich17} report observations
of 14 faint QSOs at $z\sim3$ that are consistent with much lower values for $\fescq$ 
(including zero). At slightly higher redshift,~\citet{graz18} find $0.44 < \fescq < 1.0$ 
with a mean value of $\langle \fescq \rangle = 0.74$ for 16 faint AGN at $z>4$.
Hence while it is becoming clear that $\fescq < 1$, there is not yet
observational guidance as to what one should
adopt for $\fescq$ in the case of high-redshift QSOs.  For the present, we key off
of~\citet{cris16} and assume that $\fescq= 70\%$ for QSOs as an ensemble.  In the 
long term, it may be that joint observations of QSOs and the IGM are needed to 
infer $\fescq$, just as in the case of galaxies.  

At each iteration of our radiation-ionization solver, we update the simulation's 
QSO UVB via Equation 1 of~\citet{haar12}.  We adopt for the QSO opacity the 
volume-averaged opacity as a function of frequency over our entire simulation 
volume.  The QSO UVB's ability to ionize and heat dense gas is suppressed via our 
subgrid self-shielding prescription in the same way as the galaxy UVB.

\subsection{Assumptions versus Predictions}
While the goal of cosmological simulations is to compute as much of structure 
formation as possible with as few assumptions as possible, dynamic range limitations
nonetheless force them to adopt a number of inputs, a few of which are calibrated
empirically.  As it is sometimes difficult to keep track of what cosmological 
simulations predict and what they assume, we here list the inputs into our simulations 
that we do \emph{not} tune:
\begin{enumerate}
\item The stellar initial mass function\label{param:imf}
\item The intrinsic stellar emissivity as a function of wavelength and metallicity\label{param:epsilon}
\item The QSO emissivity at 1 Ryd as a function of redshift, the QSO\label{param:qso}
continuum slope, and $\fescq$
\item The metal yield from Type II and Type Ia SNe and from AGB stars\label{param:yield}
\item The delay time distribution of Type Ia SNe\label{param:Iadtd}
\item The mass-loading factor $\etaw(\mstar)$ of galactic outflows\label{param:eta}
\end{enumerate}
In the past, $\etaw$ was treated as a relatively free parameter with its normalization
adjusted to match observations of galaxies and metal absorbers~\citep{oppe08,oppe09}.  
However, we now adopt the~\citet{mura15} dependence of $\etaw$ on $\mstar$, which in turn 
derives directly from high-resolution simulations.  Hence this scaling does not 
involve free parameters.  The following inputs are treated more freely:
\begin{enumerate}
\setcounter{enumi}{6}
\item The ionizing escape fraction from galaxies $\fescg$\label{param:tune_fesc}
\item The velocity of galactic outflows\label{param:tune_vgas}
\item The decoupling and recoupling criteria for outflowing gas\label{param:tune_recouple}
\end{enumerate}
The history of star formation is less 
sensitive to~\ref{param:tune_vgas} than to~\ref{param:eta}.  For example, in
low-resolution simulations, halving the normalization in~\ref{param:eta} 
increases the stellar mass density at $z=6$ by a factor of 2.13 whereas halving
the normalization in~\ref{param:tune_vgas} increases it by a factor of 1.53.
Meanwhile,~\ref{param:tune_recouple} has almost no effect on it at all 
(see~\citealt{scha10} for a more thorough discussion of subgrid parameters in 
cosmological simulations).  Hence \emph{by far the most significant free 
parameter in our simulation, and the only one that is tuned to match
observations in a precise way, is $\fescg$}.

\subsection{Table of Simulations}

\begin{table}
\begin{tabular}{l|cccccc}
name & RT grid$^1$ & $A^2$ & $\fescmax^2$ & $M_g/\msun^3$ & $\Gamma_{-12}^4$ & $\gamma/$H$^5$\\
\hline
n256noRT 	& HM12		& --	& --  	& 20.5  & (0.301)	&-- \\
n256RT32 	& $32^3$ 	& 1.95 	& 0.36	& 20.5  & 0.100		&2.44\\
n256RT64 	& $64^3$ 	& 3.4 	& 0.5 	& 20.5 	& 0.385		&3.02\\
n512noRT 	& HM12		& --	& --  	& 2.56 	& (0.301)	&--\\
n512RT64 	& $64^3$ 	& 3.45	& 0.5 	& 2.56 	& 0.272		&2.69\\
\end{tabular}
\caption{Our simulations. All have the same initial conditions spanning a 
cubical volume $12\hmpc$ to a side. $^1$Number of grid cells in RT grid; 
$^2$cf. Equation~\ref{eqn:fesc}; $^3$ Gas particle mass in $10^5\msun$; 
$^4$ $\hi$ photoionization rate at $z=5.75$
$^5$ Ionizing 
photons emitted per hydrogen atom when $\xhiv<1\%$.}
\label{table:sims}
\end{table}

In Table~\ref{table:sims}, we summarize our simulations.
It is critical for reionization simulations to resolve star formation in halos
down to the hydrogen cooling limit of roughly $10^8\msun$.  Star formation does
occur in the abundant population of lower-mass halos, but it is stochastic and 
inefficient~\citep{wise14}, and probably does not dominate the overall history of 
reionization~\citep{ricc04,grei06,ahn12,chenp17}.  Our highest-resolution simulations model $10^8\msun$ 
halos with roughly 62 dark matter and gas particles.  This may fall somewhat short
of completely resolving the lowest-mass hydrogen cooling halos, but it is more than
sufficient in ionized regions, where the minimum halo mass for efficient gas accretion
rapidly grows to $\sim10^9\msun$~\citep{okam08}.  All of our simulations span the 
same $12\hmpc$ cubical volume, but they use varying resolutions
for the hydrodynamic and radiation transport solvers.  Additionally, two simulations
have been run with the~\citet[][HM12]{haar12} ionizing background for comparison
with other works.  

The sixth and seventh columns present predictions that show how our simulations 
compare favorably with observational inferences from the LAF.  
In particular, the sixth column compares the volume-averaged $\hi$ 
photoionization rate at a fiducial redshift $z=5.75$; numbers in parentheses 
indicate the HM12 model.  All of our simulations are within a factor
of 3 of the HM12 model, with our high-resolution simulation closest of all.  The
seventh column reports the total number of ionizing photons per hydrogen atom 
emitted into the IGM (i.e., after $\fescg(z)$ is accounted for) 
when the volume-averaged neutral hydrogen fraction drops below 1\%; this is 
within the range 2.4--3 that is favored by observations~\citep{bolt07} and 
the updated analytical model presented by~\citet{mada17}.

\section{Results}\label{sec:results}
\subsection{Star Formation}
\begin{figure}
\centerline{
\setlength{\epsfxsize}{0.5\textwidth}
\centerline{\epsfbox{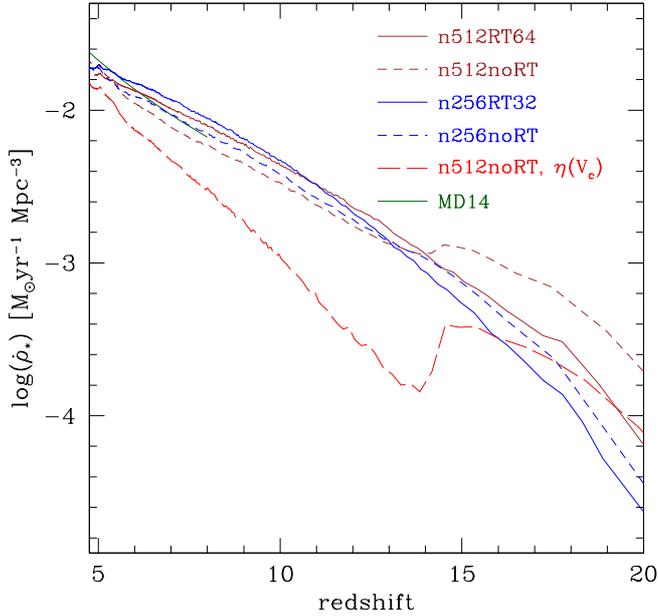}}
}
\caption{The comoving star formation rate density $\dot{\rho}_*(z)$ 
in our simulations as compared
to observations (\citealt{maddick14}; solid green curve).  Simulations that 
assume the~\citet{mura15} mass-loading factor $\etaw(\mstar)$ yield excellent 
agreement with observations out to $z=8$, with small differences driven
by photoionization heating and resolution.
}
\label{fig:madau}
\end{figure}

In Figure~\ref{fig:madau}, we compare the total simulated star formation rate
density $\dot{\rho}_*$ versus the recent synthesis of available observations 
by~\cite{maddick14}.  The simulated $\dot{\rho}_*$ is computed as
the total star formation rate over all gas regardless of the host galaxy's 
luminosity.  The good agreement between all three solid curves suggests 
that, broadly, adopting $\etaw(\mstar)$ from Equation 8 
of~\citet{mura15} leads to excellent agreement with the observationally-inferred
trend out to $z=8$, the highest redshift where observations were 
available when this fit was calibrated~\citep[see also][]{dave16,font17}.  
By contrast, comparing the red long-dashed and brown short-dashed curves 
reveals that adopting the dependence $\etaw(V_c,z)$ on circular velocity and 
redshift (Equations 4 and 5 of~\citealt{mura15}) leads to drastically 
oversuppressed star formation at early times (note that it is not surprising
that this formulation of $\etaw$ may be less accurate at early times given that 
it was calibrated from a small number of simulated halos at $z=$1--3).

Photoionization heating suppresses star formation in halos less massive than
$\sim10^9\msun$~\citep{okam08}, leading to the possibility of a rapid dip in 
$\dot{\rho}_*$ if the photoionization heating rate 
jumps~\citep{bark00}.  In simulations that assume the~\citet{haar12} UVB, 
this occurs at $z=15$.  The effect is amplified by the nonlinear coupling 
between photoionization heating and outflows~\citep{pawl09,finl11}, leading to 
a particularly dramatic drop when very strong winds are modeled at 
high resolution (red long-dashed curve).  Photoheating begins at earlier times 
in models that include a more realistic, extended reionization history because 
the UVB appears as soon as the galaxies do (solid versus short-dashed 
blue curves).  Hence while the predicted amplitude of the dip is sensitive to our 
assumptions regarding $\fescg$ and galactic outflows, we do not see evidence
for it in our most realistic calculation (solid brown curve).

\begin{figure}
\centerline{
\setlength{\epsfxsize}{0.5\textwidth}
\centerline{\epsfbox{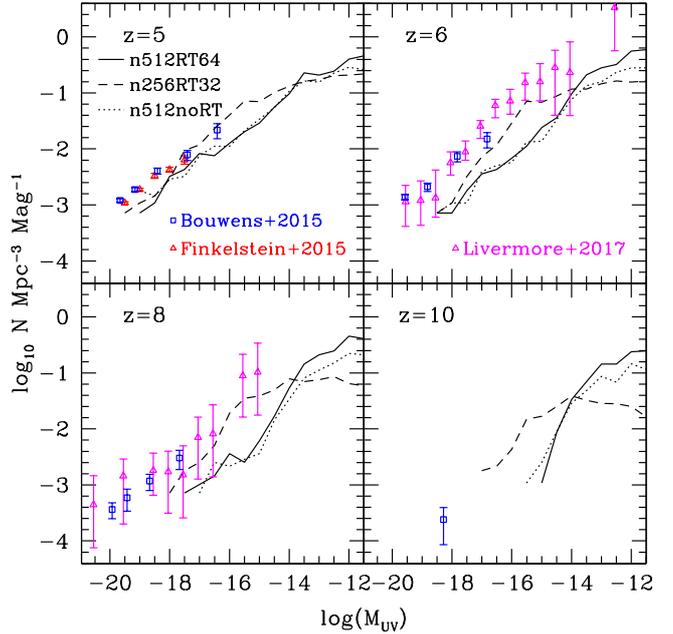}}
}
\caption{UV luminosity functions at four selected redshifts as compared to
observations.  At our lower resolution (long-dashed), agreeement is sound at all redshifts 
where measurements and simulations overlap.  At higher resolution (solid), the
simulation increasingly underproduces observations at $z>5$ owing to pre-processing
of gas in small systems.  Re-running the pl12n512 simulation using the HM12 
background (dotted) yields almost the same UV LF.
}
\label{fig:lfs}
\end{figure}

The good agreement between the simulated and observed $\dot{\rho}_*$
in Figure~\ref{fig:madau} hides an inconsistency in dynamic 
range: Whereas the~\citet{maddick14} curve results from integrating over 
all luminosities down to 0.03$L^*$ ($\MUV\approx-16.4$ at $z\sim6$), our 
simulations include (and, at $z\geq8$, are dominated by) fainter systems 
while omitting brighter ones owing to volume limitations.  A more constraining 
test is a comparison with the observed UV luminosity function (UV LF).  We 
compute the simulated LFs in a standard fashion: From each snapshot, we 
extract galaxies using 
{\sc skid}\footnote{http://www-hpcc.astro.washington.edu/tools/skid.html
}.  Each simulated galaxy consists of hundreds to thousands of star
particles, each of which has its own metallicity and age.  We compute each
star particle's rest-frame spectrum (in $\rm{ergs}\, \rm{sec}^{-1}\, \rm{Hz}^{-1}$) 
using version 2.3 of the Flexible Stellar Population Synthesis 
library~\citep{conr09}, interpolating to its age and metallicity.  Each
simulated galaxy's spectrum is then given by a sum over the spectra of its
constituent star particles.  We multiply synthetic spectra by a factor 
(0.63/0.67) in order to adjust from a~\citet{chab03} to a~\citet{krou01} 
IMF (these scaling factors are taken from Section 3.1 of~\citealt{maddick14}).
Finally, we compute the rest-frame 1500\AA~luminosity by multiplying the 
spectrum by an idealized 1500\AA~bandpass with 
a 15\% bandwidth and a full width at half-maximum of 225\AA~and integrating.  
We neglect both dust extinction and nebular continuum emission.

In Figure~\ref{fig:lfs}, we compare our simulated LFs against 
recent observations~\citep{bouw15,fink15,live17}.  
In the case of our lower-resolution, n256RT32 simulation (dashed), agreement 
is reasonable throughout $z=8\rightarrow5$.  This is a nontrivial success 
given that the most important parameterized input is $\etaw(\mstar)$, which was in
turn \emph{not} empirically-calibrated but rather adopted directly from the
high-resolution {\sc FIRE} simulations~\citep{mura15}.  At higher resolution 
(solid), strong
winds and efficient star formation in low-mass systems pre-process a great
deal of gas at early times.  This can clearly be seen in Figure~\ref{fig:madau}
as a boosted overall $\dot{\rho}_*$ at $z>11$ (compare the solid
brown and blue curves).  Additionally,
more gas condenses into small systems at high resolution, enhancing the 
faint end of the LF at $z\geq6$.  These effects both reduce 
the predicted abundance of systems with $\MUV < -14$ in our high-resolution
simulation at $z>5$.  

The dotted curve in Figure~\ref{fig:lfs} verifies that ensemble predictions 
regarding galaxies brighter than
$\MUV=-14$ are nearly unchanged if we run our simulations using the HM12 UVB 
instead of our radiation transport solver.  Such galaxies inhabit haloes whose
virial temperature exceeds the photoheated IGM's temperature, hence they
are only weakly sensitive to hydrogen reionization.

\begin{figure}
\centerline{
\setlength{\epsfxsize}{0.5\textwidth}
\centerline{\epsfbox{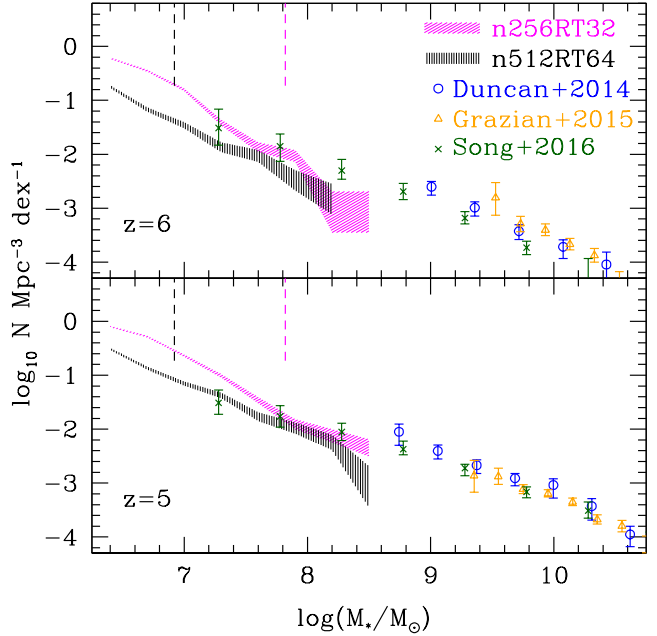}}
}
\caption{Stellar mass functions at $z=5$ and 6 in our simulations versus 
observations.  Vertical tickmarks indicate 64-star particle mass resolution 
limits for our two simulations.
While extrapolation from shallower surveys~\citep{dunc14,graz15} 
suggests that our simulations underproduce stars by up to a factor of four, 
the deeper~\citet{song16} measurements are in excellent agreement with the
simulations at $z=5$ and reasonable agreement at $z=6$.  
}
\label{fig:mfs}
\end{figure}

If the galaxies' UV luminosities are underproduced at $z>5$ as suggested 
by Figure~\ref{fig:lfs}, then their stellar masses are
also expected to be low.  We evaluate whether this is the case in 
Figure~\ref{fig:mfs}, where we compare the predicted stellar mass functions (SMFs)
at $z=5$ and 6 against observations~\citep{dunc14, graz15, song16}.  The level
of agreement is excellent at $z=5$. At $z=6$, agreement
is good although the SMF normalization may be underproduced by as much as
a factor of two.  We conclude from Figures~\ref{fig:lfs}--\ref{fig:mfs} that
our simulations produce roughly the correct number of stars at $z=5$ but may
undeproduce them at higher redshifts.

As in Figure~\ref{fig:lfs}, a tendency for the lower-resolution n256RT32 
simulation to produce more bright, massive galaxies and fewer faint, low-mass 
ones is evident.  The discrepancy diminishes above its 64-star particle 
resolution limit of $\scnot{6.56}{10}\msun$.  It occurs because, at the 
n256RT32 simulation's resolution, gas first reaches star-forming densities 
in $\sim10^9\msun$ halos.  In the n512RT64 
simulation, gas first condenses more realistically into lower-mass galaxies, 
which in turn may be comparable to the progenitors of the Milky Way dwarf 
satellites~\citep{finl17}.  This long tail extending to unobservably low masses 
is not unique to \td; it is a generic prediction also seen, for example,
in the Sherwood Simulations~\citep{keat16}.

\begin{figure}
\centerline{
\setlength{\epsfxsize}{0.5\textwidth}
\centerline{\epsfbox{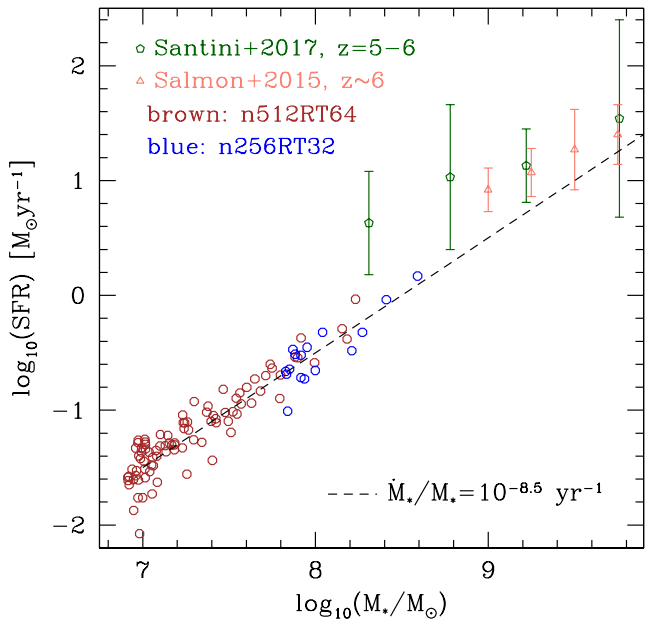}}
}
\caption{The predicted and observed relations between stellar mass and star 
formation rate at the end of the reionization epoch.  Our simulations predict a 
constant specific star formation rate of $10^{-8.5}$yr$^{-1}$ (dashed line) that
is insensitive to resolution limitations. This agrees with the more massive 
observed systems but underproduces low-mass ones at the 1--1.5$\sigma$ level.  
}
\label{fig:mstarSfr}
\end{figure}

Figures~\ref{fig:lfs}--\ref{fig:mfs} indicate that both the SFR and $\mstar$ 
are low for a given halo mass.  A simple way to verify that neither of these 
discrepancies is dominated by observational bias (for example, owing to the 
shape of the star formation history that is assumed when deriving stellar mass) 
is to compare the predicted and observed $\mstar$-SFR relationships:
if the simulated SFRs are simply oversuppressed, then their integral will be 
suppressed by the same factor, leaving the ratio intact.  We show this comparison
at $z\sim6$ with recent observations~\citep{sant17, salm15} in 
Figure~\ref{fig:mstarSfr}.  Note we plot the simulated ``current mass" on 
the x-axis; that is, the mass that remains in long-lived stars.  We include only
galaxies that contain at least 64 star particles.

The simulated loci overlap well, indicating that the ratio SFR/$\mstar$
does not suffer from resolution limitations.  Comparing to observations is difficult 
owing to the poor overlap in dynamic range, but broadly we find that the predicted 
SFR is no more than 1--1.5$\sigma$ below observations.  This is plausible agreement 
given that photometric techniques may artificially flatten the $\mstar$-SFR 
relation~\citep{salm15}.

In summary, while adopting the~\citet{mura15} outflow scalings yields 
remarkable agreement with the total $\dot{\rho}_*$, the agreement
may mask a tradeoff whereby the simulations predict a great deal of star 
formation in faint galaxies at the expense of the observable population, 
which are in turn somewhat oversuppressed at $z>5$.  
This effect does not necessarily reflect an error in the~\citet{mura15} 
scalings.  First, we have adopted only their mass loading factor; our adopted
outflow velocities are higher because our low resolution does not treat
the launch radius or velocity distribution realistically.  Second, 
the ultimate fate of ejected gas likely depends on the host halo and 
environment, whereas we tie it exclusively to the stellar mass, potentially 
permitting overly-efficient outflows.  It is easy to imagine explicitly
tuning our outflows in order to improve the level of agreement in 
Figures~\ref{fig:madau}--\ref{fig:mstarSfr}; we will return to this in
Section~\ref{ssec:disc}.

\subsection{Metal Enrichment}\label{ssec:metals}
\begin{figure}
\centerline{
\setlength{\epsfxsize}{0.5\textwidth}
\centerline{\epsfbox{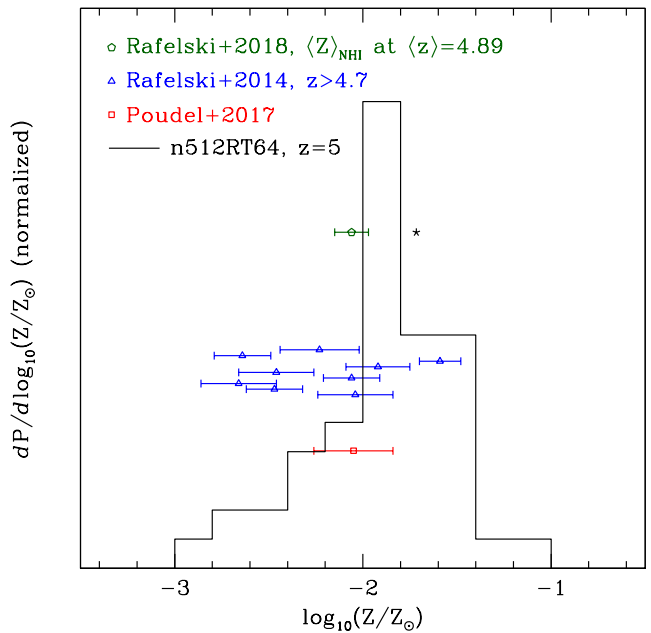}}
}
\caption{Comparison between the predicted and observed DLA metallicity
distributions at $z\sim5$.  The solid histogram gives the predicted 
probability density normalized to unit area and displayed with a linear
y-axis.  Squares and triangles represent observed DLA metallicities at 
$z>4.7$, offset arbitrarily on the y-axis for visibility.  The green pentagon 
and black star represent $N_\hi$-weighted mean metallicities 
from an observed sample at $\langle z \rangle = 4.89$ (Rafelski 2018, private 
communication) and the simulatation at $z=5$, respectively.  There is good 
overlap between the observed and predicted metallicity ranges although the 
simulation may be biased high.  
}
\label{fig:dPdZDLA}
\end{figure}

As we have updated our prescriptions for metal enrichment, it is timely to verify 
that the simulated CGM has a realistic metallicity.  
The most robust metallicity measurements come from damped Lyman-$\alpha$ absorbers 
(DLAs), which trace low-ionization gas in conditions where ionization corrections 
are negligible and metal mass fractions may be measured directly~\citep{rafe12}.  
Recent observations have uncovered DLAs out to redshifts of 
4.5--5.5~\citep{rafe12,rafe14,poud18}.  We now compare these measurements directly 
to our n512RT64 simulation at $z=5$.

We identify simulated DLAs by searching for regions along simulated
sightlines where the total $\hi$ column density $\log(N_\hi/\cm^{-2})$ within a 200$\kms$ 
segment exceeds 20.2.  This returns d$N=49$ DLAs over an absorption path length\footnote{
The ``absorption path length" $dX\equiv(1+z)^2{H_0}/{H(z)}dz$ is used to 
quantify the density of absorbers along a sightline in a way that 
factors out cosmological expansion~\citep{bahc69}: for a non-evolving population, 
the number of objects per absorption path length $dN/dX$ does not vary with 
redshift.} 
of d$X=351.01$.  
The implied number density of $dN/dX=0.14$ is consistent with the 68\%
confidence range of 0.0638--0.1435 recently measured at 
$z=4.83$--5.00~\citep[][Table A6]{bird17}.  
We then compute each simulated DLA's metallicity as the
$\hi$-weighted average of the summed mass fraction in C, O, Si, and Mg within each pixel. 
We normalize to solar by assuming a total solar mass fraction of 0.00958 for these 
elements~\citep{aspl09}.  The distribution of simulated DLA metallicities computed in this 
way is displayed as a normalized histogram in Figure~\ref{fig:dPdZDLA}.  It spans over 
two orders of magnitude, with an $\hi$-column weighted mean of -1.72 (five-pointed star).  

To test this predicted distribution, we use observed metallicities of $\hi$-selected 
systems with $\log(N_\hi)\geq20.2$ at $z>4.7$; these are shown as blue and red colored 
points with errors (the positions of observed systems on the $y$-axis is arbitrary).  
For the~\citet{poud18} system at $z=4.829$, we adopt the oxygen metallicity, whereas 
for the~\citet{rafe14} systems we adopt the bulk metallicity from the final column of 
their Table 1.  For the present, we neglect differences between the particular metal 
species that are used because the resulting metallicities, when normalized to solar, 
generally agree within $1\sigma$.

There is broad agreement between the simulated and observed DLA metallicity 
ranges.  In detail, however, the simulations may overproduce the CGM metallicity:
Computing the $\hi$-weighted mean metallicity of a sample of high-redshift DLAs
with a mean redshift $\langle z \rangle = 4.89$ yields a mean metallicity
normalized to solar of $\log(Z/Z_\odot)=-2.06\pm0.09$ (green pentagon; Rafelski 2018, 
private communication) whereas our simulation predicts -1.72 in the same units 
(black five-pointed star).
A similar discrepancy of -2.06 (observed) versus -1.86 (simulated) is obtained
 when the \emph{median} simulated metallicities at $z=5$ are compared with the 
median observed metallicities of DLAs at $z>4.7$ from~\citet{rafe14} 
and~\citet{poud18}, so the offset is robust to the choice of statistic.

The existence and sense of this offset are likely robust to systematic 
uncertainties for several reasons.  First, observations suggest that 
metal-poor DLAs are generally $\alpha$-enhanced~\citep{cook11}.  If true, and
if the observed metallicities are weighted towards $\alpha$-elements, then 
correcting for this could push the total observed metallicities down.  
Second, our fiducial n512RT64 simulation underproduces stars---and therefore
metals---at early times (Figure~\ref{fig:lfs}).  Although the effect is weak
by $z<6$, correcting for it would increase the predicted metallicities.  Finally,
the simulated UVB may be too weak at $z<6$ (Section~\ref{ssec:HReion}).  
Qualitatively, this will bring overly-rarified and presumably underenriched 
CGM gas into the simulated DLA sample; correcting for it would reduce the
predicted DLA number density and increase their metallicities.

It will be important to confirm whether the factor-of-two offset survives 
when these details are taken into account and the observed sample is larger.  If 
so, then it suggests the need for lower metal yields (perhaps by reducing the 
hypernova fraction; Section~\ref{sec:sims}) or stronger feedback at earlier times.  
Given that DLA metallicities span two orders of magnitude, however, and
given that the simulations were in no wise calibrated against this observation,
the preliminary agreement in Figure~\ref{fig:dPdZDLA} is unexpectedly good.

\subsection{The History of Hydrogen Reionization}\label{ssec:HReion}

\begin{figure}
\centerline{
\setlength{\epsfxsize}{0.45\textwidth}
\centerline{\epsfbox{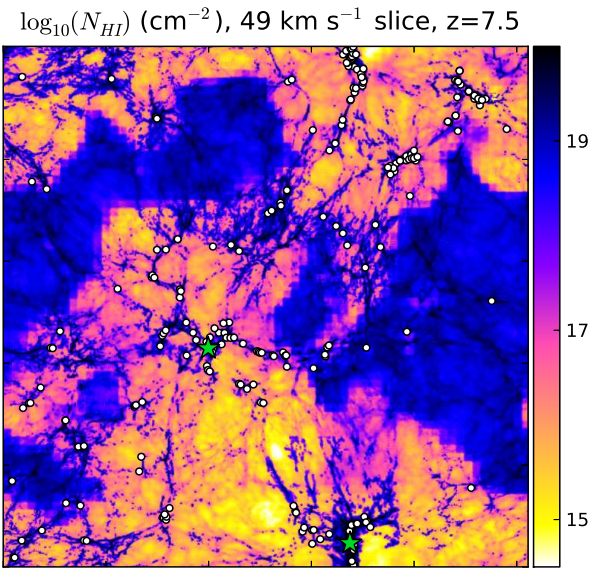}}
}
\caption{The $\hi$ column density through a $49\kms$ slice at $z=7.5$, when
the predicted neutral hydrogen volume fraction is 33\%.  The side length
is $12\hmpc$ comoving Mpc, or 2.08 proper Mpc at this redshift.  Points and
stars indicate the positions of galaxies that are fainter and brighter than
$\MUV=-15$, respectively.  Large-scale ionized regions are threaded by dense 
filaments, which clearly trace the positions of faint galaxies.  Note that 
the slightly pixellated appearance of the ionization fronts reflects the 
finite resolution of our radiation transport solver, which is coarser than 
the hydrodynamic solver.  
}
\label{fig:NHIGals}
\end{figure}

A central goal of our development efforts involves using our star 
formation and radiation transport model to generate a realistic reionization
history.  It is of obvious interest to verify that our model does so.  In
order to build intuition regarding the role of faint galaxies in driving
hydrogen reionization, we map in Figure~\ref{fig:NHIGals} the spatial 
relationship between galaxies and
$\hii$ regions at $z=7.5$ in our high-resolution simulation.  At this
redshift, the volume-averaged neutral hydrogen fraction $\xhiv$ is 33\%.  
The color bar codes for the $\hi$ column density through a slice of 
simulation $49\kms$ thick and $12\hmpc$ to a side.  The overlap of $\hii$
regions is well underway in this snapshot.  On the left and right sides
there remain Mpc-scale regions that are both partially-to-completely 
neutral and devoid 
of galaxies.  Between the neutral regions lie several $\hii$ regions that 
are about to merge.  These are threaded by filamentary Lyman-limit systems 
of dense, still-neutral gas that host the galaxies, which are indicated by
white dots or green stars.  Only two of these galaxies are observable 
($\MUV < -15$), reinforcing that, in the standard galaxy-driven reionization
scenario, most of the activity is in very faint systems.  The ratio of the 
volume-averaged and mass-averaged neutral hydrogen fractions at this redshift
is $\xhiv/\xhim = 0.87$, indicating that the voids are more ionized than the
filaments.  In other words, the reionization topology is at this point 
``outside-in", the precursor to the final stage where the filaments are
all that remain neutral~\citep{finl09}.

\begin{figure}
\centerline{
\setlength{\epsfxsize}{0.5\textwidth}
\centerline{\epsfbox{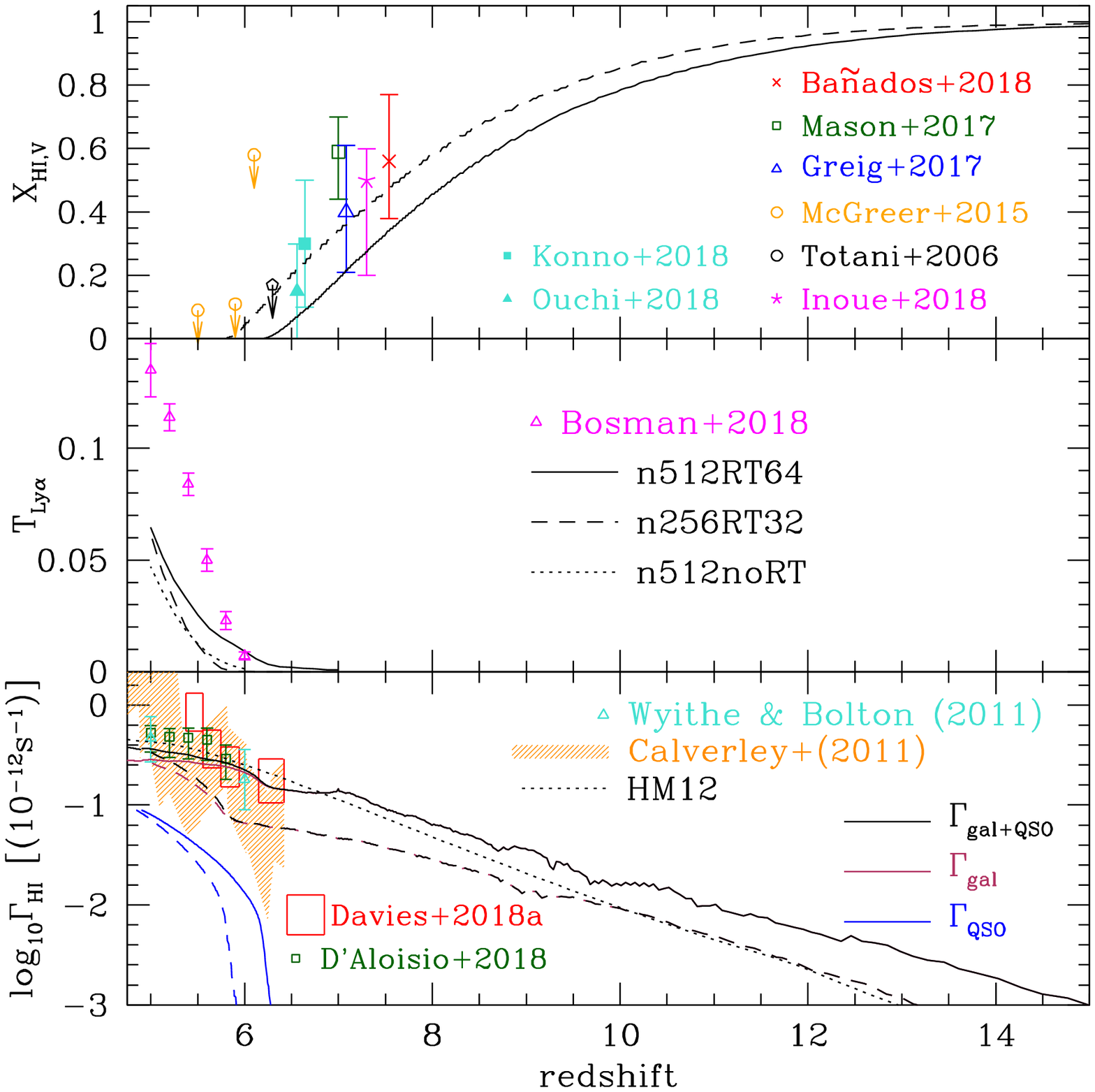}}
}
\caption{\emph{Top:} Volume-averaged neutral fraction as a function of 
redshift.  Observations: The equivalent widths of Lyman Alpha Emitters (LAEs)
~\citep[green square; magenta star:][]{maso18,inou18}; LAE 
luminosity function~\citep[turquoise square][]{konn18} and 
clustering~\citep[turquoise triangle:][]{ouch18}; 
the damping wing of a QSO at $z=7.09$
~\citep[blue triangle,][]{grei17} and $z=7.54$~\citep[red cross,][]{bana18}; 
the Lyman-$\alpha$ and Lyman-$\beta$
dark pixel fractions~\citep[orange circles,][]{mcgr14}; and a GRB damping 
wing~\citep[black pentagon,][]{tota06}. \emph{Middle:} Mean transmitted 
flux fraction in the Lyman-$\alpha$
forest versus observations~\citep{bosm18}.
\emph{Bottom:} Volume-averaged HI photoionization rate
compared with observations.  All plotted uncertainties are $1\sigma$.
}
\label{fig:HReion}
\end{figure}

Turning from the qualitative to the quantitative, we have insured a realistic 
reionization history by calibrating our two-parameter model for $\fescg(z)$ 
to match simultaneously observations of the optical depth to Thomson 
scattering $\taues$ and the inferred total ionizing photon production rate 
at $z=5$ (this is in fact our \emph{only} empirical calibration).  In 
Figure~\ref{fig:HReion}, we confirm that this
calibration yields a volume-averaged $\hi$ fraction that passes
below 50\% at $z\sim$8--8.5 and reaches 1\% around $z\sim6$ (top panel).
The corresponding predicted integrated optical depths to Thomson scattering 
are 0.05463 and 0.06048 for the low- (dashed) and high-resolution (solid) 
simulations, respectively, both in excellent agreement with the observed range 
of $0.058\pm0.012$~\citep{plan16b}.  Given that
we have independently calibrated $\fescg$ for these two resolutions, this is
not a demonstration of resolution convergence.  Instead, it shows that some
of the lack of convergence associated with incompletely modeling 
radiation transport on small scales can be compensated for by allowing $\fescg$
to vary.

Our simulated hydrogen reionization histories are consistent with the strong
upper limits on $\xhiv$ derived from the dark pixel fraction in the Lyman-$\alpha$
and Lyman-$\beta$ forests of high-redshift quasars (circles) as well as the inferred
upper limit from the damping wing of GRB 050904 (pentagon).  They are in
agreement with the neutral fraction inferred from the damping
wing of QSOs at $z=7.09$ (blue triangle) and at $z=7.54$ (red cross), although
both of these prefer the somewhat later reionization history of our n256RT32 
simulation.  We do not show the recent re-analysis of the $z>7$ QSO damping wings 
by~\cite{davi18b}, but we note that, while it yields $\xhiv$ values that are 
consistent with previous work, they are slightly higher overall, again favoring 
our n256RT32 simulation.  Our simulated $\xhiv$ are 1.5--2$\sigma$ low compared
to inferences based on the observed Lyman-$\alpha$ equivalent width
distribution of Lyman break galaxies by~\citet[][square]{maso18}, but
in agreement with the complementary analysis by~\citet[][5-pointed star]{inou18},
emphasizing the challenging nature of this measurement.  They are also consistent
with inferences from clustering of Lyman-$\alpha$ emitters~\citep{ouch18}
and their luminosity function~\citep{konn18}. Overall, we conclude that both of 
our simulations yield broadly realistic reionization histories
although current observations may show a slight systematic preference for
the slower reionization history of our n256RT32 simulation.

An even more demanding test of our model than the reionization history is
the nature of the post-reionization UVB, which is well-measured from
observations of the LAF.  In the middle panel of 
Figure~\ref{fig:HReion}, we compare the mean transmission at the 
Lyman-$\alpha$ edge predicted from 
our simulations $T_{\mathrm{Ly}\alpha}$ versus observations by~\citet{bosm18}.
We compute the predicted transmission by simply averaging over simulated 
Lyman-$\alpha$ spectra at each redshift.
For both radiation transfer simulations, the predicted transmission begins 
rising at roughly the correct redshift, but it grows more slowly 
than observations indicate.  If we adopt the HM12 UVB (n512noRT; dotted 
curve), then the mean transmission is likewise underproduced.

We do not believe that resolution limitations are the primary source of this
discrepancy.  ~\citet{bolt09} have shown that the underdensities that dominate
Lyman-$\alpha$ transmission at $z\geq5$ are rare and small, requiring both
large volumes and high resolution.  They suggest that, at our mass resolution 
and volume (see the green curve in their Figure 1), the mean Lyman-$\alpha$ 
transmission is only marginally resolved and biased somewhat low at $z=5$.
We have repeated their experiment by running simulations with a 
spatially-homogeneous HM12 UVB down to $z=5$ in volumes spanning 6, 12, and
18 $\hmpc$, each with a gas particle mass of $\scnot{2.56}{5}\msun$.  At
$z=6$, the transmissions are (0.0012, 0.0015, 0.0013). At $z=5$, they are 
(0.042, 0.047, 0.046).  Likewise, increasing the gas
particle mass within the $12\hmpc$ volume by a factor of 8 (i.e., using 
$2\times256^3$ instead of $2\times512^3$ particles) reduces the 
transmission at $z=5$ from 0.047 to 0.042.  These experiments suggest that, 
over the (limited) range of volumes and resolutions that we consider, we 
expect no more than a $\sim10\%$ impact of resolution limitations on the 
mean transmission.  

Our results are supported by~\citet{onor17}, who recently performed a 
resolution convergence study of the LAF at $z=$5--6 in 
simulations that assume an optically-thin UVB.  Their findings imply that the mean 
transmission in our n512RT64 simulation may be biased low by $\sim5\%$ owing to
mass resolution limitations while being biased high at the $\sim$<1\% level
owing to limited simulation volume.  Likewise, the tests reported in Appendix
A of~\citet{dalo18} indicate that our simulated mean transmission is
converged to 10\% with respect to resolution.

We speculate that, instead, the UVB may be strengthening too slowly in our
simulations despite the agreement in the UVB amplitude (bottom panel), 
reflecting either uncertainty in $\fescg(z)$ or systematic error in the 
observed UVB amplitude.  Our 
inability to reproduce the rapid evolution in the observed mean transmission 
even when effects such as UVB fluctuations and scatter in the $\rho-T$ 
distribution (Figure~\ref{fig:rhoT}) are treated accurately echoes the findings 
of~\citet{bosm18}, and seems to be a generic problem afflicting
cosmological simulations in which the adopted UVB amplitude matches 
observations.

Our results also echo superficially those of \citet{gned17}, who performed a 
detailed comparison between the predicted statistical properties of the 
LAF from the
pathbreaking {\sc CROCS} simulation suite versus observations.  They also 
found that their simulations yielded an IGM that was somewhat too 
opaque (as quantified by the widths of Lyman-$\alpha$ transmission peaks), 
though likely for very different reasons.  While their spatial resolution is 
comparable to ours ($\leq100$ physical parsecs at all $z>6$), their 
simulation boxes are 2--3.3$\times$ as wide and their predicted reionization
history ends at $z\sim7$; both of these effects could boost the predicted
mean transmission with respect to ours.  They concluded that the source of
the discrepancy was unclear, although proximity zones around foreground
quasars were one possible unaccounted-for issue.  To the extent that this
is true, we should expect our simulations to have the same problem.

In the bottom panel, we compare the predicted and observed volume-averaged 
hydrogen photoionization rate $\Gamma_{HI}$ and its contributions from galaxies 
and quasars.   Quasars contribute only $\sim$a few \% of the hydrogen-ionization 
photons at $z>5$, consistent with the consensus that is emerging in the 
literature~\citep{finl16,dalo17,pars18,qin17,hass18,mcgr18}.
For both simulations, $\Gamma_{HI}=$0.1--0.5$\times10^{-12}$s$^{-1}$ at $z\geq5$, 
in excellent agreement with most observations~\citep{calv11,wyitbolt11,davi18a,dalo18}.  
In particular, the most recent measurements of the post-reionization 
photoionization rate~\citep{davi18a,dalo18} seem clearly to prefer the 
n512RT64 model, which 
in turn closely tracks the~\citet{haar12} model.  None of the models
reproduce the sharp decline in $\Gamma_{HI}$ from $z=5.5\rightarrow6$
deduced by~\citet{davi18a}. Given that the~\citet{davi18a} measurements are
presented as a proof-of-concept based on a single sightline, and given 
that~\citet{dalo18} do not recover this jump, it is not clear whether 
this behavior is real.

In summary, the comparisons in Figure~\ref{fig:HReion} show that, maugre 
dynamic range limitations and lingering uncertainty regarding $\fescg(z)$, 
our simulations yield a reionization history \textit{and} an early UVB 
evolution that are in encouraging agreement with observations.  However,
we find interesting tensions between the implications of the different 
observations: Whereas observations of $\xhiv$ tend to prefer the slower 
reionization history of our n256RT32 simulation, observations of $\Gamma_{HI}$ 
prefer the more intense ionizing background of our n512RT64 simulation.  
Meanwhile, both simulations yield a mean transmission in the LAF
that is low compared to observations.  This is also true of the HM12
simulation, whose predicted mean transmission tracks the n256RT32
simulation even though its $\Gamma_{HI}$ tracks the n512RT64 simulation.
We will show below that the HM12 UVB and, perhaps more importantly, the
assumption of ionization equilibrium, leads to a colder IGM at
$z<6$ (see also~\citealt{puch15} and Figure~\ref{fig:rhoT}), which undoubtedly 
contributes to its lower transmission.

\subsection{The Simulated UVB}\label{ssec:uvb}

\begin{figure}
\centerline{
\setlength{\epsfxsize}{0.45\textwidth}
\centerline{\epsfbox{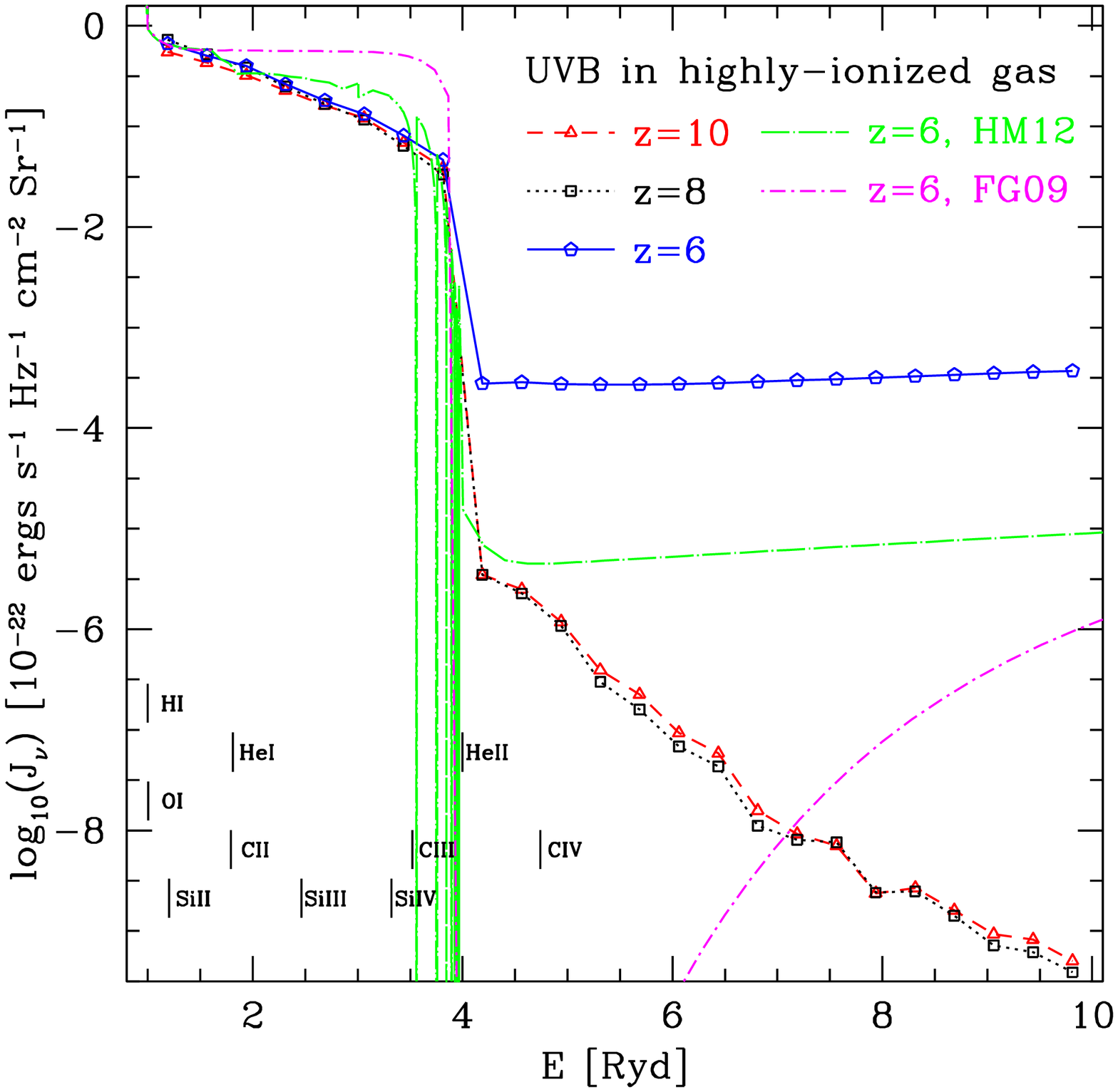}}
}
\caption{The volume-averaged mean UVB in voxels where the $\hi$ neutral 
fraction is less than 0.5\% at three representative redshifts.  The galaxy
contribution to the UVB in ionized regions is nearly redshift-independent,
while the quasar contribution dominates above 4 Ryd after $z=8$.  The
UVBs predicted by~\citet{haar12} and~\citet{fauc09} at $z=6$ are included
for comparison (long dashed-dotted green and short dash-dotted magenta, 
respectively). The ionization potentials of common high-redshift CGM 
ions are indicated at the bottom of the figure.
}
\label{fig:UVBs}
\end{figure}

As the normalization, slope, and spatial fluctuations in the 
reionization-epoch UVB remain a major source of uncertainty in large-scale 
models of reionization, it is of interest to study how it evolves and to
compare it with complementary models.  In Figure~\ref{fig:UVBs}, we show 
the UVB predicted by our n512RT64 simulation in highly-ionized regions (local 
$\xhiv < 0.005$) at three representative reionization-epoch redshifts.  
Remarkably, it is invariant at $z>6$ between the $\hi$ and $\heii$ ionization 
edges.  At higher energies, the UVB is dominated by galaxies
prior to $z=8$ and quasars afterwards.

Comparing our simulated UVB with two complementary models reveals weak
disagreement for $h\nu < 4$ Ryd and dramatic disagreement at higher
energies.  The HM12 model (long dash-dotted green) is in good
agreement with \td~below 4 Ryd although its inclusion of bound-bound
$\heii$ transitions leads to sharp absorption features between 3--4 Ryd.
By contrast, it predicts a $\heii$-ionizing background that is 
$\approx30\times$ weaker.  Meanwhile, the~\citet{fauc09} 
background (dash-dotted magenta)\footnote{\url{http://galaxies.northwestern.edu/uvb/}}
has a bluer spectral slope below 4 Ryd and is much weaker than the 
other models at higher energies.  From this figure alone, it is easy to see
that the expected $\civ$ fraction in the CGM remains subject to
much larger systematic uncertainty than, for example, $\siiv$ owing
to lack of theoretical understanding of the early stages of 
$\heii$ reionization.  Put differently, this re-iterates that 
$\civ$ offers unique insight into the early stages of $\heii$ 
reionization just as $\oi$ tracks the early stages of $\hi$ 
reionization~\citep{oh02,keat14}.

\begin{figure}
\centerline{
\setlength{\epsfxsize}{0.45\textwidth}
\centerline{\epsfbox{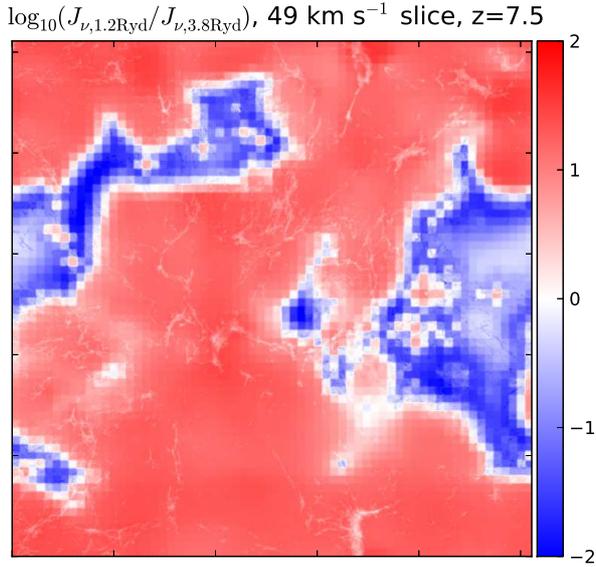}}
}
\caption{The UVB's spectral hardness as measured by the ratio of the mean
specific intensities $J_\nu$ in energy bins centered at 1.1875 and 3.8125
Ryd in the same slice as in Figure~\ref{fig:NHIGals}.  Blue and red regions
have blue and red spectral slopes between 1--4 Ryd, respectively.  The 
impact of spectral filtering is clearly seen both in HI regions and in 
self-shielded filaments.
}
\label{fig:JslopeGals}
\end{figure}

An additional strength of the \td~simulations is their ability to capture
spectral filtering of the UVB by the IGM.  This effect is potentially important
as it gives rise to small-scale fluctuations in the UVB's spectral hardness, which
impacts the IGM's $\rho-T$ relation as well as metal absorber characteristics.
We demonstrate that our simulations account for spectral filtering
in Figure~\ref{fig:JslopeGals}, which maps the ratio of the UVB strength $J_\nu$
in frequency bins centered at 1.1875 and 3.8125 Ryd.  In reionized regions, 
$J_\nu$ is flat (white) or declines with energy (red), whereas in filaments and 
neutral regions it increases (blue).  Importantly, the effect is accounted for 
both by the radiation transport solver itself and by the subgrid self-shielding
prescription, which is frequency-dependent and preferentially shields dense
regions from softer photons.

\subsection{The IGM Temperature}\label{ssec:IGMTemp}

\begin{figure}
\centerline{
\setlength{\epsfxsize}{0.45\textwidth}
\centerline{\epsfbox{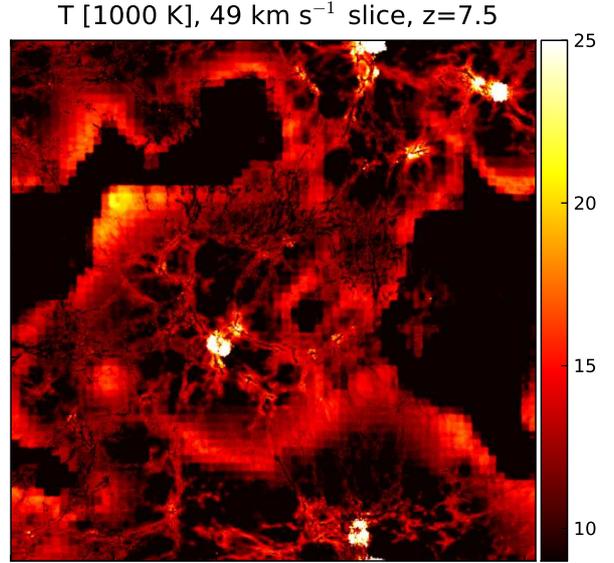}}
}
\caption{The mass-weighted mean gas temperature through the same slice as in
Figure~\ref{fig:NHIGals}.  Gas is heated by ionization fronts as well as in
dense regions near star-forming galaxies where adiabatic compression and 
outflows are in play.}
\label{fig:TGals}
\end{figure}

How hot is the reionization-epoch IGM? Its temperature sources a factor-of-two
uncertainty in reionization calculations owing to the temperature
dependence of the recombination rate: colder gas recombines faster, consuming
more photons, whereas hotter gas requires fewer photons and a generally harder
ionizing continuum.  Additionally, the reionized IGM's cooling time is longer 
than its recombination time, leading to long-lived ``footprints" of 
inhomogeneous reionization in the form of temperature 
fluctuations that source extra inhomogeneity in the IGM opacity and suppress
small-scale fluctuations in the Lyman-$\alpha$ forest flux power 
spectrum~\citep[Section~\ref{ssec:LyaPS};][]{dalo15,keat18}. A key 
advantage to using our multifrequency, on-the-fly radiation transport 
solver is that the IGM temperature evolution and its spectral filtering 
impact on the UVB are faithfully modeled.  

The maximum temperature of post-reionization gas may be estimated 
from the UVB as the mean heat deposited per photoelectron divided 
by the total number of particles per photoelectron in reionized gas.  
In the optically-thin approximation, the photoionization heating 
rates for hydrogen and helium are
\begin{eqnarray}\label{eqn:heatPerIon}
\mathcal{H}_{\hi} & = & n_\hi \int d\nu \frac{4\pi J_\nu}{h\nu} \sigma_{\nu, \hi}h (\nu - \nu_{i,\hi}) \\
\mathcal{H}_{\hei} & = & n_\hei \int d\nu \frac{4\pi J_\nu}{h\nu} \sigma_{\nu,\hei}h (\nu - \nu_{i,\hei})
\end{eqnarray}
Likewise, the photionization rates are
\begin{eqnarray}\label{eqn:ionRate}
\Gamma_{\hi} & = & n_\hi \int d\nu \frac{4\pi J_\nu}{h\nu} \sigma_{\nu, \hi} \\
\Gamma_{\hei} & = & n_\hei \int d\nu \frac{4\pi J_\nu}{h\nu} \sigma_{\nu,\hei}
\end{eqnarray}
The maximum post-reionization temperature is then
\begin{eqnarray}\label{eqn:heating}
\frac{3}{2}k_B\Delta T = \frac{1}{2}\frac{\mathcal{H}_{\hi} + \mathcal{H}_{\hei}}{\Gamma_{\hi} + \Gamma_{\hei}},
\end{eqnarray}
where the leading factor of $1/2$ on the right-hand side accounts for the 
fact that there are twice as many particles as electrons in ionized gas.  
Applying this to the simulated UVBs in Figure~\ref{fig:UVBs} leads to 
$\Delta T = $18,000--19,000K, depending on the redshift.  This corresponds 
to mean photon energies of 18.3--18.6 eV, consistent with the range that 
is required in order to match observations of the post-reionization IGM 
temperature~\citep{keat18}.  Photoionization of $\hei$ increases the 
heating rate by 6--7\% and the ionization rate (that is, total photoelectron 
production rate) by 5--6\%.  This level of heating is significantly less 
than what is required to reproduce the observed distribution of optical depths 
in the LAF (30,000 K;~\citealt{dalo15}).  Moreover, gas in 
and behind ionization fronts 
cools efficiently owing to bound-bound transitions and Compton 
cooling~\citep{mcqu16}, so recently-reionized gas will generally be even 
cooler.  Quantifying this effect is nontrivial, but a crude investigation of 
our $\rho-T$ diagrams from $z=11$--6 (not shown) reveals maximal 
post-reionization temperatures rising smoothly from 14,000K at $z=11$ to 
19,000K at $z=6$, indicating slightly more efficient cooling at higher 
redshifts as has previously been noted~\citep{keat18}.  In short, our 
multifrequency UVB is inconsistent with post-reionization IGM temperatures 
above 20,000 K, and when cooling is taken account, even lower temperatures 
may be more realistic.

While our simulations do not directly predict post-reionization IGM temperatures
above 20,000 K, for completeness, we note that two resolution limitations impact
the predicted heating rate within ionization fronts.
First, our radiation transport solver's finite \textit{spatial} resolution 
smears out ionization fronts, causing gas to spend too much time 
in a partially-ionized state where collisional excitation cooling is 
effective.  Avoiding this problem requires a spatial resolution that
is several times smaller than the size of ionization fronts.  For photons with
energies $h\nu=$1--4 Ryd, the mean free path at mean density in a neutral IGM
at $z=6$ is 5.5--133 comoving kpc, as compared to our simulation's 
277 kpc.  Resolution limitations are less severe at higher energies and in
voids, which in fact dominate the LAF for $z>5$~\citep{bolt09}. Nonetheless,
our simulations may suffer from overcooling.

The second way in which our heating rate may be unconverged relates to our
finite \textit{energy} resolution.  Our heating module assumes that all of
the photons within a particular energy bin have an energy 
corresponding to the middle of that bin.  This may lead to overheating or 
underheating depending on the local spectral slope.  By computing the heating
owing to a trial UVB that has a fiducial $\nu^{-2}$ spectral slope from 1--4 Ryd
using increasingly finer energy sampling, we estimate that the post-reionization
temperatures with energy bins spaced 0.375 Ryd apart may be up to 23\% high in the
ionized IGM (where Equations~\ref{eqn:heatPerIon}--\ref{eqn:heating} apply) and 
up to 4\% low within ionization fronts (where nearly all photons are 
locally-absorbed).

Equation~\ref{eqn:heating} is idealized in that it applies to an 
optically-thin medium and ignores cooling.  In order to explore how heating 
proceeds when these effects are accounted for in an inhomogeneous reionization
scenario, we show in Figure~\ref{fig:TGals} that the reionized 
IGM at $z=7.5$ is indeed heated to 15,000--20,000 K,
with the hottest diffuse regions located immediately behind ionization fronts.  
In fact, comparison of Figures~\ref{fig:NHIGals} and~\ref{fig:TGals} indicates
that heating (correctly) precedes reionization owing to the somewhat longer 
mean free path of high-energy photons.  Isolated areas roughly $\sim100\kpc$ 
across can be seen at the top and bottom where the gas is shock-heated to much 
higher temperatures by galactic outflows.  Surrounding these star-forming regions
are ``webs" of filaments and voids where ongoing reionization is heating the
filaments while the voids have already cooled.  Meanwhile, outside of $\hii$ 
regions and within self-shielded filaments, the gas is characteristically 
colder than $10^4$ K.

\subsection{The Lyman-$\alpha$ Flux Power Spectrum}\label{ssec:LyaPS}

\begin{figure}
\centerline{
\setlength{\epsfxsize}{0.45\textwidth}
\centerline{\epsfbox{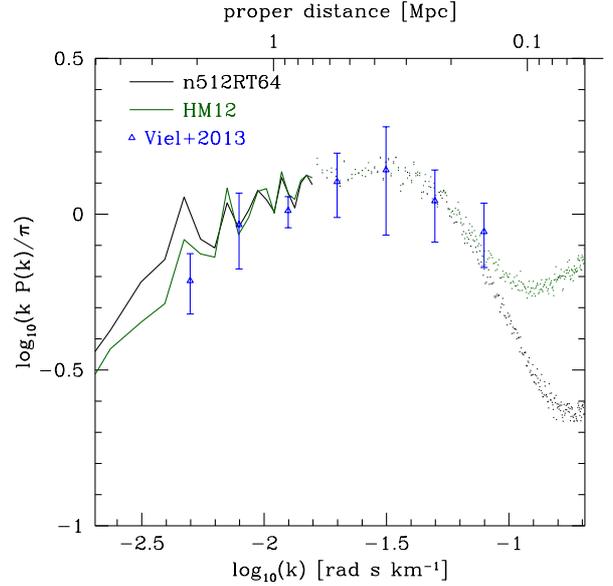}}
}
\caption{The Lyman-$\alpha$ flux power spectrum at $z=5.375$ as compared 
to observations at $z=5.4$~\citep{viel13a}.  Black and green points and 
curves show the simulated flux power spectra for our radiation transport
and HM12 models, respectively, after rescaling to the observed mean 
transmission (see text).  The top axis maps wavenumber to proper distance
($\Mpc$).  The simulation volume's proper side length $L_p$
at $z=5.4$ corresponds to a wavenumber $\log[2\pi/(H(z)L_p)] = -2.4$, which
is comparable to the largest measured scale.
}
\label{fig:HPowerSpec}
\end{figure}

\begin{figure}
\centerline{
\setlength{\epsfxsize}{0.45\textwidth}
\centerline{\epsfbox{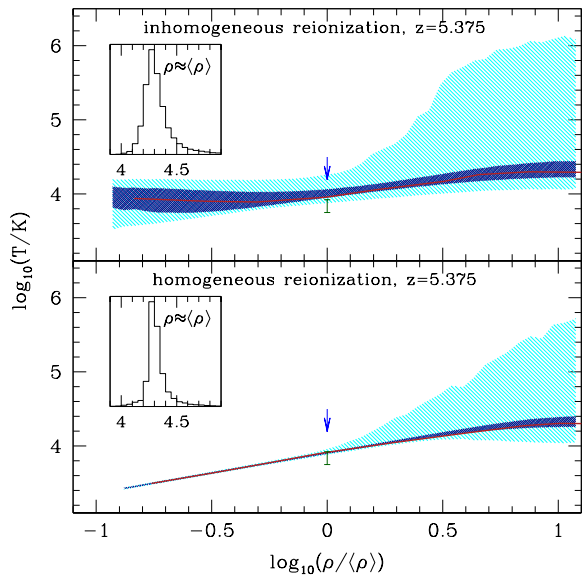}}
}
\caption{The predicted relationship between temperature and density
for underdense to mildly-overdense gas at $z=5.375$ in our
n512RT64 (top) and n512noRT (bottom) simulations.  The red curve
represents a running median $T$ at fixed $\rho$ while the inner and
outer hashed regions enclose 68\% and 99\% of gas at each density.  
The green point and blue limit at the mean density 
from~\citet{viel13a} and~\citet{garz17}, respectively.  Inset panels 
show the normalized distribution of $\log(T/K)$ at the mean density 
(see text).
}
\label{fig:rhoT}
\end{figure}

The middle panel of Figure~\ref{fig:HReion} suggests that the mean 
transmission in the LAF is reasonable near $z\sim6$ but
then rises too slowly with time.  However, this conclusion is quite
sensitive to the way in which the continuum level of the observed
LAF is inferred.  A complementary statistic that
is sensitive to reionization while circumventing uncertainty in continuum
corrections is the Lyman-$\alpha$ flux power 
spectrum~\citep[for example,][]{onor17}.  We compute the simulated power spectra
in a way that mimics observations: First, we rescale the simulated 
optical depth at each pixel by a constant factor so that the normalized
mean transmission (including pixels where it 
is negative owing to our error model) matches the continuum-corrected 
mean flux $\exp(-\tau_{\mathrm{eff}})$.  Simulated fluxes that are negative
are rescaled in the same way.  For consistency with~\citet{viel13a}, we 
choose $\tau_{\mathrm{eff}}(z=5.4)=2.64$.  For our n512RT64 and n512noRT 
simulations, matching this requires us to rescale all optical depths by 
factors of 0.56 and 0.41, respectively, consistent with the result that 
the n512RT64 simulation yields a slightly higher mean transmission.  The
simulated volume-averaged hydrogen photoionization rates at $z=5.375$ 
are $\scnot{3.21}{-13}$s$^{-1}$ and $\scnot{3.78}{-13}$s$^{-1}$ for 
the n512RT64 and HM12 UVBs, respectively, so interpreting these rescaling 
factors purely as a modification to the UVB amplitude yields ``measurements" 
of $\Gamma_{HI}$ at $z=5.375$ of $\scnot{5.73}{-13}$s$^{-1}$ and 
$\scnot{9.22}{-13}$s$^{-1}$.  The nearly factor-of-two difference between 
these inferences reflects the systematically hotter IGM in the \td~simulations 
(Figure~\ref{fig:rhoT}) and re-iterates that measuring the UVB amplitude 
from the observed mean transmission in this way is 
model-dependent~\citep{gned02}.

Having normalized our simulated LAF spectrum, we 
break it up into chunks of length 8000$\kms$.  We then extract the 
dimensionless power spectrum at each wavenumber $k$ from each chunk 
following~\citet[][Equation 12]{luki15}.  Finally, we average in 
bins of $k$.

We compare the predicted Lyman-$\alpha$ flux power spectra with 
observations at $z=5.4$ by~\citet{viel13a} in Figure~\ref{fig:HPowerSpec}.  
As long as they are rescaled to a common mean transmission, both our n512RT64 
and n512noRT simulations predict similar power spectra over the observed range. 
Moreover, they are within $1\sigma$ of observations at all scales.  In detail, 
the n512noRT model (which assumes a spatially-uniform HM12 UVB) predicts 
significantly more power at very small scales ($k > 0.06$ rad s km$^{-1}$).  
At the largest observed scale ($k < 0.01$ rad s km$^{-1}$), the RT model 
predicts somewhat more power owing presumably to residual large-scale 
fluctuations in the 
UVB.\footnote{The impact of UVB fluctuations in our model is small 
compared to the ``fluctuating UVB model" in~\citet[][Figure 16]{viel13a}
for two reasons.  First, our galaxy-dominated UVB is generally more 
homogeneous than their QSO-dominated case.  Second, our model includes both
temperature and UVB amplitude fluctuations, which tend to have opposing
effects on the Lyman-$\alpha$ transmission~\citep[][Figure 6]{davi17}.} 
It is worth noting that, 
whereas~\citet{viel13a} allowed the mean transmission in their simulated
spectra to vary about the observed value (from their Equation 4) in order to 
improve the fit with the observed power spectrum (with a best fit at $z=5.4$ 
of 3.09; see their Table II), we find reasonable agreement with the 
directly-observed optical depth.

What causes the dramatic suppression of small-scale power ($k > 0.1$ rad s km$^{-1}$)
in our reionization simulations? It could reflect either a reionization 
history that begins earlier, yielding increased pressure smoothing, or 
one that has heated the IGM more efficiently or recently, yielding 
increased Doppler broadening~\citep{onor17}.  As our radiation transport 
simulations reionize later than the HM12 UVB (cf.\ HM12, figure 14), 
the IGM temperature is the most likely culprit.  To emphasize this point,
we compare in Figure~\ref{fig:rhoT} the predicted relationship between
temperature and density for underdense to mildly-overdense gas
at $z=5.375$ in our n512RT64 (top) and n512noRT (bottom)
simulations.  The only difference between these two simulations is
the reionization history: the top panel reflects a mostly galaxy-driven,
inhomogeneous reionization scenario that is only 50\% complete at $z=8$
and includes nonequilibrium ionizations and self-shielding.  By contrast,
the bottom reflects the spatially-homogeneous HM12 UVB, which heats the
gas significantly already at $z=15$, and the assumption of ionization 
equilibrium.  Without an inhomogeneous 
reionization history, we reproduce the well-known power-law $\rho$-$T$ 
relation~\citep{hui97} by $z=5.375$.  By contrast, our inhomogeneous 
reionization model sources significant scatter in $T$ at fixed $\rho$ 
at the same redshift, yielding a slightly ``inverted" $\rho$-T relation,
as expected in the immediate aftermath of 
reionization~\citep[for example,][]{trac08,furl09}.

The data points at the mean density indicate that our simulations are
consistent with available measurements.  Both are below the
1-$\sigma$ upper limit of~\citet[][blue arrow]{garz17}.  Possibly more 
constraining is the green data point, which represents the inferred 
temperature at the mean density at $z=5.378$ from~\citet[][Figure 10]{viel13a}.  
This measurement, while marginally consistent with both simulations, is
systematically lower.  The discrepancy may reflect differences 
in the underlying simulation methodology such as the assumption that gas is 
in ionization equilibrium with a uniform UVB or the efficiency of 
photoionization heating.  Differences in the reionization history itself 
are also likely to contribute.  For example, their modeling prefers an 
instantaneous hydrogen reionization redshift of $z=11.2$ whereas our 
radiation transport model reaches a 50\% neutral fraction around $z\sim8$.
Another difference involves the uncertain timing of $\heii$ reionization,
the initial stages of which likely boost the temperature scatter in our 
radiation transport simulation over models in which the UVB first hardens 
at a lower redshift~\citep{mcqu09}.  Finally, differences in the choice
of atomic cooling rates can lead to a $\sim10\%$ difference in 
$T(\langle\rho\rangle)$~\citep{luki15}, although any correction to a
cooler IGM would only exacerbate the discrepancy with the observed mean 
transmission.  Given these differences, the marginal agreement is, if 
anything, surprisingly encouraging.

This impact of reionization on the $\rho-T$ relation is strongly 
density-dependent.  The inset panels of Figure~\ref{fig:rhoT}
show the predicted temperature distribution in each model for gas near
the mean density ($0.95 < \rho/\langle\rho\rangle < 1.05$).  The median
temperatures are quite similar, but the scatter about this median is
larger in the case of inhomogeneous reionization.  The median temperature
and scatter grow systematially to lower overdensities, which may play
a larger role in shaping the LAF at $z>5$~\citep{bolt09}.

The boosted $\rho-T$ scatter is likely complemented
by slight adjustments to the gas density distribution owing
to the fact that regions that reionized and were pressurized earlier 
have longer to relax hydrodynamically.  Together, these effects erase 
the tight relationship between overdensity and transmission that
obtains in the n512noRT simulation, modulating the filtering scale
and suppressing small-scale power. 

The suppression of small-scale power in our radiation transfer
model may be degenerate with other effects
such as the free streaming of warm dark matter 
particles~\citep{viel13a}.  Our simulations therefore suggest 
that accounting for inhomogeneous reionization will \textit{tighten} 
constraints on such processes (for example, by increasing the 
minimum permitted warm dark matter particle mass).  The slight 
preference in the smallest-scale bin that is observed for the 
homogeneous reionization history, if confirmed, could point to
the need for a more sudden reionization history or else 
significantly more uniform pre-reionization heating, perhaps owing
to X-ray heating (which we do not currently account for).

The comparison in Figure~\ref{fig:HPowerSpec} suffers from 
limitations associated with our finite simulation 
volume ($12\hmpc$) and limited mass resolution, both of which are
known to grow increasingly important at high redshift~\citep{bolt09}.
Fortunately, published studies suggest that resolution and volume 
limitations are likely at the $\sim$5\% level in our 
case~\citep{viel13a,onor17}.  Perhaps more worryingly, our
simulated mean transmission at $z=5.4$ requires a significant 
correction in order to match the observed $\tau_{\mathrm{eff}}$;
this correction cannot remove changes to the filtering 
scale~\citep{luki15}.  If the simulated end-game to reionization 
is indeed too slow, then accelerating it (by boosting the ionizing 
emissivity at $z<6$) would likely suppress large-scale power.

We note additionally that, while it is possible that galactic 
winds boost large-scale power at $z<4$ by modulating the distribution
of gas densities within the density range that the LAF
probes, they are not expected to be important 
at $z>5$.  This is because the LAF at higher
redshifts traces underdense gas, which is located predominantly 
far from the galaxies~\citep{viel13b}.

\subsection{CGM}\label{ssec:cgm}
Observations of the CGM complement the IGM at $z>5$, where the LAF
traces predominantly underdense gas.  The CGM is moreover a strong 
observational probe of UVB fluctuations~\citep{oppe09,finl15}.  On spatial 
scales larger than a galaxy, the UVB amplitude at the Lyman limit is 
believed to be relatively homogeneous below $z<5$, but it was inhomogeneous at 
earlier times~\citep{wors14,beck15a,beck18,bosm18}.  Spatial fluctuations on scales 
larger than a galaxy may boost the fraction of CGM metals that are in high 
ionization states such as $\civ$ and $\siiv$.   On scales that are smaller than
Lyman-limit systems ($\sim10\kpc$;~\citealt{scha01}), the UVB is suppressed
by self-shielding in overdense systems.  Our simulations account for both
effects in a frequency-dependent way.  We have previously shown that 
local UVB enhancements boost the $\civ$ abundance while self-shielding
boosts the abundance of low-ionization ions such as $\oi$, $\cii$, and 
$\siii$~\citep{finl13,finl15}.  Here, we revisit these observational 
comparisons in the light of our newer model, which contains numerous
physical improvements, reproduces observations of the galaxy abundance 
more faithfully, and sports an enhanced dynamic range.

We generate mock spectra by
passing a sightline through our simulation volume that is oblique
to its boundaries and wraps around them until it 
subtends an absorption path length of roughly 420.  The temperature,
density, metallicity, and proper velocities of gas particles are
smoothed onto the sightline into pixels with a proper velocity width
of $2\kms$.  The temperature of multi-phase star-forming gas particles
is set to 5000 K\footnote{Varying this temperature within the range
1000--20,000 K has negligible impact on the predicted $\cii$ and $\civ$ 
column density distributions because these species do not trace 
star-forming gas.}.  The ionization state for each metal species is 
computed using an ionization equilibrium calculation that takes all 
relevant 
reactions into account (see~\citealt{finl15} for details).  The UVB 
is either the local UVB computed directly by the simulation including
our self-shielding treatment, or else the spatially-homogeneous~\citet{haar12} 
UVB.  We then compute the simulated optical depth along the sightline 
following~\citet{theu98}.  Finally, we smooth the spectrum with a 
gaussian response function with a full-width at half-maximum
of $10\kms$ and add gaussian noise corresponding to a 
signal-to-noise of 50 per pixel.  The simulated spectra therefore have 
an effective resolution of $R=$30,000.

We extract catalogs of simulated metal absorbers by searching for
regions where the flux over at least three consecutive pixels drops
$5\sigma$ below the continuum.  The total column density for each
simulated absorber is then computed following~\citet{sava91}.
We merge the column densities and equivalent widths of individual 
absorbers whose flux-weighted mean velocities are closer than $50\kms$ 
into absorption systems to produce model catalogs~\citep[e.g.,][]{song01}.

\begin{figure}
\centerline{
\setlength{\epsfxsize}{0.5\textwidth}
\centerline{\epsfbox{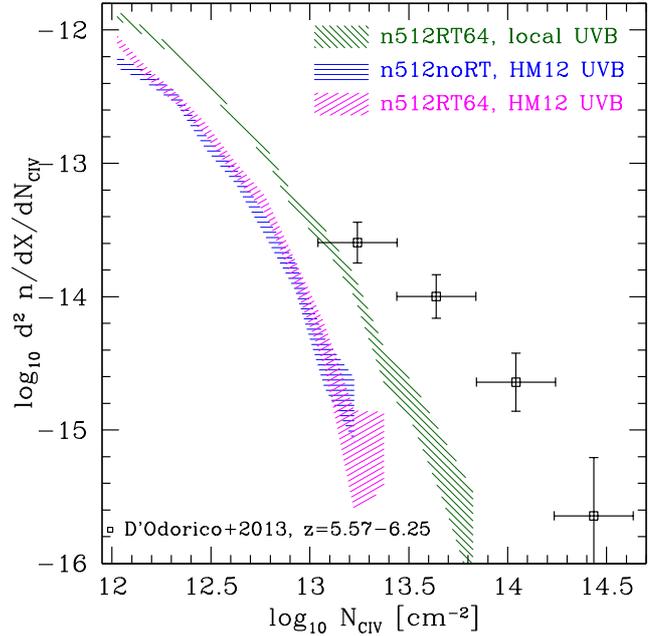}}
}
\caption{Simulated $\civ$ column density distribution at $z=5.75$ as
compared to observations.  The more intense UVB at $>4$ Ryd predicted
by \td~boosts the predicted $\civ$ abundance by $\approx0.3$ dex, 
alleviating but not removing tension with observations.  The simulated
distributions are not adjusted to account for observational
completeness, which is $>60\%$ for $\log(N_\civ) > 13.3$~\citep{dodo13}.
}
\label{fig:dNdXdN_CIV}
\end{figure}

We begin in Figure~\ref{fig:dNdXdN_CIV} by examining the $\civ$ column 
density distribution (CDD) from a conventional simulation (n512noRT) in which 
the HM12 UVB is assumed both during the simulation and in post-processing
(blue horizontal hatches).  This underpredicts the observed $\civ$ at all
column densities, suggesting either that too few metals are ejected or
that the UVB is too weak or too soft.  Next, we consider our radiation
hydrodynamic simulation with a realistic, spatially-inhomogeneous UVB
(n512RT64; green downward hatches).  This simulation generates fewer stars (and hence
fewer metals) after $z<15$ (Figure~\ref{fig:madau}) and its UVB
amplitude at the Lyman limit is slightly lower than the HM12 UVB at $z=5.75$ 
(Table~\ref{table:sims}).  Nevertheless, it yields 0.3--0.4 dex \emph{more}
$\civ$ at all columns, bringing the prediction at the highest columns within
a factor of $\approx$six of observations.  In order to verify that it is the
UVB that dominates this difference, we 
re-extract metal absorbers from the n512RT64 simulation after substituting
the HM12 UVB in place of the simulated one (magenta upward hatches).  
This yields essentially the same absorber population as the n512noRT model, 
confirming that enhanced flux at $>4$ Ryd dominates the $\civ$ enhancement 
in our n512RT64 simulation.

Recently, several works have pointed out that conventional hydrodynamic 
simulations have difficulty reproducing the observed $\civ$ abundance 
at high redshift~\citep{keat16,bird16,rahm16}.  Our results are qualitatively 
similar but with important differences.  Whereas complementary efforts such 
as the Sherwood and Illustris simulations reproduce the observed 
SMF~\citep[for example, Figure 1 of][]{keat16} but adopt the weak $\heii$-ionizing
background in the HM12 model, our simulations produce somewhat fewer stars
but predict a harder UVB (figure~\ref{fig:UVBs}).  This raises the question 
as to whether simulations predict a diffuse CGM that is underenriched, 
underionized, or both.  

It is not likely that the gap can be eliminated through further boosts to 
the outflow velocities because the assumed outflows already contain roughly 
six times as much kinetic energy as occurs naturally in zoom 
simulations~\citep{mura15}.  On the other hand, \textit{weakening} outflows 
in order to improve agreement with the SMF at
$z>5$ would boost the CGM metallicity~\citep{rahm16}.  Even this may not
be enough though: Figure~\ref{fig:mfs} suggests that the stellar mass density
and hence the overall metallicity may be low by as $\approx$a factor of two,
but a factor of two increase in the abundance of strong $\civ$ systems would 
not bring the prediction within the $1\sigma$ confidence interval.  Nor can 
we improve agreement through simple adjustments to the metal yield: boosting 
the carbon yield would exacerbate tension with the observed abundance of weak 
$\cii$ absorbers (although it would help with the strong ones; Figure~\ref{fig:dNdEWdX_CII})
 as well as with observations of DLA metallicities (Figure~\ref{fig:dPdZDLA}).
In short, if underenrichment is the problem, it is not easy to see how it can 
be resolved through adjustments to our subgrid models for feedback and 
enrichment.

The alternative explanation is that our model underproduces $\civ$ because the 
CGM's carbon remains underionized, even when we account for a stronger background
at $> 4$ Ryd from quasars (Figure~\ref{fig:UVBs}) as well as small-scale UVB 
fluctuations.  If the predicted CGM is underionized, this cannot owe to resolution
limitations as our simulations resolve the relevant UVB fluctuations spatially: 
the mean free path at $z=6$ for photons with energies greater than 4 Ryd in gas 
that it is at least 99\% ionized exceeds $1 \Mpc$ (proper), while the RT voxels 
in our n512RT64 simulation are $\approx40\kpc$ wide.  On the other hand, it
is possible that the true UVB is brighter in $\heii$-ionizing energies than we
predict owing to a stronger contribution from low-metallicity stars; we will 
return to this point in Section~\ref{ssec:disc}.  

Finally, we note a possible contribution from the fact that our post-processing 
calculations assume ionization equilibrium.  Relaxing this assumption would 
change the $\civ$ abundance, although it is difficult to predict in what 
direction~\citep{oppe13}.

\begin{figure}
\centerline{
\setlength{\epsfxsize}{0.5\textwidth}
\centerline{\epsfbox{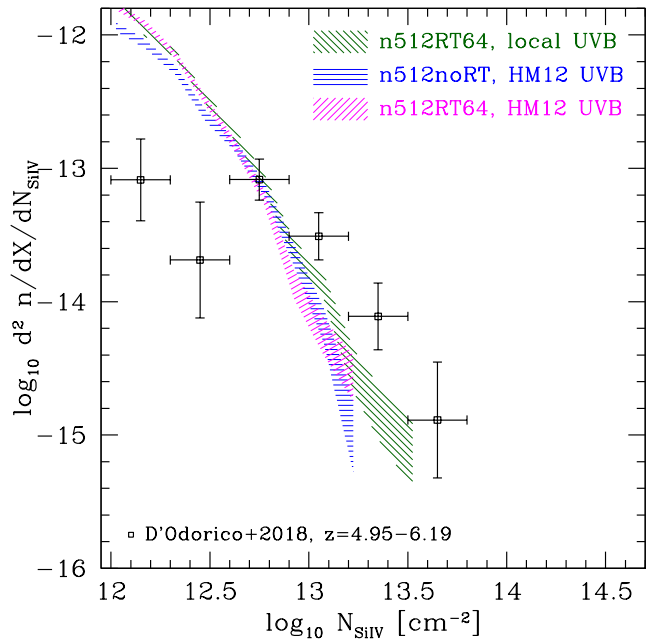}}
}
\caption{The simulated $\siiv$ column density distribution is in reasonable
agreement with observations at $z=5.75$ irrespective of the choice of UVB.
The flattening in the observed distribution at $\log(N_\siiv) < 12.9$ likely
reflects observational incompleteness, which is not modeled (see text).
}
\label{fig:dNdXdN_SiIV}
\end{figure}

$\siiv$ is a useful complement to $\civ$ because its first ionization 
potential is only 3.3 Ryd as compared to 4.7 Ryd for $\civ$ 
(Figure~\ref{fig:UVBs}).  Consequently, its evolution is only weakly 
sensitive to the progress of $\heii$ reionization and to the contribution 
of quasars.  In Figure~\ref{fig:dNdXdN_SiIV}, we show the $\siiv$ CDD at 
$z=5.75$ using the simulated and HM12 UVBs as in Figure~\ref{fig:dNdXdN_CIV}.  
Observations are from D'Odorico et al.\ 2018 (in preparation).

The observed CDD shows nonmonotonic behavior for $\log_{10}(N_\siiv) < 12.7$,
which we attribute to observational completeness.  Although rigorous 
evaluation of completeness is in-progress, we estimate it by assuming
that the data probe to a similar optical depth as the $\civ$ measurements
presented in~\citet{dodo13}, which are in turn 60\% complete for 
$\log_{10}(N_\civ) = 13.3$.  By computing the $\siiv$ column density
for which the $\lambda$1393\AA~line yields the same optical 
depth as the 1548\AA~line from an absorber with 
$\log_{10}(N_\civ)=13.3$, we estimate that the $\siiv$ data are $60\%$ 
complete at $\log_{10}(N_\siiv) = 12.9$, roughly the column density 
below which the observed CDD flattens.

Encouragingly, the predicted $\siiv$ CDD agrees with observations at the 
$2\sigma$ level wherever the observations are reasonably complete and 
irrespective of the choice of UVB.  The insensitivity to the choice of
UVB confirms that the progress of $\heii$ reionization does not impact 
$\siiv$, which is instead a joint tracer of metal production and the growth 
of the UVB at $\hi$-ionizing energies.  Although we do not demonstrate this,
adjusting our feedback scheme to boost the overall star formation rate
efficiency by $\sim50\%$, which is observationally permitted 
(Figures~\ref{fig:lfs}--\ref{fig:mfs}), would eliminate the outstanding 
discrepancy with observations.

\begin{figure}
\centerline{
\setlength{\epsfxsize}{0.5\textwidth}
\centerline{\epsfbox{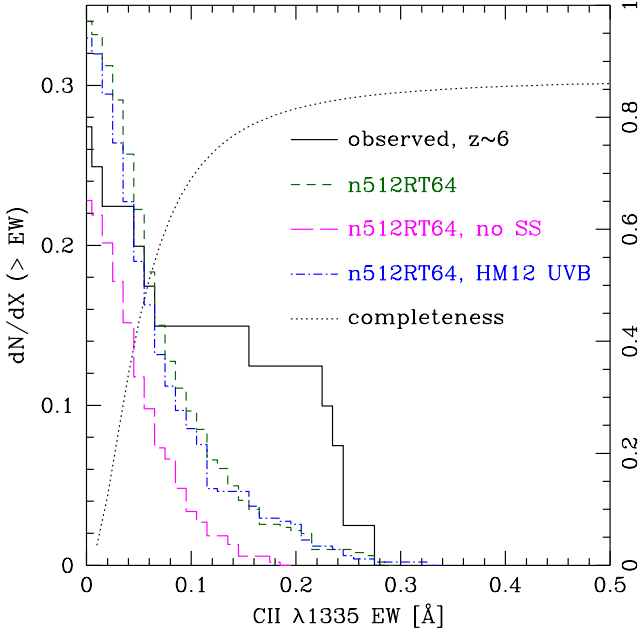}}
}
\caption{The simulated $\cii$ equivalent width distribution at $z=5.75$
as compared to observations~\citep{beck11a,bosm17}.  Agreement is reasonable
at low equivalent widths, but the model underproduces strong $\cii$ absorbers.
}
\label{fig:dNdEWdX_CII}
\end{figure}

A thorough assessment of our simulated CGM requires us to consider both
high- and low-ionization absorbers such as $\cii$, $\siii$, and $\oi$ 
because they trace different gas phases and different portions of the 
UVB. For example, $\cii$ complements $\civ$ in that it is sensitive 
primarily to photons with energies of $<4$ Ryd (Figure~\ref{fig:UVBs}).  
Moreover, it is sensitive
to the progress of hydrogen reionization via charge-exchange reactions
with neutral hydrogen.  An array of theoretical studies have found that,
while the abundance of high-redshift $\civ$ absorbers is difficult to 
reproduce in cosmological simulations, the predicted abundance of 
low-ionization absorbers is relatively insensitive to the details
of star formation feedback and generally in better agreement with 
observations than the $\civ$ CDD~\citep[for example,][]{keat16}.  

\begin{figure}
\centerline{
\setlength{\epsfxsize}{0.5\textwidth}
\centerline{\epsfbox{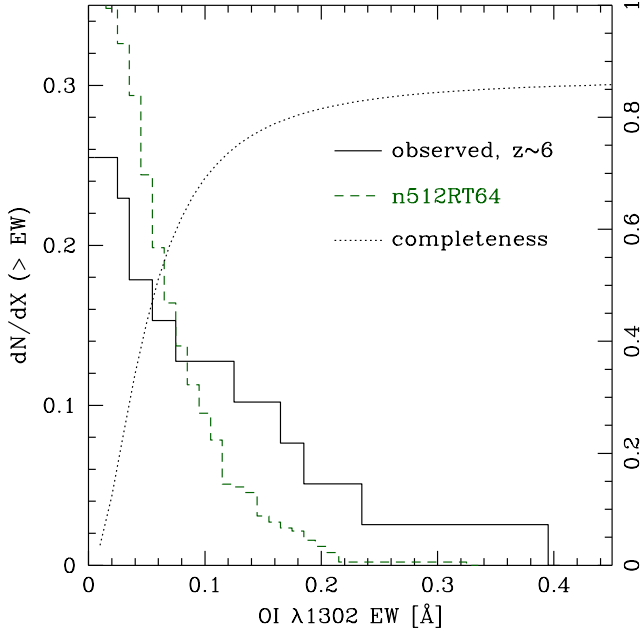}}
}
\caption{The simulated $\oi$ equivalent width distribution at 
$z=5.75$ is in tolerable agreement with observations of weak
absorbers $z\sim6$~\citep{beck11a,bosm17} but may underproduce strong
systems.  The observed and simulated curves were generated in
a similar way to Figure~\ref{fig:dNdEWdX_CII}.
}
\label{fig:dNdEWdX_OI}
\end{figure}

We evaluate in Figure~\ref{fig:dNdEWdX_CII} how our simulated $\cii$ 
equivalent width distributions compare to observations.  In order 
to generate the observed distribution (solid black), we correct 
the reported survey's absorption path length by~\citet{beck11a} of 39.5 to our cosmology, 
yielding 37.2.  We also fold in the results from~\citet{bosm17}, who found 
one $\cii$ absorber with a rest-frame equivalent width of 0.0368~\AA~along 
the line of sight to a quasar at $z=7.084$ that surveyed an absorption 
path length of $\Delta X=2.906$.  We correct the \emph{simulated} absorber 
abundances (at $z=5.75$) down in a way that mimics observational 
incompleteness following~\citet{keat16}: we first fit a function of 
the form 
\begin{equation}\label{eqn:comp}
f(x) = \frac{L}{1 + \exp(-k(x - x_0))}
\end{equation}
to the results displayed in Figure 11 of~\citet{beck11a} and take
$x$ as $\log_{10}(\mathrm{EW}_\oi)$.  Our fit parameters are
$(L,k,x_0) = (87\%,4.5,-1.3)$.  We then weight each simulated absorber
by the observational completeness at its equivalent width based
on Equation~\ref{eqn:comp}.  This fit is given by the dotted line and
the right y-axis in Figures~\ref{fig:dNdEWdX_CII}--\ref{fig:dNdEWdX_OI}.

Our simulations (green short-dashed) reproduce the observed abundance 
of absorbers to within a factor of 3 at all equivalent widths where measurements
are available.  Given that the factors that govern these predictions
such as metal yields, galactic outflow rates, initial mass function, and 
quasar emissivities were not tuned to match these observations, this level
of agreement is remarkable.  In detail, our simulations overproduce the 
abundance of absorbers that are weaker than the estimated 50\% completeness 
limit and underproduce the stronger systems.  

We evaluate the impact of local UVB fluctuations on $\cii$ as in the
case of $\civ$, by replacing the simulated UVB with the HM12 UVB (blue 
dot-dashed) in post-processing.  Surprisingly, the resulting predicted 
distribution is  largely unchanged.  While this agreement is nominally expected 
given the close correspondence between the HM12 and the volume-averaged 
simulated UVB near the Lyman limit (Figure~\ref{fig:HReion}), it hides a near-complete 
cancellation between short-range UVB amplifications and self-shielding.  If
we neglect self-shielding in post-processing (magenta long-dashed), then there 
is significantly less $\cii$ at all equivalent widths.  This, in turn, implies 
that modeling $\cii$ in a way that accounts for self-shielding of dense gas 
but \textit{neglects} UVB amplifications may artificially inflate the 
predicted $\cii$ abundance.

Neutral oxygen traces a denser gas phase than the other ions, hence
its abundance is a more direct probe of the gradual encroachment of
the UVB into the CGM~\citep{keat14}.  In Figure~\ref{fig:dNdEWdX_OI},
we compare our simulation versus the observed cumulative distribution
of $\oi$ equivalent widths at $z\sim6$~\citep{beck11a,bosm17}.  In order 
to generate the observed distribution (solid black), we treat the upper 
limit on the $\oi$ equivalent width of the system along the line of sight 
to J1630+4012 as a detection.  We also add $\Delta X = 2.00926$
to the absorption path length in order to account for the fact that~\citet{bosm17}
uncovered no $\oi$ absorbers along the path to a quasar at $z=7.084$.

The result of this comparison is quite similar
to Figure~\ref{fig:dNdEWdX_CII}: our simulation overproduces weak
absorbers, produces roughly the correct number at the 50\% observational
completeness limit, and underproduces the strong 
ones.\footnote{Note that it is not useful to vary the UVB in
post-processing with $\oi$ as we did with $\cii$ because it is 
charge-locked to $\hi$, whose nonequilibrium ionization state is 
adopted directly from the simulation.}

We now consider why the predicted shape of our low-ionization absorber
equivalent width distributions may be too steep.  The comparison in 
Figure~\ref{fig:dNdEWdX_OI} does not obviously imply 
that the simulated oxygen yields are low: boosting the metal yield would, 
to first order, shift the entire curve to higher equivalent widths without 
changing its shape.  Agreement at strong equivalent widths would come at 
the cost of overproducing the abundance of well-sampled weak systems.
Moreover, our metal yields
are, in a sense, already boosted by our assumption that 50\% of 
SNe explode as hypernovae~\citep{nomo06}, and further boosts would,
as before, exacerbate tension with observations of DLA metallicities.

Another possible explanation for the overabundance of weak low-ionization
systems may be inaccuracy in our adopted model for galactic outflows.
To address this, we refer to~\citet{keat16}, who explored how a number of 
models for hydrodynamic feedback, UVB amplitude, and self-shielding impact 
the predicted absorber abundances.  They found that weak $\cii$ absorbers 
(EW $< 0.1$ \AA) are strongly overproduced in their \textit{HVEL} model, which 
imposes a minimum outflow velocity of $600\kms$.  This suggests that
our simulated outflows may be too fast (Equation~\ref{eqn:vwind}).  
Suppressing the outflow velocities may reduce the predicted $\cii$ 
abundance into better agreement with observations---but at the cost of
cutting the numbers of $\civ$ and $\siiv$ absorbers, which are already low.  
The impact of changing
outflow velocities on the predicted $\oi$ abundance will be generally
weaker~\citep{keat16}, although it will likely improve agreement.  Likewise,
it would improve agreement with the observed UV LF (Figure~\ref{fig:lfs}).

Alternatively, the overabundance of weak low-ionization absorbers 
may owe to the UVB, which disproprortionately affects weak systems.  
Strengthening it would preferentially suppress weak $\cii$ and $\oi$
systems while boosting weak $\civ$ systems~\citep[][Figure 10]{keat16}.

Finally, it is always possible that the observationally-inferred completeness
is too high owing to an inaccurate model for weak absorbers.  To this
end, it would be interesting to explore how the predicted profiles of weak 
absorbers compare with conventional assumptions; however, we defer this to 
future work.

In summary, Figures~\ref{fig:dNdXdN_CIV}--\ref{fig:dNdEWdX_OI} show that 
our simulations roughly reproduce the observed abundance of weak $\civ$, 
$\siiv$, $\cii$, and $\oi$ absorbers at $z\sim6$ while underproducing the 
strong systems in the same ions. We are thus confronted with the odd result 
that, while $\civ$ seems to require that the CGM near galaxies be more 
highly-ionized, $\cii$ and $\oi$ require it to be more neutral!  The
overall good agreement among the weak systems corroborates the relatively
good agreement in Figure~\ref{fig:dPdZDLA} in suggesting that the CGM's 
metal enrichment is realistic.  In detail, however, the tendency for the
simulated equivalent width distributions to be too steep irrespective of
the ion suggests non-trivial issues with wind speed, UVB, and overall
CGM metallicity.

\section{Summary and Discussion}\label{sec:sumDisc}
\subsection{Summary}\label{ssec:summary}

The \td~simulations combine a model for star- and galaxy- formation that is 
anchored in high-resolution simulations with an on-the-fly multifrequency 
radiation transport solver so that galaxies, their CGM, the IGM, the UVB, 
and reionization 
are all followed simultaneously.  Their predictions are subject to only one 
parameter that has been tuned to match observations in a precise
way, $\fescg(z)$, 
hence they represent a unified statement of how well the reigning 
galaxy-driven reionization scenario can account for a wide variety of 
observations of the reionization epoch and its immediate aftermath.  We
have, of course, explored only a subset of the available observations,
but our results so far already raise some interesting questions.

The predicted history of star formation is consistent with 
observations, which is encouraging given that the agreement does not
represent empirical calibration.  In detail, comparisons with the 
observed UV LF and SMF indicate that star formation is oversuppressed 
by up to a factor of 2 at $z>5$, with larger discrepancies at higher 
resolution.  The predicted $\mstar$-SFR relationship is reasonably 
robust to resolution effects~\citep[although see][]{scha15}, and is 
consistent with observations at $z\sim6$ within the uncertainty, 
suggesting that the UV LF and SMF discrepancies share a common origin.  
These results are largely 
insensitive to the progress of reionization because observed galaxies 
live in haloes that are too massive for photoionization feedback.  

Turning to reionization, the model for $\fescg(z)$ that we adopt falls 
from 35--50\% at early times to $<20\%$ by $z=5$.  The values at $z>6$ 
are higher than observations and some complementary models indicate, but 
have been observed directly in isolated galaxies.  Adopting this model 
yields a reionization history that is in
broad agreement with observations of the volume-averaged neutral hydrogen
fraction and the hydrogen ionization rate at the close of the hydrogen
reionization epoch.  Curiously, our low-resolution n256RT32 
simulation yields marginally better agreement with observations of $\xhiv(z)$
while underproducing $\Gamma_{HI}$, whilst our high-resolution n512RT64
simulation is in excellent agreement with $\Gamma_{HI}$ and marginally 
underproduces $\xhiv(z)$.  The mean transmission in the n512RT64 simulation 
is systematically low for $z<5.7$, though not as low as in the 
case of the HM12 UVB.  This problem, which appears generic to hydrodynamic 
simulations of reionization, could indicate the need for very large-scale
UVB fluctuations~\citep{beck18}, a hotter IGM, or systematic uncertainties 
in the observations.

The UVB that drives reionization is spectrally-soft,
corresponding to less than 5 eV of IGM heating per photoionization.  
Owing to this tepid heating, the observed Lyman-$\alpha$ flux power 
spectrum is reproduced at all but the largest $k$-bin, where it may be
$1.5\sigma$ high.  In detail, however, a tendency
for inhomogeneous reionization to boost large-scale power and suppress small-scale
power suggests that current measurements contain untapped clues into large-scale 
UVB fluctuations, small-scale scatter in the $\rho-T$ distribution, and possibly
even the early stages of $\heii$ reionization; simulations such as the \td~suite 
will be crucial in unlocking these secrets.

The simulated $\rho$-T relationship at $z=5.4$ 
shows significantly enhanced scatter with respect to a uniform-UVB 
model, as expected in the immediate aftermath of reionization. This 
scatter suppresses small-scale LAF fluctuations, although the largest differences with respect to a uniform-UVB
model are at scales that have not yet been measured directly.

Finally, simulated galactic outflows disperse enough metals into the 
CGM to yield DLA metallicities that are consistent with 
observations at $z\sim5$, although there may be evidence that
simulated DLAs are systematically overenriched at the factor-of-two 
level.  There is an underabundance of strong $\civ$, 
$\cii$, and $\oi$ absorbers and a possible excess of weak $\cii$ and 
$\oi$ absorbers.  The predicted $\civ$ abundance is significantly 
enhanced by the harder \td~UVB, though not enough to match 
observations.  $\cii$ is unaffected by UVB fluctuations owing to a near-perfect 
cancellation between the effects of living near ionizing sources
(i.e., galaxies) and originating in self-shielded gas.  
The $\siiv$ CDD is $\approx1\sigma$ low at the highest observed
columns and in agreement at lower columns, significantly better 
than $\civ$.

\subsection{Discussion}\label{ssec:disc}

The results that the \td~simulations underproduce stars and
strong $\oi$, $\cii$, and $\siiv$ absorbers could 
all point to the same problem, namely that our subgrid prescription for 
generating galactic outflows (Equations~\ref{eqn:mlf} and~\ref{eqn:vwind}) 
is too vigorous. Moderating either the mass-loading factor or the wind
launch velocity would both boost star formation and enhance the 
abundance of strong metal absorbers~\citep{rahm16}.

Such tuning would not be inconsistent with results from high-resolution 
simulations:~\citet{mura15} assign to the normalization of 
Equation~\ref{eqn:mlf} an uncertainty of 0.2 dex. Reducing it
by this factor would significantly alleviate discrepancies.  Reducing the
wind launch velocities would additionally reduce the predicted abundance
of weak low-ionization absorbers~\citep{keat16}, improving
agreement with observations.  At the same time, the extra star formation 
would allow us to lower $\fescg(z)$ by a similar factor, bringing it 
closer to the 5--15\% range that seems favored overall by complementary 
works.

More broadly, however, if the underproduction of stars and strong metal 
absorbers can be ameliorated
by moderating galactic outflows, then they highlight the problems that 
arise from tying outflow scalings in cosmological simulations too closely 
to results from high-resolution simulations.  In the latter, most of the
material that is ejected from galaxies returns on a short timescale while
only a fraction escapes to the virial radius~\citep{mura15,chri16}.
This broad range of outcomes for ejected gas is not treated realistically 
by our simulations, which assume a relatively narrow range of launch
velocities and use hydrodynamic de-coupling to guarantee that most outflowing
gas escapes to the virial radius.  These simplifications may contribute to 
the oversuppression of star formation and metal generation, and calibrating
them more carefully would improve agreement with observations.

However, this tuning, by itself, will not bring the predicted
abundance of strong $\civ$ absorbers into agreement with observations:
With the HM12 UVB, the predicted abundance for $N_\civ>10^{13.5}\cm^{-2}$
is low by a factor of $\approx 10$.  Adopting the harder \td~UVB,
strong $\civ$ absorbers remain low by a factor of 3--6.  Closing 
the remaining gap through further enhancements to metal production would
inevitably overproduce strong $\cii$, $\oi$, and $\siiv$ as well as
DLA metallicities.  Hence while enhancing star formation to the point that 
the SMF is clearly reproduced will alleviate the discrepancy, it may 
well leave the need for additional mechanisms for boosting the $\civ$ 
fraction.

One possibility is that conventional stellar population synthesis models
simply underpredict the $\heii$-ionizing continuum from massive, 
low-metallicity stars.~~\citet{senc17} have shown that low-redshift 
galaxies with metallicities below $Z/Z_\odot = 0.2$ produce
significantly stronger $\heii$ recombination lines than can be explained
by the expected ionizing flux from stellar population synthesis models.
Likewise, high-resolution observations of high-redshift galaxies indicate
rest-frame ultraviolet emission lines including $\civ 1548,1550$ and
$\heii 1640$ that have unaccountably high equivalent 
widths~\citep[for example,][]{smit17,berg18}.  Given that our most 
metal-enriched galaxy at $z=6$ has $Z_*/Z_\odot = 0.1$, \emph{all} of 
our simulated reionization-epoch galaxies may be too faint at 
$\heii$-ionizing energies.  Accounting for effects associated with binary 
star evolution would harden the predicted UVB~\citep{stan16}.  
Additionally, it would increase the overall ionizing 
emssivity~\citep{rosd18}, enabling reionization constraints to be matched 
with a reduced $\fescg(z)$.

It is worth noting that boosting the emissivity from galaxies at $> 4$ 
Ryd will accelerate the early stages of $\heii$ reionization.  This 
``second stage" to reionization is conventionally considered to have begun 
after $z=5$~\citep{mcqu09,haar12}.  However, In the \td~simulations, the 
$\heiii$ fraction is predicted to reach 30\% by $z=5$ 
(see also~\citealt{ciar12}).  This is qualitatively consistent with 
observational suggestions that $\heii$ reionization begins too early 
to be purely quasar-driven~\citep{beck11b,wors16}, and it underscores
the need for further investigation into the possible role of galaxies
in jump-starting $\heii$ reionization.

A final possible method for boosting the CGM's ionization state involves
appealing to the inherently stochastic nature of star formation in
dwarf galaxies.  Currently, \td~bases the emissivity on the gas 
particles' instantaneous star formation rates by assuming that they have
been forming stars constantly at the current rate for 100 Myr 
(Section~\ref{sec:sims}).  While this can effectively increase the 
radiation transport solver's spatial resolution~\citep{rahm13}, it misses 
the dramatic temporal fluctuations that occur when the emissivity is tied 
to the star particles~\citep{ma15}.  These fluctuations could
enhance the CGM's $\civ$ fraction, particularly if the assumption of
ionization equilibrium were relaxed~\citep{oppe13}.

The tendency for the \td~simulations to underproduce the mean transmission
in the LAF suggests that some measurements of the UVB amplitude may be biased 
low.  At $z=5.4$, observations indicate 
that the IGM's $\hi$ photoionization rate $\Gamma_{\hi,-13}=$ 
4.7~\citep{dalo18}, 1.3~\citep{calv11}, 1.8--4.7~\citep{wyitbolt11}, or 
8.5~\citep{davi18a}, where $\Gamma_{\hi,-13}$
is in units of $10^{-13}$s$^{-1}$. The n512RT64 simulation predicts an 
ionization rate of 3.2 at the same redshift, within the lower end of the
observed range, while underproducing the mean transmission (see also 
Figure~\ref{fig:HReion}).  By contrast, the $40\hmpc$ radiation transport 
simulation of~\citet{keat18} predicts $\Gamma_{\hi,-13}=$6.67 and 
\emph{over}produces the mean transmission~\citep[Table 4 of][]{bosm18}.  
This indicates that, at least qualitatively, a stronger UVB at $z=$5--6
may solve the problem.  At the same time, a stronger UVB would alleviate 
tension between the simulated and observed $\civ$ CDDs and reduce the 
predicted DLA abundance, which currently lies at the top of the observed
68\% confidence interval (Section~\ref{ssec:metals}).  These
considerations all raise the question of whether a stronger $\Gamma_{\hi}$ 
is observationally permitted, as recently suggested by~\citet{davi18a}. 

Systematic uncertainties could readily compromise the comparison between 
the predicted and inferred UVB amplitude because we quote the actual 
simulated UVB amplitude rather than the one that would be inferred from 
our simulated observables.  The measured UVB amplitude is model-dependent 
at the $\approx$factor-of-two level owing, for example, to uncertainties in 
the assumed IGM temperature (Section~\ref{ssec:LyaPS}) or role of ionizing 
photons released during recombinations, and we have not attempted to correct 
for these issues.  It would be interesting to evaluate this possibility by 
applying techniques for measuring the IGM temperature and UVB amplitude 
using realistic, inhomogeneous reionization models.  For the present,
however, we regard the discrepancy in the mean transmission as the more 
robust indicator of a problem.

We may condense this discussion into the following questions:

\begin{enumerate}
\item Can the underabundance of strong $\oi$, $\cii$, and $\siiv$ absorbers and the oversuppression of star formation both be resolved by moderating galactic outflows, or do these discrepancies indicate small-scale spatial fluctuations in the CGM temperature and UVB that are not yet numerically resolved?
\item If galactic outflows are weakened, does the $\fescg(z)$ that is required for matching the history of reionization fall within agreement with observations and high-resolution simulations?
\item Can reionization models simultaneously reproduce the observed mean transmission in the LAF without overproducing the observed $\hi$ photoionization rate?
\item Do observations that young, low-metallicity galaxies are unaccountably bright in $\heii$ emission while their CGM are unaccountably weak in $\civ$ absorption represent the same problem? Can both problems be resolved through enhanced $\heii$-ionizing emission from young galaxies without overheating the IGM? If so, what are the implications for the post-reionization $\hi$ LAF and the history of $\heii$ reionization, which is conventially assumed to be entirely quasar-driven~\citep{furl08,haar12}?
\end{enumerate}

These considerations emphasize that the conventional, galaxy-driven $\hi$ 
reionization hypothesis contains many unsolved problems.  They also 
illustrate how coupling multifrequency radiation transport solvers to
realistic models for galaxy evolution will allow the synthesis 
of observations that are normally considered in isolation, likely 
unveiling qualitatively new synergies and tensions.

\section*{Acknowledgements}
Our simulations were run on NMSU's {\sc joker} and XSEDE's {\sc comet} 
supercomputers; for advice and support we thank the NMSU ICT department and
XSEDE.  We thank Moire Prescott, Caitlin Doughty, Valentina D'Odorico, 
Zheng Cai, Lise Christensen, and Marc Rafelski for helpful comments and 
conversations. George Becker we thank particularly 
warmly for encouragement and helpful comments on the draft.  KF thanks Claudia
Scarlata for organizing the conference ``Distant Galaxies from the Far South",
where the thinking for this work matured significantly.  We also thank
the anonymous referee for a sharp-eyed and supportive report that improved the paper.  
Our work made use of the WebPlotDigitizer tool (https://automeris.io/WebPlotDigitizer),
for which we thank A.\ Rohatgi.  This research would also have been quite 
impossible without the NASA Astrophysics Data System and the arXiv eprint 
service.  EZ acknowledges funding from the Swedish National Space Board, and
LK acknowledges support from a fellowship from the Canadian Institute
for Theoretical Astrophysics. The Cosmic Dawn Center is funded by the Danish 
National Research Foundation.
\onecolumn
\begin{table}
\small
\setlength\tabcolsep{2.1pt}
\begin{tabular}{l|cccccccccc}
$Z$ & C & O & Si & Fe & N & Ne & Mg & S & Ca & Ti\\
\hline
0 & 0.00191541 & 0.0213646 & 0.00179892 & 0.000844243 & 5.49083e-06 & 0.00457574 & 0.00153244 & 0.000708791 & 8.15213e-05 & 1.49848e-06 \\
0.001 & 0.00111629 & 0.0222635 & 0.0016386 & 0.000891533 & 9.13731e-05 & 0.00431286 & 0.00199825 & 0.000627048 & 7.49162e-05 & 1.60633e-06 \\
0.004 & 0.00119844 & 0.0153524 & 0.00150934 & 0.000693353 & 0.000220067 & 0.00343836 & 0.00121643 & 0.000591956 & 6.75242e-05 & 1.21226e-06 \\
0.02 & 0.00193115 & 0.0148044 & 0.00121749 & 0.000768802 & 0.000680037 & 0.004222 & 0.00124573 & 0.000549918 & 6.05825e-05 & 1.60292e-06 \\
\end{tabular}
\caption{Our assumed metal yields, computed from~\citet{nomo06} assuming a~\citet{krou01} IMF
from 0.1--100$\msun$ and a 50\% hypernova fraction.}
\label{table:yields}
\end{table}

\begin{table}
\small
\setlength\tabcolsep{2.1pt}
\begin{tabular}{l|cccccccccc}
$f_\mathrm{HNe}$ & C & O & Si & Fe & N & Ne & Mg & S & Ca & Ti\\
\hline
0 & 0.00108745 & 0.0186467 & 0.00161469 & 0.000703623 & 7.41348e-05 & 0.00362083 & 0.0015565 & 0.00066296 & 8.10135e-05 & 1.21059e-06 \\
0.5 & 0.00111629 & 0.0222635 & 0.0016386 & 0.000891533 & 9.13731e-05 & 0.00431286 & 0.00199825 & 0.000627048 & 7.49162e-05 & 1.60633e-06 \\
1.0 & 0.00114513 & 0.0258802 & 0.00166252 & 0.00107944 & 0.000108611 & 0.00500489 & 0.00244 & 0.000591136 & 6.88188e-05 & 2.00206e-06 \\
\end{tabular}
\caption{The dependence of the metal yield on HNe fraction $f_\mathrm{HNe}$ for the same IMF as in Table~\ref{table:yieldDep} and a 
stellar population with $Z=0.001$.  The second row is reproduced from Table~\ref{table:yields}}
\label{table:yieldDep}
\end{table}

\twocolumn



\bibliographystyle{mnras}
\bibliography{td} 

\begin{thebibliography}{}
\makeatletter
\relax
\def\mn@urlcharsother{\let\do\@makeother \do\$\do\&\do\#\do\^\do\_\do\%\do\~}
\def\mn@doi{\begingroup\mn@urlcharsother \@ifnextchar [ {\mn@doi@}
  {\mn@doi@[]}}
\def\mn@doi@[#1]#2{\def\@tempa{#1}\ifx\@tempa\@empty \href
  {http://dx.doi.org/#2} {doi:#2}\else \href {http://dx.doi.org/#2} {#1}\fi
  \endgroup}
\def\mn@eprint#1#2{\mn@eprint@#1:#2::\@nil}
\def\mn@eprint@arXiv#1{\href {http://arxiv.org/abs/#1} {{\tt arXiv:#1}}}
\def\mn@eprint@dblp#1{\href {http://dblp.uni-trier.de/rec/bibtex/#1.xml}
  {dblp:#1}}
\def\mn@eprint@#1:#2:#3:#4\@nil{\def\@tempa {#1}\def\@tempb {#2}\def\@tempc
  {#3}\ifx \@tempc \@empty \let \@tempc \@tempb \let \@tempb \@tempa \fi \ifx
  \@tempb \@empty \def\@tempb {arXiv}\fi \@ifundefined
  {mn@eprint@\@tempb}{\@tempb:\@tempc}{\expandafter \expandafter \csname
  mn@eprint@\@tempb\endcsname \expandafter{\@tempc}}}

\bibitem[\protect\citeauthoryear{{Ahn}, {Iliev}, {Shapiro}, {Mellema}, {Koda}
  \& {Mao}}{{Ahn} et~al.}{2012}]{ahn12}
{Ahn} K.,  {Iliev} I.~T.,  {Shapiro} P.~R.,  {Mellema} G.,  {Koda} J.,   {Mao}
  Y.,  2012, \mn@doi [\apjl] {10.1088/2041-8205/756/1/L16}, \href
  {http://adsabs.harvard.edu/abs/2012ApJ...756L..16A} {756, L16}

\bibitem[\protect\citeauthoryear{{Alvarez}, {Finlator}  \& {Trenti}}{{Alvarez}
  et~al.}{2012}]{alva12}
{Alvarez} M.~A.,  {Finlator} K.,   {Trenti} M.,  2012, \mn@doi [\apjl]
  {10.1088/2041-8205/759/2/L38}, \href
  {http://adsabs.harvard.edu/abs/2012ApJ...759L..38A} {759, L38}

\bibitem[\protect\citeauthoryear{{Angel}, {Poole}, {Ludlow}, {Duffy}, {Geil},
  {Mutch}, {Mesinger}  \& {Wyithe}}{{Angel} et~al.}{2016}]{ange16}
{Angel} P.~W.,  {Poole} G.~B.,  {Ludlow} A.~D.,  {Duffy} A.~R.,  {Geil} P.~M.,
  {Mutch} S.~J.,  {Mesinger} A.,   {Wyithe} J.~S.~B.,  2016, \mn@doi [\mnras]
  {10.1093/mnras/stw737}, \href
  {http://adsabs.harvard.edu/abs/2016MNRAS.459.2106A} {459, 2106}

\bibitem[\protect\citeauthoryear{{Asplund}, {Grevesse}, {Sauval}  \&
  {Scott}}{{Asplund} et~al.}{2009}]{aspl09}
{Asplund} M.,  {Grevesse} N.,  {Sauval} A.~J.,   {Scott} P.,  2009, \mn@doi
  [\araa] {10.1146/annurev.astro.46.060407.145222}, \href
  {http://adsabs.harvard.edu/abs/2009ARA%26A..47..481A} {47, 481}

\bibitem[\protect\citeauthoryear{{Ba{\~n}ados} et~al.,}{{Ba{\~n}ados}
  et~al.}{2018}]{bana18}
{Ba{\~n}ados} E.,  et~al., 2018, \mn@doi [\nat] {10.1038/nature25180}, \href
  {http://adsabs.harvard.edu/abs/2018Natur.553..473B} {553, 473}

\bibitem[\protect\citeauthoryear{{Bahcall} \& {Peebles}}{{Bahcall} \&
  {Peebles}}{1969}]{bahc69}
{Bahcall} J.~N.,  {Peebles} P.~J.~E.,  1969, \mn@doi [\apjl] {10.1086/180337},
  \href {http://adsabs.harvard.edu/abs/1969ApJ...156L...7B} {156, L7}

\bibitem[\protect\citeauthoryear{{Barkana} \& {Loeb}}{{Barkana} \&
  {Loeb}}{2000}]{bark00}
{Barkana} R.,  {Loeb} A.,  2000, \mn@doi [\apj] {10.1086/309229}, \href
  {http://adsabs.harvard.edu/abs/2000ApJ...539...20B} {539, 20}

\bibitem[\protect\citeauthoryear{{Barkana} \& {Loeb}}{{Barkana} \&
  {Loeb}}{2004}]{bark04}
{Barkana} R.,  {Loeb} A.,  2004, \mn@doi [\apj] {10.1086/421079}, \href
  {http://adsabs.harvard.edu/abs/2004ApJ...609..474B} {609, 474}

\bibitem[\protect\citeauthoryear{{Becker}, {Bolton}, {Haehnelt}  \&
  {Sargent}}{{Becker} et~al.}{2011a}]{beck11b}
{Becker} G.~D.,  {Bolton} J.~S.,  {Haehnelt} M.~G.,   {Sargent} W.~L.~W.,
  2011a, \mn@doi [\mnras] {10.1111/j.1365-2966.2010.17507.x}, \href
  {http://adsabs.harvard.edu/abs/2011MNRAS.410.1096B} {410, 1096}

\bibitem[\protect\citeauthoryear{{Becker}, {Sargent}, {Rauch}  \&
  {Calverley}}{{Becker} et~al.}{2011b}]{beck11a}
{Becker} G.~D.,  {Sargent} W.~L.~W.,  {Rauch} M.,   {Calverley} A.~P.,  2011b,
  \mn@doi [\apj] {10.1088/0004-637X/735/2/93}, \href
  {http://adsabs.harvard.edu/abs/2011ApJ...735...93B} {735, 93}

\bibitem[\protect\citeauthoryear{{Becker}, {Bolton}  \& {Lidz}}{{Becker}
  et~al.}{2015a}]{beck15b}
{Becker} G.~D.,  {Bolton} J.~S.,   {Lidz} A.,  2015a, \mn@doi [\pasa]
  {10.1017/pasa.2015.45}, \href
  {http://adsabs.harvard.edu/abs/2015PASA...32...45B} {32, e045}

\bibitem[\protect\citeauthoryear{{Becker}, {Bolton}, {Madau}, {Pettini},
  {Ryan-Weber}  \& {Venemans}}{{Becker} et~al.}{2015b}]{beck15a}
{Becker} G.~D.,  {Bolton} J.~S.,  {Madau} P.,  {Pettini} M.,  {Ryan-Weber}
  E.~V.,   {Venemans} B.~P.,  2015b, \mn@doi [\mnras] {10.1093/mnras/stu2646},
  \href {http://adsabs.harvard.edu/abs/2015MNRAS.447.3402B} {447, 3402}

\bibitem[\protect\citeauthoryear{{Becker}, {Davies}, {Furlanetto}, {Malkan},
  {Boera}  \& {Douglass}}{{Becker} et~al.}{2018}]{beck18}
{Becker} G.~D.,  {Davies} F.~B.,  {Furlanetto} S.~R.,  {Malkan} M.~A.,  {Boera}
  E.,   {Douglass} C.,  2018, preprint, \href
  {http://adsabs.harvard.edu/abs/2018arXiv180308932B} {} (\mn@eprint {arXiv}
  {1803.08932})

\bibitem[\protect\citeauthoryear{{Berg}, {Erb}, {Auger}, {Pettini}  \&
  {Brammer}}{{Berg} et~al.}{2018}]{berg18}
{Berg} D.~A.,  {Erb} D.~K.,  {Auger} M.~W.,  {Pettini} M.,   {Brammer} G.~B.,
  2018, preprint, \href {http://adsabs.harvard.edu/abs/2018arXiv180302340B} {}
  (\mn@eprint {arXiv} {1803.02340})

\bibitem[\protect\citeauthoryear{{Bird}, {Rubin}, {Suresh}  \&
  {Hernquist}}{{Bird} et~al.}{2016}]{bird16}
{Bird} S.,  {Rubin} K.~H.~R.,  {Suresh} J.,   {Hernquist} L.,  2016, \mn@doi
  [\mnras] {10.1093/mnras/stw1582}, \href
  {http://adsabs.harvard.edu/abs/2016MNRAS.462..307B} {462, 307}

\bibitem[\protect\citeauthoryear{{Bird}, {Garnett}  \& {Ho}}{{Bird}
  et~al.}{2017}]{bird17}
{Bird} S.,  {Garnett} R.,   {Ho} S.,  2017, \mn@doi [\mnras]
  {10.1093/mnras/stw3246}, \href
  {http://adsabs.harvard.edu/abs/2017MNRAS.466.2111B} {466, 2111}

\bibitem[\protect\citeauthoryear{{Bolton} \& {Becker}}{{Bolton} \&
  {Becker}}{2009}]{bolt09}
{Bolton} J.~S.,  {Becker} G.~D.,  2009, \mn@doi [\mnras]
  {10.1111/j.1745-3933.2009.00700.x}, \href
  {http://adsabs.harvard.edu/abs/2009MNRAS.398L..26B} {398, L26}

\bibitem[\protect\citeauthoryear{{Bolton} \& {Haehnelt}}{{Bolton} \&
  {Haehnelt}}{2007}]{bolt07}
{Bolton} J.~S.,  {Haehnelt} M.~G.,  2007, \mn@doi [\mnras]
  {10.1111/j.1365-2966.2007.12372.x}, \href
  {http://adsabs.harvard.edu/abs/2007MNRAS.382..325B} {382, 325}

\bibitem[\protect\citeauthoryear{{Bosman}, {Becker}, {Haehnelt}, {Hewett},
  {McMahon}, {Mortlock}, {Simpson}  \& {Venemans}}{{Bosman}
  et~al.}{2017}]{bosm17}
{Bosman} S.~E.~I.,  {Becker} G.~D.,  {Haehnelt} M.~G.,  {Hewett} P.~C.,
  {McMahon} R.~G.,  {Mortlock} D.~J.,  {Simpson} C.,   {Venemans} B.~P.,  2017,
  \mn@doi [\mnras] {10.1093/mnras/stx1305}, \href
  {http://adsabs.harvard.edu/abs/2017MNRAS.470.1919B} {470, 1919}

\bibitem[\protect\citeauthoryear{{Bosman}, {Fan}, {Jiang}, {Reed}, {Matsuoka},
  {Becker}  \& {Haehnelt}}{{Bosman} et~al.}{2018}]{bosm18}
{Bosman} S.~E.~I.,  {Fan} X.,  {Jiang} L.,  {Reed} S.~L.,  {Matsuoka} Y.,
  {Becker} G.~D.,   {Haehnelt} M.~G.,  2018, preprint, \href
  {http://adsabs.harvard.edu/abs/2018arXiv180208177B} {} (\mn@eprint {arXiv}
  {1802.08177})

\bibitem[\protect\citeauthoryear{{Bouwens} et~al.,}{{Bouwens}
  et~al.}{2015}]{bouw15}
{Bouwens} R.~J.,  et~al., 2015, \mn@doi [\apj] {10.1088/0004-637X/803/1/34},
  \href {http://adsabs.harvard.edu/abs/2015ApJ...803...34B} {803, 34}

\bibitem[\protect\citeauthoryear{{Bryan} \& {Norman}}{{Bryan} \&
  {Norman}}{1998}]{brya98}
{Bryan} G.~L.,  {Norman} M.~L.,  1998, \mn@doi [\apj] {10.1086/305262}, \href
  {http://adsabs.harvard.edu/abs/1998ApJ...495...80B} {495, 80}

\bibitem[\protect\citeauthoryear{{Cai}, {Fan}, {Dave}, {Finlator}  \&
  {Oppenheimer}}{{Cai} et~al.}{2017}]{cai17}
{Cai} Z.,  {Fan} X.,  {Dave} R.,  {Finlator} K.,   {Oppenheimer} B.,  2017,
  \mn@doi [\apjl] {10.3847/2041-8213/aa8fc6}, \href
  {http://adsabs.harvard.edu/abs/2017ApJ...849L..18C} {849, L18}

\bibitem[\protect\citeauthoryear{{Calverley}, {Becker}, {Haehnelt}  \&
  {Bolton}}{{Calverley} et~al.}{2011}]{calv11}
{Calverley} A.~P.,  {Becker} G.~D.,  {Haehnelt} M.~G.,   {Bolton} J.~S.,  2011,
  \mn@doi [\mnras] {10.1111/j.1365-2966.2010.18072.x}, \href
  {http://adsabs.harvard.edu/abs/2011MNRAS.412.2543C} {412, 2543}

\bibitem[\protect\citeauthoryear{{Chabrier}}{{Chabrier}}{2003}]{chab03}
{Chabrier} G.,  2003, \mn@doi [\pasp] {10.1086/376392}, \href
  {http://adsabs.harvard.edu/abs/2003PASP..115..763C} {115, 763}

\bibitem[\protect\citeauthoryear{{Chardin}, {Haehnelt}, {Aubert}  \&
  {Puchwein}}{{Chardin} et~al.}{2015}]{char15}
{Chardin} J.,  {Haehnelt} M.~G.,  {Aubert} D.,   {Puchwein} E.,  2015, \mn@doi
  [\mnras] {10.1093/mnras/stv1786}, \href
  {http://adsabs.harvard.edu/abs/2015MNRAS.453.2943C} {453, 2943}

\bibitem[\protect\citeauthoryear{{Chen}, {Norman}, {Xu}  \& {Wise}}{{Chen}
  et~al.}{2017a}]{chenp17}
{Chen} P.,  {Norman} M.~L.,  {Xu} H.,   {Wise} J.~H.,  2017a, preprint, \href
  {http://adsabs.harvard.edu/abs/2017arXiv170500026C} {} (\mn@eprint {arXiv}
  {1705.00026})

\bibitem[\protect\citeauthoryear{{Chen} et~al.,}{{Chen}
  et~al.}{2017b}]{chensf17}
{Chen} S.-F.~S.,  et~al., 2017b, \mn@doi [\apj] {10.3847/1538-4357/aa9707},
  \href {http://adsabs.harvard.edu/abs/2017ApJ...850..188C} {850, 188}

\bibitem[\protect\citeauthoryear{{Chisholm}, {Tremonti}, {Leitherer}, {Chen},
  {Wofford}  \& {Lundgren}}{{Chisholm} et~al.}{2015}]{chis15}
{Chisholm} J.,  {Tremonti} C.~A.,  {Leitherer} C.,  {Chen} Y.,  {Wofford} A.,
  {Lundgren} B.,  2015, \mn@doi [\apj] {10.1088/0004-637X/811/2/149}, \href
  {http://adsabs.harvard.edu/abs/2015ApJ...811..149C} {811, 149}

\bibitem[\protect\citeauthoryear{{Christensen}, {Dav{\'e}}, {Governato},
  {Pontzen}, {Brooks}, {Munshi}, {Quinn}  \& {Wadsley}}{{Christensen}
  et~al.}{2016}]{chri16}
{Christensen} C.~R.,  {Dav{\'e}} R.,  {Governato} F.,  {Pontzen} A.,  {Brooks}
  A.,  {Munshi} F.,  {Quinn} T.,   {Wadsley} J.,  2016, \mn@doi [\apj]
  {10.3847/0004-637X/824/1/57}, \href
  {http://adsabs.harvard.edu/abs/2016ApJ...824...57C} {824, 57}

\bibitem[\protect\citeauthoryear{{Ciardi}, {Bolton}, {Maselli}  \&
  {Graziani}}{{Ciardi} et~al.}{2012}]{ciar12}
{Ciardi} B.,  {Bolton} J.~S.,  {Maselli} A.,   {Graziani} L.,  2012, \mn@doi
  [\mnras] {10.1111/j.1365-2966.2012.20902.x}, \href
  {http://adsabs.harvard.edu/abs/2012MNRAS.423..558C} {423, 558}

\bibitem[\protect\citeauthoryear{{Conroy}, {Gunn}  \& {White}}{{Conroy}
  et~al.}{2009}]{conr09}
{Conroy} C.,  {Gunn} J.~E.,   {White} M.,  2009, \mn@doi [\apj]
  {10.1088/0004-637X/699/1/486}, \href
  {http://adsabs.harvard.edu/abs/2009ApJ...699..486C} {699, 486}

\bibitem[\protect\citeauthoryear{{Cooke}, {Pettini}, {Steidel}, {Rudie}  \&
  {Nissen}}{{Cooke} et~al.}{2011}]{cook11}
{Cooke} R.,  {Pettini} M.,  {Steidel} C.~C.,  {Rudie} G.~C.,   {Nissen} P.~E.,
  2011, \mn@doi [\mnras] {10.1111/j.1365-2966.2011.19365.x}, \href
  {http://adsabs.harvard.edu/abs/2011MNRAS.417.1534C} {417, 1534}

\bibitem[\protect\citeauthoryear{{Cristiani}, {Serrano}, {Fontanot}, {Vanzella}
   \& {Monaco}}{{Cristiani} et~al.}{2016}]{cris16}
{Cristiani} S.,  {Serrano} L.~M.,  {Fontanot} F.,  {Vanzella} E.,   {Monaco}
  P.,  2016, \mn@doi [\mnras] {10.1093/mnras/stw1810}, \href
  {http://adsabs.harvard.edu/abs/2016MNRAS.462.2478C} {462, 2478}

\bibitem[\protect\citeauthoryear{{Crowther} et~al.,}{{Crowther}
  et~al.}{2016}]{crow16}
{Crowther} P.~A.,  et~al., 2016, \mn@doi [\mnras] {10.1093/mnras/stw273}, \href
  {http://adsabs.harvard.edu/abs/2016MNRAS.458..624C} {458, 624}

\bibitem[\protect\citeauthoryear{{D'Aloisio}, {McQuinn}  \& {Trac}}{{D'Aloisio}
  et~al.}{2015}]{dalo15}
{D'Aloisio} A.,  {McQuinn} M.,   {Trac} H.,  2015, \mn@doi [\apjl]
  {10.1088/2041-8205/813/2/L38}, \href
  {http://adsabs.harvard.edu/abs/2015ApJ...813L..38D} {813, L38}

\bibitem[\protect\citeauthoryear{{D'Aloisio}, {Upton Sanderbeck}, {McQuinn},
  {Trac}  \& {Shapiro}}{{D'Aloisio} et~al.}{2017}]{dalo17}
{D'Aloisio} A.,  {Upton Sanderbeck} P.~R.,  {McQuinn} M.,  {Trac} H.,
  {Shapiro} P.~R.,  2017, \mn@doi [\mnras] {10.1093/mnras/stx711}, \href
  {http://adsabs.harvard.edu/abs/2017MNRAS.468.4691D} {468, 4691}

\bibitem[\protect\citeauthoryear{{D'Aloisio}, {McQuinn}, {Davies}  \&
  {Furlanetto}}{{D'Aloisio} et~al.}{2018}]{dalo18}
{D'Aloisio} A.,  {McQuinn} M.,  {Davies} F.~B.,   {Furlanetto} S.~R.,  2018,
  \mn@doi [\mnras] {10.1093/mnras/stx2341}, \href
  {http://adsabs.harvard.edu/abs/2018MNRAS.473..560D} {473, 560}

\bibitem[\protect\citeauthoryear{{D'Odorico} et~al.,}{{D'Odorico}
  et~al.}{2013}]{dodo13}
{D'Odorico} V.,  et~al., 2013, \mn@doi [\mnras] {10.1093/mnras/stt1365}, \href
  {http://adsabs.harvard.edu/abs/2013MNRAS.435.1198D} {435, 1198}

\bibitem[\protect\citeauthoryear{{Dav{\'e}}, {Thompson}  \&
  {Hopkins}}{{Dav{\'e}} et~al.}{2016}]{dave16}
{Dav{\'e}} R.,  {Thompson} R.,   {Hopkins} P.~F.,  2016, \mn@doi [\mnras]
  {10.1093/mnras/stw1862}, \href
  {http://adsabs.harvard.edu/abs/2016MNRAS.462.3265D} {462, 3265}

\bibitem[\protect\citeauthoryear{{Davies}, {Becker}  \& {Furlanetto}}{{Davies}
  et~al.}{2017}]{davi17}
{Davies} F.~B.,  {Becker} G.~D.,   {Furlanetto} S.~R.,  2017, preprint, \href
  {http://adsabs.harvard.edu/abs/2017arXiv170808927D} {} (\mn@eprint {arXiv}
  {1708.08927})

\bibitem[\protect\citeauthoryear{{Davies} et~al.,}{{Davies}
  et~al.}{2018a}]{davi18b}
{Davies} F.~B.,  et~al., 2018a, preprint, \href
  {http://adsabs.harvard.edu/abs/2018arXiv180206066D} {} (\mn@eprint {arXiv}
  {1802.06066})

\bibitem[\protect\citeauthoryear{{Davies}, {Hennawi}, {Eilers}  \&
  {Luki{\'c}}}{{Davies} et~al.}{2018b}]{davi18a}
{Davies} F.~B.,  {Hennawi} J.~F.,  {Eilers} A.-C.,   {Luki{\'c}} Z.,  2018b,
  \mn@doi [\apj] {10.3847/1538-4357/aaaf70}, \href
  {http://adsabs.harvard.edu/abs/2018ApJ...855..106D} {855, 106}

\bibitem[\protect\citeauthoryear{{D{\'{\i}}az}, {Koyama}, {Ryan-Weber},
  {Cooke}, {Ouchi}, {Shimasaku}  \& {Nakata}}{{D{\'{\i}}az}
  et~al.}{2014}]{diaz14}
{D{\'{\i}}az} C.~G.,  {Koyama} Y.,  {Ryan-Weber} E.~V.,  {Cooke} J.,  {Ouchi}
  M.,  {Shimasaku} K.,   {Nakata} F.,  2014, \mn@doi [\mnras]
  {10.1093/mnras/stu914}, \href
  {http://adsabs.harvard.edu/abs/2014MNRAS.442..946D} {442, 946}

\bibitem[\protect\citeauthoryear{{Doussot}, {Trac}  \& {Cen}}{{Doussot}
  et~al.}{2017}]{dous17}
{Doussot} A.,  {Trac} H.,   {Cen} R.,  2017, preprint, \href
  {http://adsabs.harvard.edu/abs/2017arXiv171204464D} {} (\mn@eprint {arXiv}
  {1712.04464})

\bibitem[\protect\citeauthoryear{{Duncan} et~al.,}{{Duncan}
  et~al.}{2014}]{dunc14}
{Duncan} K.,  et~al., 2014, \mn@doi [\mnras] {10.1093/mnras/stu1622}, \href
  {http://adsabs.harvard.edu/abs/2014MNRAS.444.2960D} {444, 2960}

\bibitem[\protect\citeauthoryear{{Faisst}}{{Faisst}}{2016}]{fais16}
{Faisst} A.~L.,  2016, \mn@doi [\apj] {10.3847/0004-637X/829/2/99}, \href
  {http://adsabs.harvard.edu/abs/2016ApJ...829...99F} {829, 99}

\bibitem[\protect\citeauthoryear{{Faucher-Gigu{\`e}re}, {Lidz}, {Zaldarriaga}
  \& {Hernquist}}{{Faucher-Gigu{\`e}re} et~al.}{2009}]{fauc09}
{Faucher-Gigu{\`e}re} C.-A.,  {Lidz} A.,  {Zaldarriaga} M.,   {Hernquist} L.,
  2009, \mn@doi [\apj] {10.1088/0004-637X/703/2/1416}, \href
  {http://adsabs.harvard.edu/abs/2009ApJ...703.1416F} {703, 1416}

\bibitem[\protect\citeauthoryear{{Finkelstein} et~al.,}{{Finkelstein}
  et~al.}{2015}]{fink15}
{Finkelstein} S.~L.,  et~al., 2015, \mn@doi [\apj]
  {10.1088/0004-637X/810/1/71}, \href
  {http://adsabs.harvard.edu/abs/2015ApJ...810...71F} {810, 71}

\bibitem[\protect\citeauthoryear{{Finlator}, {{\"O}zel}  \&
  {Dav{\'e}}}{{Finlator} et~al.}{2009}]{finl09}
{Finlator} K.,  {{\"O}zel} F.,   {Dav{\'e}} R.,  2009, \mn@doi [\mnras]
  {10.1111/j.1365-2966.2008.14190.x}, \href
  {http://adsabs.harvard.edu/abs/2009MNRAS.393.1090F} {393, 1090}

\bibitem[\protect\citeauthoryear{{Finlator}, {Dav{\'e}}  \&
  {{\"O}zel}}{{Finlator} et~al.}{2011}]{finl11}
{Finlator} K.,  {Dav{\'e}} R.,   {{\"O}zel} F.,  2011, \mn@doi [\apj]
  {10.1088/0004-637X/743/2/169}, \href
  {http://adsabs.harvard.edu/abs/2011ApJ...743..169F} {743, 169}

\bibitem[\protect\citeauthoryear{{Finlator}, {Mu{\~n}oz}, {Oppenheimer}, {Oh},
  {{\"O}zel}  \& {Dav{\'e}}}{{Finlator} et~al.}{2013}]{finl13}
{Finlator} K.,  {Mu{\~n}oz} J.~A.,  {Oppenheimer} B.~D.,  {Oh} S.~P.,
  {{\"O}zel} F.,   {Dav{\'e}} R.,  2013, \mn@doi [\mnras]
  {10.1093/mnras/stt1697}, \href
  {http://adsabs.harvard.edu/abs/2013MNRAS.436.1818F} {436, 1818}

\bibitem[\protect\citeauthoryear{{Finlator}, {Thompson}, {Huang}, {Dav{\'e}},
  {Zackrisson}  \& {Oppenheimer}}{{Finlator} et~al.}{2015}]{finl15}
{Finlator} K.,  {Thompson} R.,  {Huang} S.,  {Dav{\'e}} R.,  {Zackrisson} E.,
  {Oppenheimer} B.~D.,  2015, \mn@doi [\mnras] {10.1093/mnras/stu2668}, \href
  {http://adsabs.harvard.edu/abs/2015MNRAS.447.2526F} {447, 2526}

\bibitem[\protect\citeauthoryear{{Finlator}, {Oppenheimer}, {Dav{\'e}},
  {Zackrisson}, {Thompson}  \& {Huang}}{{Finlator} et~al.}{2016}]{finl16}
{Finlator} K.,  {Oppenheimer} B.~D.,  {Dav{\'e}} R.,  {Zackrisson} E.,
  {Thompson} R.,   {Huang} S.,  2016, \mn@doi [\mnras] {10.1093/mnras/stw805},
  \href {http://adsabs.harvard.edu/abs/2016MNRAS.459.2299F} {459, 2299}

\bibitem[\protect\citeauthoryear{{Finlator} et~al.,}{{Finlator}
  et~al.}{2017}]{finl17}
{Finlator} K.,  et~al., 2017, \mn@doi [\mnras] {10.1093/mnras/stw2433}, \href
  {http://adsabs.harvard.edu/abs/2017MNRAS.464.1633F} {464, 1633}

\bibitem[\protect\citeauthoryear{{Fontanot}, {Hirschmann}  \& {De
  Lucia}}{{Fontanot} et~al.}{2017}]{font17}
{Fontanot} F.,  {Hirschmann} M.,   {De Lucia} G.,  2017, \mn@doi [\apjl]
  {10.3847/2041-8213/aa74bd}, \href
  {http://adsabs.harvard.edu/abs/2017ApJ...842L..14F} {842, L14}

\bibitem[\protect\citeauthoryear{{Furlanetto} \& {Oh}}{{Furlanetto} \&
  {Oh}}{2008}]{furl08}
{Furlanetto} S.~R.,  {Oh} S.~P.,  2008, \mn@doi [\apj] {10.1086/588546}, \href
  {http://adsabs.harvard.edu/abs/2008ApJ...681....1F} {681, 1}

\bibitem[\protect\citeauthoryear{{Furlanetto} \& {Oh}}{{Furlanetto} \&
  {Oh}}{2009}]{furl09}
{Furlanetto} S.~R.,  {Oh} S.~P.,  2009, \mn@doi [\apj]
  {10.1088/0004-637X/701/1/94}, \href
  {http://adsabs.harvard.edu/abs/2009ApJ...701...94F} {701, 94}

\bibitem[\protect\citeauthoryear{{Garzilli}, {Boyarsky}  \&
  {Ruchayskiy}}{{Garzilli} et~al.}{2017}]{garz17}
{Garzilli} A.,  {Boyarsky} A.,   {Ruchayskiy} O.,  2017, \mn@doi [Physics
  Letters B] {10.1016/j.physletb.2017.08.022}, \href
  {http://adsabs.harvard.edu/abs/2017PhLB..773..258G} {773, 258}

\bibitem[\protect\citeauthoryear{{Gnedin}}{{Gnedin}}{2014}]{gned14}
{Gnedin} N.~Y.,  2014, \mn@doi [\apj] {10.1088/0004-637X/793/1/29}, \href
  {http://adsabs.harvard.edu/abs/2014ApJ...793...29G} {793, 29}

\bibitem[\protect\citeauthoryear{{Gnedin} \& {Abel}}{{Gnedin} \&
  {Abel}}{2001}]{gned01}
{Gnedin} N.~Y.,  {Abel} T.,  2001, \mn@doi [\na]
  {10.1016/S1384-1076(01)00068-9}, \href
  {http://adsabs.harvard.edu/abs/2001NewA....6..437G} {6, 437}

\bibitem[\protect\citeauthoryear{{Gnedin} \& {Hamilton}}{{Gnedin} \&
  {Hamilton}}{2002}]{gned02}
{Gnedin} N.~Y.,  {Hamilton} A.~J.~S.,  2002, \mn@doi [\mnras]
  {10.1046/j.1365-8711.2002.05490.x}, \href
  {http://adsabs.harvard.edu/abs/2002MNRAS.334..107G} {334, 107}

\bibitem[\protect\citeauthoryear{{Gnedin} \& {Ostriker}}{{Gnedin} \&
  {Ostriker}}{1997}]{gned97}
{Gnedin} N.~Y.,  {Ostriker} J.~P.,  1997, \mn@doi [\apj] {10.1086/304548},
  \href {http://adsabs.harvard.edu/abs/1997ApJ...486..581G} {486, 581}

\bibitem[\protect\citeauthoryear{{Gnedin}, {Becker}  \& {Fan}}{{Gnedin}
  et~al.}{2017}]{gned17}
{Gnedin} N.~Y.,  {Becker} G.~D.,   {Fan} X.,  2017, \mn@doi [\apj]
  {10.3847/1538-4357/aa6c24}, \href
  {http://adsabs.harvard.edu/abs/2017ApJ...841...26G} {841, 26}

\bibitem[\protect\citeauthoryear{{Grazian} et~al.,}{{Grazian}
  et~al.}{2015}]{graz15}
{Grazian} A.,  et~al., 2015, \mn@doi [\aap] {10.1051/0004-6361/201424750},
  \href {http://adsabs.harvard.edu/abs/2015A%26A...575A..96G} {575, A96}

\bibitem[\protect\citeauthoryear{{Grazian} et~al.,}{{Grazian}
  et~al.}{2018}]{graz18}
{Grazian} A.,  et~al., 2018, \mn@doi [\aap] {10.1051/0004-6361/201732385},
  \href {http://adsabs.harvard.edu/abs/2018A%26A...613A..44G} {613, A44}

\bibitem[\protect\citeauthoryear{{Greif} \& {Bromm}}{{Greif} \&
  {Bromm}}{2006}]{grei06}
{Greif} T.~H.,  {Bromm} V.,  2006, \mn@doi [\mnras]
  {10.1111/j.1365-2966.2006.11017.x}, \href
  {http://adsabs.harvard.edu/abs/2006MNRAS.373..128G} {373, 128}

\bibitem[\protect\citeauthoryear{{Greig}, {Mesinger}, {Haiman}  \&
  {Simcoe}}{{Greig} et~al.}{2017}]{grei17}
{Greig} B.,  {Mesinger} A.,  {Haiman} Z.,   {Simcoe} R.~A.,  2017, \mn@doi
  [\mnras] {10.1093/mnras/stw3351}, \href
  {http://adsabs.harvard.edu/abs/2017MNRAS.466.4239G} {466, 4239}

\bibitem[\protect\citeauthoryear{{Haardt} \& {Madau}}{{Haardt} \&
  {Madau}}{2012}]{haar12}
{Haardt} F.,  {Madau} P.,  2012, \mn@doi [\apj] {10.1088/0004-637X/746/2/125},
  \href {http://adsabs.harvard.edu/abs/2012ApJ...746..125H} {746, 125}

\bibitem[\protect\citeauthoryear{{Haehnelt}, {Steinmetz}  \&
  {Rauch}}{{Haehnelt} et~al.}{1998}]{haeh98}
{Haehnelt} M.~G.,  {Steinmetz} M.,   {Rauch} M.,  1998, \mn@doi [\apj]
  {10.1086/305323}, \href {http://adsabs.harvard.edu/abs/1998ApJ...495..647H}
  {495, 647}

\bibitem[\protect\citeauthoryear{{Hahn} \& {Abel}}{{Hahn} \&
  {Abel}}{2011}]{hahn11}
{Hahn} O.,  {Abel} T.,  2011, \mn@doi [\mnras]
  {10.1111/j.1365-2966.2011.18820.x}, \href
  {http://adsabs.harvard.edu/abs/2011MNRAS.415.2101H} {415, 2101}

\bibitem[\protect\citeauthoryear{{Hassan}, {Dav{\'e}}, {Mitra}, {Finlator},
  {Ciardi}  \& {Santos}}{{Hassan} et~al.}{2018}]{hass18}
{Hassan} S.,  {Dav{\'e}} R.,  {Mitra} S.,  {Finlator} K.,  {Ciardi} B.,
  {Santos} M.~G.,  2018, \mn@doi [\mnras] {10.1093/mnras/stx2194}, \href
  {http://adsabs.harvard.edu/abs/2018MNRAS.473..227H} {473, 227}

\bibitem[\protect\citeauthoryear{{Heringer}, {Pritchet}, {Kezwer}, {Graham},
  {Sand}  \& {Bildfell}}{{Heringer} et~al.}{2017}]{heri17}
{Heringer} E.,  {Pritchet} C.,  {Kezwer} J.,  {Graham} M.~L.,  {Sand} D.,
  {Bildfell} C.,  2017, \mn@doi [\apj] {10.3847/1538-4357/834/1/15}, \href
  {http://adsabs.harvard.edu/abs/2017ApJ...834...15H} {834, 15}

\bibitem[\protect\citeauthoryear{{Hopkins}}{{Hopkins}}{2013}]{hopk13}
{Hopkins} P.~F.,  2013, \mn@doi [\mnras] {10.1093/mnras/sts210}, \href
  {http://adsabs.harvard.edu/abs/2013MNRAS.428.2840H} {428, 2840}

\bibitem[\protect\citeauthoryear{{Huang} et~al.,}{{Huang}
  et~al.}{2017}]{huan17}
{Huang} K.-H.,  et~al., 2017, \mn@doi [\apj] {10.3847/1538-4357/aa62a6}, \href
  {http://adsabs.harvard.edu/abs/2017ApJ...838....6H} {838, 6}

\bibitem[\protect\citeauthoryear{{Hui} \& {Gnedin}}{{Hui} \&
  {Gnedin}}{1997}]{hui97}
{Hui} L.,  {Gnedin} N.~Y.,  1997, \mn@doi [\mnras] {10.1093/mnras/292.1.27},
  \href {http://adsabs.harvard.edu/abs/1997MNRAS.292...27H} {292, 27}

\bibitem[\protect\citeauthoryear{{Iliev}, {Mellema}, {Ahn}, {Shapiro}, {Mao}
  \& {Pen}}{{Iliev} et~al.}{2014}]{ilie14}
{Iliev} I.~T.,  {Mellema} G.,  {Ahn} K.,  {Shapiro} P.~R.,  {Mao} Y.,   {Pen}
  U.-L.,  2014, \mn@doi [\mnras] {10.1093/mnras/stt2497}, \href
  {http://adsabs.harvard.edu/abs/2014MNRAS.439..725I} {439, 725}

\bibitem[\protect\citeauthoryear{{Inoue} et~al.,}{{Inoue}
  et~al.}{2018}]{inou18}
{Inoue} A.~K.,  et~al., 2018, preprint, \href
  {http://adsabs.harvard.edu/abs/2018arXiv180100067I} {} (\mn@eprint {arXiv}
  {1801.00067})

\bibitem[\protect\citeauthoryear{{Izotov}, {Schaerer}, {Worseck}, {Guseva},
  {Thuan}, {Verhamme}, {Orlitov{\'a}}  \& {Fricke}}{{Izotov}
  et~al.}{2018}]{izot18}
{Izotov} Y.~I.,  {Schaerer} D.,  {Worseck} G.,  {Guseva} N.~G.,  {Thuan} T.~X.,
   {Verhamme} A.,  {Orlitov{\'a}} I.,   {Fricke} K.~J.,  2018, \mn@doi [\mnras]
  {10.1093/mnras/stx3115}, \href
  {http://adsabs.harvard.edu/abs/2018MNRAS.474.4514I} {474, 4514}

\bibitem[\protect\citeauthoryear{{Kakiichi} et~al.,}{{Kakiichi}
  et~al.}{2018}]{kaki18}
{Kakiichi} K.,  et~al., 2018, preprint, \href
  {http://adsabs.harvard.edu/abs/2018arXiv180302981K} {} (\mn@eprint {arXiv}
  {1803.02981})

\bibitem[\protect\citeauthoryear{{Katz}, {Weinberg}  \& {Hernquist}}{{Katz}
  et~al.}{1996}]{katz96}
{Katz} N.,  {Weinberg} D.~H.,   {Hernquist} L.,  1996, \mn@doi [\apjs]
  {10.1086/192305}, \href {http://adsabs.harvard.edu/abs/1996ApJS..105...19K}
  {105, 19}

\bibitem[\protect\citeauthoryear{{Katz}, {Kimm}, {Haehnelt}, {Sijacki},
  {Rosdahl}  \& {Blaizot}}{{Katz} et~al.}{2018}]{katz18}
{Katz} H.,  {Kimm} T.,  {Haehnelt} M.,  {Sijacki} D.,  {Rosdahl} J.,
  {Blaizot} J.,  2018, \mn@doi [\mnras] {10.1093/mnras/sty1225}, \href
  {http://adsabs.harvard.edu/abs/2018MNRAS.478.4986K} {478, 4986}

\bibitem[\protect\citeauthoryear{{Keating}, {Haehnelt}, {Becker}  \&
  {Bolton}}{{Keating} et~al.}{2014}]{keat14}
{Keating} L.~C.,  {Haehnelt} M.~G.,  {Becker} G.~D.,   {Bolton} J.~S.,  2014,
  \mn@doi [\mnras] {10.1093/mnras/stt2324}, \href
  {http://adsabs.harvard.edu/abs/2014MNRAS.438.1820K} {438, 1820}

\bibitem[\protect\citeauthoryear{{Keating}, {Puchwein}, {Haehnelt}, {Bird}  \&
  {Bolton}}{{Keating} et~al.}{2016}]{keat16}
{Keating} L.~C.,  {Puchwein} E.,  {Haehnelt} M.~G.,  {Bird} S.,   {Bolton}
  J.~S.,  2016, \mn@doi [\mnras] {10.1093/mnras/stw1306}, \href
  {http://adsabs.harvard.edu/abs/2016MNRAS.461..606K} {461, 606}

\bibitem[\protect\citeauthoryear{{Keating}, {Puchwein}  \&
  {Haehnelt}}{{Keating} et~al.}{2018}]{keat18}
{Keating} L.~C.,  {Puchwein} E.,   {Haehnelt} M.~G.,  2018, \mn@doi [\mnras]
  {10.1093/mnras/sty968}, \href
  {http://adsabs.harvard.edu/abs/2018MNRAS.477.5501K} {477, 5501}

\bibitem[\protect\citeauthoryear{{Khaire}, {Srianand}, {Choudhury}  \&
  {Gaikwad}}{{Khaire} et~al.}{2016}]{khai16}
{Khaire} V.,  {Srianand} R.,  {Choudhury} T.~R.,   {Gaikwad} P.,  2016, \mn@doi
  [\mnras] {10.1093/mnras/stw192}, \href
  {http://adsabs.harvard.edu/abs/2016MNRAS.457.4051K} {457, 4051}

\bibitem[\protect\citeauthoryear{{Konno} et~al.,}{{Konno}
  et~al.}{2018}]{konn18}
{Konno} A.,  et~al., 2018, \mn@doi [\pasj] {10.1093/pasj/psx131}, \href
  {http://adsabs.harvard.edu/abs/2018PASJ...70S..16K} {70, S16}

\bibitem[\protect\citeauthoryear{{Kroupa}}{{Kroupa}}{2001}]{krou01}
{Kroupa} P.,  2001, \mn@doi [\mnras] {10.1046/j.1365-8711.2001.04022.x}, \href
  {http://adsabs.harvard.edu/abs/2001MNRAS.322..231K} {322, 231}

\bibitem[\protect\citeauthoryear{{Kuhlen} \& {Faucher-Gigu{\`e}re}}{{Kuhlen} \&
  {Faucher-Gigu{\`e}re}}{2012}]{kuhl12}
{Kuhlen} M.,  {Faucher-Gigu{\`e}re} C.-A.,  2012, \mn@doi [\mnras]
  {10.1111/j.1365-2966.2012.20924.x}, \href
  {http://adsabs.harvard.edu/abs/2012MNRAS.423..862K} {423, 862}

\bibitem[\protect\citeauthoryear{{Leitherer} et~al.,}{{Leitherer}
  et~al.}{1999}]{leit99}
{Leitherer} C.,  et~al., 1999, \mn@doi [\apjs] {10.1086/313233}, \href
  {http://adsabs.harvard.edu/abs/1999ApJS..123....3L} {123, 3}

\bibitem[\protect\citeauthoryear{{Livermore}, {Finkelstein}  \&
  {Lotz}}{{Livermore} et~al.}{2017}]{live17}
{Livermore} R.~C.,  {Finkelstein} S.~L.,   {Lotz} J.~M.,  2017, \mn@doi [\apj]
  {10.3847/1538-4357/835/2/113}, \href
  {http://adsabs.harvard.edu/abs/2017ApJ...835..113L} {835, 113}

\bibitem[\protect\citeauthoryear{{Luki{\'c}}, {Stark}, {Nugent}, {White},
  {Meiksin}  \& {Almgren}}{{Luki{\'c}} et~al.}{2015}]{luki15}
{Luki{\'c}} Z.,  {Stark} C.~W.,  {Nugent} P.,  {White} M.,  {Meiksin} A.~A.,
  {Almgren} A.,  2015, \mn@doi [\mnras] {10.1093/mnras/stu2377}, \href
  {http://adsabs.harvard.edu/abs/2015MNRAS.446.3697L} {446, 3697}

\bibitem[\protect\citeauthoryear{{Lusso}, {Worseck}, {Hennawi}, {Prochaska},
  {Vignali}, {Stern}  \& {O'Meara}}{{Lusso} et~al.}{2015}]{luss15}
{Lusso} E.,  {Worseck} G.,  {Hennawi} J.~F.,  {Prochaska} J.~X.,  {Vignali} C.,
   {Stern} J.,   {O'Meara} J.~M.,  2015, \mn@doi [\mnras]
  {10.1093/mnras/stv516}, \href
  {http://adsabs.harvard.edu/abs/2015MNRAS.449.4204L} {449, 4204}

\bibitem[\protect\citeauthoryear{{Ma}, {Kasen}, {Hopkins},
  {Faucher-Gigu{\`e}re}, {Quataert}, {Kere{\v s}}  \& {Murray}}{{Ma}
  et~al.}{2015}]{ma15}
{Ma} X.,  {Kasen} D.,  {Hopkins} P.~F.,  {Faucher-Gigu{\`e}re} C.-A.,
  {Quataert} E.,  {Kere{\v s}} D.,   {Murray} N.,  2015, \mn@doi [\mnras]
  {10.1093/mnras/stv1679}, \href
  {http://adsabs.harvard.edu/abs/2015MNRAS.453..960M} {453, 960}

\bibitem[\protect\citeauthoryear{{Madau}}{{Madau}}{2017}]{mada17}
{Madau} P.,  2017, \mn@doi [\apj] {10.3847/1538-4357/aa9715}, \href
  {http://adsabs.harvard.edu/abs/2017ApJ...851...50M} {851, 50}

\bibitem[\protect\citeauthoryear{{Madau} \& {Dickinson}}{{Madau} \&
  {Dickinson}}{2014}]{maddick14}
{Madau} P.,  {Dickinson} M.,  2014, \mn@doi [\araa]
  {10.1146/annurev-astro-081811-125615}, \href
  {http://adsabs.harvard.edu/abs/2014ARA%26A..52..415M} {52, 415}

\bibitem[\protect\citeauthoryear{{Madau} \& {Haardt}}{{Madau} \&
  {Haardt}}{2015}]{mada15}
{Madau} P.,  {Haardt} F.,  2015, \mn@doi [\apjl] {10.1088/2041-8205/813/1/L8},
  \href {http://adsabs.harvard.edu/abs/2015ApJ...813L...8M} {813, L8}

\bibitem[\protect\citeauthoryear{{Maeda}, {R{\"o}pke}, {Fink}, {Hillebrandt},
  {Travaglio}  \& {Thielemann}}{{Maeda} et~al.}{2010}]{maed10}
{Maeda} K.,  {R{\"o}pke} F.~K.,  {Fink} M.,  {Hillebrandt} W.,  {Travaglio} C.,
    {Thielemann} F.-K.,  2010, \mn@doi [\apj] {10.1088/0004-637X/712/1/624},
  \href {http://adsabs.harvard.edu/abs/2010ApJ...712..624M} {712, 624}

\bibitem[\protect\citeauthoryear{{Manti}, {Gallerani}, {Ferrara}, {Greig}  \&
  {Feruglio}}{{Manti} et~al.}{2017}]{mant17}
{Manti} S.,  {Gallerani} S.,  {Ferrara} A.,  {Greig} B.,   {Feruglio} C.,
  2017, \mn@doi [\mnras] {10.1093/mnras/stw3168}, \href
  {http://adsabs.harvard.edu/abs/2017MNRAS.466.1160M} {466, 1160}

\bibitem[\protect\citeauthoryear{{Mason}, {Treu}, {Dijkstra}, {Mesinger},
  {Trenti}, {Pentericci}, {de Barros}  \& {Vanzella}}{{Mason}
  et~al.}{2018}]{maso18}
{Mason} C.~A.,  {Treu} T.,  {Dijkstra} M.,  {Mesinger} A.,  {Trenti} M.,
  {Pentericci} L.,  {de Barros} S.,   {Vanzella} E.,  2018, \mn@doi [\apj]
  {10.3847/1538-4357/aab0a7}, \href
  {http://adsabs.harvard.edu/abs/2018ApJ...856....2M} {856, 2}

\bibitem[\protect\citeauthoryear{McGreer, Mesinger  \& D'Odorico}{McGreer
  et~al.}{2015}]{mcgr14}
McGreer I.~D.,  Mesinger A.,   D'Odorico V.,  2015, \mn@doi [Monthly Notices of
  the Royal Astronomical Society] {10.1093/mnras/stu2449}, 447, 499

\bibitem[\protect\citeauthoryear{{McGreer}, {Fan}, {Jiang}  \& {Cai}}{{McGreer}
  et~al.}{2018}]{mcgr18}
{McGreer} I.~D.,  {Fan} X.,  {Jiang} L.,   {Cai} Z.,  2018, \mn@doi [\aj]
  {10.3847/1538-3881/aaaab4}, \href
  {http://adsabs.harvard.edu/abs/2018AJ....155..131M} {155, 131}

\bibitem[\protect\citeauthoryear{{McQuinn}}{{McQuinn}}{2016}]{mcqu16}
{McQuinn} M.,  2016, \mn@doi [\araa] {10.1146/annurev-astro-082214-122355},
  \href {http://adsabs.harvard.edu/abs/2016ARA%26A..54..313M} {54, 313}

\bibitem[\protect\citeauthoryear{{McQuinn}, {Lidz}, {Zaldarriaga}, {Hernquist},
  {Hopkins}, {Dutta}  \& {Faucher-Gigu{\`e}re}}{{McQuinn}
  et~al.}{2009}]{mcqu09}
{McQuinn} M.,  {Lidz} A.,  {Zaldarriaga} M.,  {Hernquist} L.,  {Hopkins} P.~F.,
   {Dutta} S.,   {Faucher-Gigu{\`e}re} C.-A.,  2009, \mn@doi [\apj]
  {10.1088/0004-637X/694/2/842}, \href
  {http://adsabs.harvard.edu/abs/2009ApJ...694..842M} {694, 842}

\bibitem[\protect\citeauthoryear{{Micheva}, {Iwata}  \& {Inoue}}{{Micheva}
  et~al.}{2017}]{mich17}
{Micheva} G.,  {Iwata} I.,   {Inoue} A.~K.,  2017, \mn@doi [\mnras]
  {10.1093/mnras/stw1329}, \href
  {http://adsabs.harvard.edu/abs/2017MNRAS.465..302M} {465, 302}

\bibitem[\protect\citeauthoryear{{Mitra}, {Choudhury}  \& {Ferrara}}{{Mitra}
  et~al.}{2015}]{mitr15}
{Mitra} S.,  {Choudhury} T.~R.,   {Ferrara} A.,  2015, \mn@doi [\mnras]
  {10.1093/mnrasl/slv134}, \href
  {http://adsabs.harvard.edu/abs/2015MNRAS.454L..76M} {454, L76}

\bibitem[\protect\citeauthoryear{{Muratov}, {Kere{\v s}},
  {Faucher-Gigu{\`e}re}, {Hopkins}, {Quataert}  \& {Murray}}{{Muratov}
  et~al.}{2015}]{mura15}
{Muratov} A.~L.,  {Kere{\v s}} D.,  {Faucher-Gigu{\`e}re} C.-A.,  {Hopkins}
  P.~F.,  {Quataert} E.,   {Murray} N.,  2015, \mn@doi [\mnras]
  {10.1093/mnras/stv2126}, \href
  {http://adsabs.harvard.edu/abs/2015MNRAS.454.2691M} {454, 2691}

\bibitem[\protect\citeauthoryear{{Navarro}, {Frenk}  \& {White}}{{Navarro}
  et~al.}{1996}]{nfw96}
{Navarro} J.~F.,  {Frenk} C.~S.,   {White} S.~D.~M.,  1996, \mn@doi [\apj]
  {10.1086/177173}, \href {http://adsabs.harvard.edu/abs/1996ApJ...462..563N}
  {462, 563}

\bibitem[\protect\citeauthoryear{{Nomoto}, {Tominaga}, {Umeda}, {Kobayashi}  \&
  {Maeda}}{{Nomoto} et~al.}{2006}]{nomo06}
{Nomoto} K.,  {Tominaga} N.,  {Umeda} H.,  {Kobayashi} C.,   {Maeda} K.,  2006,
  \mn@doi [Nuclear Physics A] {10.1016/j.nuclphysa.2006.05.008}, \href
  {http://adsabs.harvard.edu/abs/2006NuPhA.777..424N} {777, 424}

\bibitem[\protect\citeauthoryear{{O{\~n}orbe}, {Hennawi}, {Luki{\'c}}  \&
  {Walther}}{{O{\~n}orbe} et~al.}{2017}]{onor17}
{O{\~n}orbe} J.,  {Hennawi} J.~F.,  {Luki{\'c}} Z.,   {Walther} M.,  2017,
  \mn@doi [\apj] {10.3847/1538-4357/aa898d}, \href
  {http://adsabs.harvard.edu/abs/2017ApJ...847...63O} {847, 63}

\bibitem[\protect\citeauthoryear{{Oh}}{{Oh}}{2002}]{oh02}
{Oh} S.~P.,  2002, \mn@doi [\mnras] {10.1046/j.1365-8711.2002.05859.x}, \href
  {http://adsabs.harvard.edu/abs/2002MNRAS.336.1021O} {336, 1021}

\bibitem[\protect\citeauthoryear{{Okamoto}, {Gao}  \& {Theuns}}{{Okamoto}
  et~al.}{2008}]{okam08}
{Okamoto} T.,  {Gao} L.,   {Theuns} T.,  2008, \mn@doi [\mnras]
  {10.1111/j.1365-2966.2008.13830.x}, \href
  {http://adsabs.harvard.edu/abs/2008MNRAS.390..920O} {390, 920}

\bibitem[\protect\citeauthoryear{{Oppenheimer} \& {Dav{\'e}}}{{Oppenheimer} \&
  {Dav{\'e}}}{2008}]{oppe08}
{Oppenheimer} B.~D.,  {Dav{\'e}} R.,  2008, \mn@doi [\mnras]
  {10.1111/j.1365-2966.2008.13280.x}, \href
  {http://adsabs.harvard.edu/abs/2008MNRAS.387..577O} {387, 577}

\bibitem[\protect\citeauthoryear{{Oppenheimer} \& {Schaye}}{{Oppenheimer} \&
  {Schaye}}{2013}]{oppe13}
{Oppenheimer} B.~D.,  {Schaye} J.,  2013, \mn@doi [\mnras]
  {10.1093/mnras/stt1150}, \href
  {http://adsabs.harvard.edu/abs/2013MNRAS.434.1063O} {434, 1063}

\bibitem[\protect\citeauthoryear{{Oppenheimer}, {Dav{\'e}}  \&
  {Finlator}}{{Oppenheimer} et~al.}{2009}]{oppe09}
{Oppenheimer} B.~D.,  {Dav{\'e}} R.,   {Finlator} K.,  2009, \mn@doi [\mnras]
  {10.1111/j.1365-2966.2009.14771.x}, \href
  {http://adsabs.harvard.edu/abs/2009MNRAS.396..729O} {396, 729}

\bibitem[\protect\citeauthoryear{{Ouchi} et~al.,}{{Ouchi}
  et~al.}{2018}]{ouch18}
{Ouchi} M.,  et~al., 2018, \mn@doi [\pasj] {10.1093/pasj/psx074}, \href
  {http://adsabs.harvard.edu/abs/2018PASJ...70S..13O} {70, S13}

\bibitem[\protect\citeauthoryear{{Paardekooper}, {Khochfar}  \& {Dalla
  Vecchia}}{{Paardekooper} et~al.}{2015}]{paar15}
{Paardekooper} J.-P.,  {Khochfar} S.,   {Dalla Vecchia} C.,  2015, \mn@doi
  [\mnras] {10.1093/mnras/stv1114}, \href
  {http://adsabs.harvard.edu/abs/2015MNRAS.451.2544P} {451, 2544}

\bibitem[\protect\citeauthoryear{{Parsa}, {Dunlop}  \& {McLure}}{{Parsa}
  et~al.}{2018}]{pars18}
{Parsa} S.,  {Dunlop} J.~S.,   {McLure} R.~J.,  2018, \mn@doi [\mnras]
  {10.1093/mnras/stx2887}, \href
  {http://adsabs.harvard.edu/abs/2018MNRAS.474.2904P} {474, 2904}

\bibitem[\protect\citeauthoryear{{Pawlik} \& {Schaye}}{{Pawlik} \&
  {Schaye}}{2009}]{pawl09}
{Pawlik} A.~H.,  {Schaye} J.,  2009, \mn@doi [\mnras]
  {10.1111/j.1745-3933.2009.00659.x}, \href
  {http://adsabs.harvard.edu/abs/2009MNRAS.396L..46P} {396, L46}

\bibitem[\protect\citeauthoryear{{Planck Collaboration} et~al.,}{{Planck
  Collaboration} et~al.}{2016a}]{plan16a}
{Planck Collaboration} et~al., 2016a, \mn@doi [\aap]
  {10.1051/0004-6361/201525830}, \href
  {http://adsabs.harvard.edu/abs/2016A%26A...594A..13P} {594, A13}

\bibitem[\protect\citeauthoryear{{Planck Collaboration} et~al.,}{{Planck
  Collaboration} et~al.}{2016b}]{plan16b}
{Planck Collaboration} et~al., 2016b, \mn@doi [\aap]
  {10.1051/0004-6361/201628897}, \href
  {http://adsabs.harvard.edu/abs/2016A%26A...596A.108P} {596, A108}

\bibitem[\protect\citeauthoryear{{Poudel}, {Kulkarni}, {Morrison},
  {P{\'e}roux}, {Som}, {Rahmani}  \& {Quiret}}{{Poudel} et~al.}{2018}]{poud18}
{Poudel} S.,  {Kulkarni} V.~P.,  {Morrison} S.,  {P{\'e}roux} C.,  {Som} D.,
  {Rahmani} H.,   {Quiret} S.,  2018, \mn@doi [\mnras] {10.1093/mnras/stx2607},
  \href {http://adsabs.harvard.edu/abs/2018MNRAS.473.3559P} {473, 3559}

\bibitem[\protect\citeauthoryear{{Price}, {Trac}  \& {Cen}}{{Price}
  et~al.}{2016}]{pric16}
{Price} L.~C.,  {Trac} H.,   {Cen} R.,  2016, preprint, \href
  {http://adsabs.harvard.edu/abs/2016arXiv160503970P} {} (\mn@eprint {arXiv}
  {1605.03970})

\bibitem[\protect\citeauthoryear{{Puchwein}, {Bolton}, {Haehnelt}, {Madau},
  {Becker}  \& {Haardt}}{{Puchwein} et~al.}{2015}]{puch15}
{Puchwein} E.,  {Bolton} J.~S.,  {Haehnelt} M.~G.,  {Madau} P.,  {Becker}
  G.~D.,   {Haardt} F.,  2015, \mn@doi [\mnras] {10.1093/mnras/stv773}, \href
  {http://adsabs.harvard.edu/abs/2015MNRAS.450.4081P} {450, 4081}

\bibitem[\protect\citeauthoryear{{Qin} et~al.,}{{Qin} et~al.}{2017}]{qin17}
{Qin} Y.,  et~al., 2017, \mn@doi [\mnras] {10.1093/mnras/stx1909}, \href
  {http://adsabs.harvard.edu/abs/2017MNRAS.472.2009Q} {472, 2009}

\bibitem[\protect\citeauthoryear{{Rafelski}, {Wolfe}, {Prochaska}, {Neeleman}
  \& {Mendez}}{{Rafelski} et~al.}{2012}]{rafe12}
{Rafelski} M.,  {Wolfe} A.~M.,  {Prochaska} J.~X.,  {Neeleman} M.,   {Mendez}
  A.~J.,  2012, \mn@doi [\apj] {10.1088/0004-637X/755/2/89}, \href
  {http://adsabs.harvard.edu/abs/2012ApJ...755...89R} {755, 89}

\bibitem[\protect\citeauthoryear{{Rafelski}, {Neeleman}, {Fumagalli}, {Wolfe}
  \& {Prochaska}}{{Rafelski} et~al.}{2014}]{rafe14}
{Rafelski} M.,  {Neeleman} M.,  {Fumagalli} M.,  {Wolfe} A.~M.,   {Prochaska}
  J.~X.,  2014, \mn@doi [\apjl] {10.1088/2041-8205/782/2/L29}, \href
  {http://adsabs.harvard.edu/abs/2014ApJ...782L..29R} {782, L29}

\bibitem[\protect\citeauthoryear{{Rahmati}, {Schaye}, {Pawlik}  \&
  {Rai\v{c}evi\`{c}}}{{Rahmati} et~al.}{2013}]{rahm13}
{Rahmati} A.,  {Schaye} J.,  {Pawlik} A.~H.,   {Rai\v{c}evi\`{c}} M.,  2013,
  \mn@doi [\mnras] {10.1093/mnras/stt324}, \href
  {http://adsabs.harvard.edu/abs/2013MNRAS.431.2261R} {431, 2261}

\bibitem[\protect\citeauthoryear{{Rahmati}, {Schaye}, {Crain}, {Oppenheimer},
  {Schaller}  \& {Theuns}}{{Rahmati} et~al.}{2016}]{rahm16}
{Rahmati} A.,  {Schaye} J.,  {Crain} R.~A.,  {Oppenheimer} B.~D.,  {Schaller}
  M.,   {Theuns} T.,  2016, \mn@doi [\mnras] {10.1093/mnras/stw453}, \href
  {http://adsabs.harvard.edu/abs/2016MNRAS.459..310R} {459, 310}

\bibitem[\protect\citeauthoryear{{Raiter}, {Schaerer}  \& {Fosbury}}{{Raiter}
  et~al.}{2010}]{rait10}
{Raiter} A.,  {Schaerer} D.,   {Fosbury} R.~A.~E.,  2010, \mn@doi [\aap]
  {10.1051/0004-6361/201015236}, \href
  {http://adsabs.harvard.edu/abs/2010A%26A...523A..64R} {523, A64}

\bibitem[\protect\citeauthoryear{{Ricotti} \& {Ostriker}}{{Ricotti} \&
  {Ostriker}}{2004}]{ricc04}
{Ricotti} M.,  {Ostriker} J.~P.,  2004, \mn@doi [\mnras]
  {10.1111/j.1365-2966.2004.07662.x}, \href
  {http://adsabs.harvard.edu/abs/2004MNRAS.350..539R} {350, 539}

\bibitem[\protect\citeauthoryear{{Rosdahl} et~al.,}{{Rosdahl}
  et~al.}{2018}]{rosd18}
{Rosdahl} J.,  et~al., 2018, preprint, \href
  {http://adsabs.harvard.edu/abs/2018arXiv180107259R} {} (\mn@eprint {arXiv}
  {1801.07259})

\bibitem[\protect\citeauthoryear{{Rutkowski} et~al.,}{{Rutkowski}
  et~al.}{2017}]{rutk17}
{Rutkowski} M.~J.,  et~al., 2017, \mn@doi [\apjl] {10.3847/2041-8213/aa733b},
  \href {http://adsabs.harvard.edu/abs/2017ApJ...841L..27R} {841, L27}

\bibitem[\protect\citeauthoryear{{Salmon} et~al.,}{{Salmon}
  et~al.}{2015}]{salm15}
{Salmon} B.,  et~al., 2015, \mn@doi [\apj] {10.1088/0004-637X/799/2/183}, \href
  {http://adsabs.harvard.edu/abs/2015ApJ...799..183S} {799, 183}

\bibitem[\protect\citeauthoryear{{Santini} et~al.,}{{Santini}
  et~al.}{2017}]{sant17}
{Santini} P.,  et~al., 2017, \mn@doi [\apj] {10.3847/1538-4357/aa8874}, \href
  {http://adsabs.harvard.edu/abs/2017ApJ...847...76S} {847, 76}

\bibitem[\protect\citeauthoryear{{Savage} \& {Sembach}}{{Savage} \&
  {Sembach}}{1991}]{sava91}
{Savage} B.~D.,  {Sembach} K.~R.,  1991, \mn@doi [\apj] {10.1086/170498}, \href
  {http://adsabs.harvard.edu/abs/1991ApJ...379..245S} {379, 245}

\bibitem[\protect\citeauthoryear{{Schaerer}}{{Schaerer}}{2002}]{scha02}
{Schaerer} D.,  2002, \mn@doi [\aap] {10.1051/0004-6361:20011619}, \href
  {http://adsabs.harvard.edu/abs/2002A%26A...382...28S} {382, 28}

\bibitem[\protect\citeauthoryear{{Schaye}}{{Schaye}}{2001}]{scha01}
{Schaye} J.,  2001, \mn@doi [\apj] {10.1086/322421}, \href
  {http://adsabs.harvard.edu/abs/2001ApJ...559..507S} {559, 507}

\bibitem[\protect\citeauthoryear{{Schaye} et~al.,}{{Schaye}
  et~al.}{2010}]{scha10}
{Schaye} J.,  et~al., 2010, \mn@doi [\mnras]
  {10.1111/j.1365-2966.2009.16029.x}, \href
  {http://adsabs.harvard.edu/abs/2010MNRAS.402.1536S} {402, 1536}

\bibitem[\protect\citeauthoryear{{Schaye} et~al.,}{{Schaye}
  et~al.}{2015}]{scha15}
{Schaye} J.,  et~al., 2015, \mn@doi [\mnras] {10.1093/mnras/stu2058}, \href
  {http://adsabs.harvard.edu/abs/2015MNRAS.446..521S} {446, 521}

\bibitem[\protect\citeauthoryear{{Schneider} et~al.,}{{Schneider}
  et~al.}{2018}]{schn18}
{Schneider} F.~R.~N.,  et~al., 2018, \mn@doi [Science]
  {10.1126/science.aan0106}, \href
  {http://adsabs.harvard.edu/abs/2018Sci...359...69S} {359, 69}

\bibitem[\protect\citeauthoryear{{Senchyna} et~al.,}{{Senchyna}
  et~al.}{2017}]{senc17}
{Senchyna} P.,  et~al., 2017, \mn@doi [\mnras] {10.1093/mnras/stx2059}, \href
  {http://adsabs.harvard.edu/abs/2017MNRAS.472.2608S} {472, 2608}

\bibitem[\protect\citeauthoryear{{Sharma}, {Theuns}, {Frenk}, {Bower}, {Crain},
  {Schaller}  \& {Schaye}}{{Sharma} et~al.}{2016}]{shar16}
{Sharma} M.,  {Theuns} T.,  {Frenk} C.,  {Bower} R.,  {Crain} R.,  {Schaller}
  M.,   {Schaye} J.,  2016, \mn@doi [\mnras] {10.1093/mnrasl/slw021}, \href
  {http://adsabs.harvard.edu/abs/2016MNRAS.458L..94S} {458, L94}

\bibitem[\protect\citeauthoryear{{Smit}, {Swinbank}, {Massey}, {Richard},
  {Smail}  \& {Kneib}}{{Smit} et~al.}{2017}]{smit17}
{Smit} R.,  {Swinbank} A.~M.,  {Massey} R.,  {Richard} J.,  {Smail} I.,
  {Kneib} J.-P.,  2017, \mn@doi [\mnras] {10.1093/mnras/stx245}, \href
  {http://adsabs.harvard.edu/abs/2017MNRAS.467.3306S} {467, 3306}

\bibitem[\protect\citeauthoryear{{Somerville} \& {Dav{\'e}}}{{Somerville} \&
  {Dav{\'e}}}{2015}]{somd15}
{Somerville} R.~S.,  {Dav{\'e}} R.,  2015, \mn@doi [\araa]
  {10.1146/annurev-astro-082812-140951}, \href
  {http://adsabs.harvard.edu/abs/2015ARA%26A..53...51S} {53, 51}

\bibitem[\protect\citeauthoryear{{Song} et~al.,}{{Song} et~al.}{2016}]{song16}
{Song} M.,  et~al., 2016, \mn@doi [\apj] {10.3847/0004-637X/825/1/5}, \href
  {http://adsabs.harvard.edu/abs/2016ApJ...825....5S} {825, 5}

\bibitem[\protect\citeauthoryear{{Songaila}}{{Songaila}}{2001}]{song01}
{Songaila} A.,  2001, \mn@doi [\apjl] {10.1086/324761}, \href
  {http://adsabs.harvard.edu/abs/2001ApJ...561L.153S} {561, L153}

\bibitem[\protect\citeauthoryear{{Springel}}{{Springel}}{2005}]{spri05}
{Springel} V.,  2005, \mn@doi [\mnras] {10.1111/j.1365-2966.2005.09655.x},
  \href {http://adsabs.harvard.edu/abs/2005MNRAS.364.1105S} {364, 1105}

\bibitem[\protect\citeauthoryear{{Springel} \& {Hernquist}}{{Springel} \&
  {Hernquist}}{2003}]{spri03}
{Springel} V.,  {Hernquist} L.,  2003, \mn@doi [\mnras]
  {10.1046/j.1365-8711.2003.06206.x}, \href
  {http://adsabs.harvard.edu/abs/2003MNRAS.339..289S} {339, 289}

\bibitem[\protect\citeauthoryear{{Stanway}, {Eldridge}  \& {Becker}}{{Stanway}
  et~al.}{2016}]{stan16}
{Stanway} E.~R.,  {Eldridge} J.~J.,   {Becker} G.~D.,  2016, \mn@doi [\mnras]
  {10.1093/mnras/stv2661}, \href
  {http://adsabs.harvard.edu/abs/2016MNRAS.456..485S} {456, 485}

\bibitem[\protect\citeauthoryear{{Steidel}, {Strom}, {Pettini}, {Rudie},
  {Reddy}  \& {Trainor}}{{Steidel} et~al.}{2016}]{stei16}
{Steidel} C.~C.,  {Strom} A.~L.,  {Pettini} M.,  {Rudie} G.~C.,  {Reddy} N.~A.,
    {Trainor} R.~F.,  2016, \mn@doi [\apj] {10.3847/0004-637X/826/2/159}, \href
  {http://adsabs.harvard.edu/abs/2016ApJ...826..159S} {826, 159}

\bibitem[\protect\citeauthoryear{{Sutherland} \& {Dopita}}{{Sutherland} \&
  {Dopita}}{1993}]{suth93}
{Sutherland} R.~S.,  {Dopita} M.~A.,  1993, \mn@doi [\apjs] {10.1086/191823},
  \href {http://adsabs.harvard.edu/abs/1993ApJS...88..253S} {88, 253}

\bibitem[\protect\citeauthoryear{{Theuns}, {Leonard}, {Efstathiou}, {Pearce}
  \& {Thomas}}{{Theuns} et~al.}{1998}]{theu98}
{Theuns} T.,  {Leonard} A.,  {Efstathiou} G.,  {Pearce} F.~R.,   {Thomas}
  P.~A.,  1998, \mn@doi [\mnras] {10.1046/j.1365-8711.1998.02040.x}, \href
  {http://adsabs.harvard.edu/abs/1998MNRAS.301..478T} {301, 478}

\bibitem[\protect\citeauthoryear{{Totani}, {Kawai}, {Kosugi}, {Aoki}, {Yamada},
  {Iye}, {Ohta}  \& {Hattori}}{{Totani} et~al.}{2006}]{tota06}
{Totani} T.,  {Kawai} N.,  {Kosugi} G.,  {Aoki} K.,  {Yamada} T.,  {Iye} M.,
  {Ohta} K.,   {Hattori} T.,  2006, \mn@doi [\pasj] {10.1093/pasj/58.3.485},
  \href {http://adsabs.harvard.edu/abs/2006PASJ...58..485T} {58, 485}

\bibitem[\protect\citeauthoryear{{Trac} \& {Gnedin}}{{Trac} \&
  {Gnedin}}{2011}]{trac11}
{Trac} H.~Y.,  {Gnedin} N.~Y.,  2011, \mn@doi [Advanced Science Letters]
  {10.1166/asl.2011.1214}, \href
  {http://adsabs.harvard.edu/abs/2011ASL.....4..228T} {4, 228}

\bibitem[\protect\citeauthoryear{{Trac}, {Cen}  \& {Loeb}}{{Trac}
  et~al.}{2008}]{trac08}
{Trac} H.,  {Cen} R.,   {Loeb} A.,  2008, \mn@doi [\apjl] {10.1086/595678},
  \href {http://adsabs.harvard.edu/abs/2008ApJ...689L..81T} {689, L81}

\bibitem[\protect\citeauthoryear{{Turner}, {Schaye}, {Crain}, {Theuns}  \&
  {Wendt}}{{Turner} et~al.}{2016}]{turn16}
{Turner} M.~L.,  {Schaye} J.,  {Crain} R.~A.,  {Theuns} T.,   {Wendt} M.,
  2016, \mn@doi [\mnras] {10.1093/mnras/stw1816}, \href
  {http://adsabs.harvard.edu/abs/2016MNRAS.462.2440T} {462, 2440}

\bibitem[\protect\citeauthoryear{{Vanzella} et~al.,}{{Vanzella}
  et~al.}{2018}]{vanz18}
{Vanzella} E.,  et~al., 2018, \mn@doi [\mnras] {10.1093/mnrasl/sly023}, \href
  {http://adsabs.harvard.edu/abs/2018MNRAS.tmpL..26V} {}

\bibitem[\protect\citeauthoryear{{Viel}, {Becker}, {Bolton}  \&
  {Haehnelt}}{{Viel} et~al.}{2013a}]{viel13a}
{Viel} M.,  {Becker} G.~D.,  {Bolton} J.~S.,   {Haehnelt} M.~G.,  2013a,
  \mn@doi [\prd] {10.1103/PhysRevD.88.043502}, \href
  {http://adsabs.harvard.edu/abs/2013PhRvD..88d3502V} {88, 043502}

\bibitem[\protect\citeauthoryear{{Viel}, {Schaye}  \& {Booth}}{{Viel}
  et~al.}{2013b}]{viel13b}
{Viel} M.,  {Schaye} J.,   {Booth} C.~M.,  2013b, \mn@doi [\mnras]
  {10.1093/mnras/sts465}, \href
  {http://adsabs.harvard.edu/abs/2013MNRAS.429.1734V} {429, 1734}

\bibitem[\protect\citeauthoryear{{Wiersma}, {Schaye}, {Theuns}, {Dalla Vecchia}
   \& {Tornatore}}{{Wiersma} et~al.}{2009}]{wier09}
{Wiersma} R.~P.~C.,  {Schaye} J.,  {Theuns} T.,  {Dalla Vecchia} C.,
  {Tornatore} L.,  2009, \mn@doi [\mnras] {10.1111/j.1365-2966.2009.15331.x},
  \href {http://adsabs.harvard.edu/abs/2009MNRAS.399..574W} {399, 574}

\bibitem[\protect\citeauthoryear{{Wise}, {Demchenko}, {Halicek}, {Norman},
  {Turk}, {Abel}  \& {Smith}}{{Wise} et~al.}{2014}]{wise14}
{Wise} J.~H.,  {Demchenko} V.~G.,  {Halicek} M.~T.,  {Norman} M.~L.,  {Turk}
  M.~J.,  {Abel} T.,   {Smith} B.~D.,  2014, \mn@doi [\mnras]
  {10.1093/mnras/stu979}, \href
  {http://adsabs.harvard.edu/abs/2014MNRAS.442.2560W} {442, 2560}

\bibitem[\protect\citeauthoryear{{Worseck} et~al.,}{{Worseck}
  et~al.}{2014}]{wors14}
{Worseck} G.,  et~al., 2014, \mn@doi [\mnras] {10.1093/mnras/stu1827}, \href
  {http://adsabs.harvard.edu/abs/2014MNRAS.445.1745W} {445, 1745}

\bibitem[\protect\citeauthoryear{{Worseck}, {Prochaska}, {Hennawi}  \&
  {McQuinn}}{{Worseck} et~al.}{2016}]{wors16}
{Worseck} G.,  {Prochaska} J.~X.,  {Hennawi} J.~F.,   {McQuinn} M.,  2016,
  \mn@doi [\apj] {10.3847/0004-637X/825/2/144}, \href
  {http://adsabs.harvard.edu/abs/2016ApJ...825..144W} {825, 144}

\bibitem[\protect\citeauthoryear{{Wyithe} \& {Bolton}}{{Wyithe} \&
  {Bolton}}{2011}]{wyitbolt11}
{Wyithe} J.~S.~B.,  {Bolton} J.~S.,  2011, \mn@doi [\mnras]
  {10.1111/j.1365-2966.2010.18030.x}, \href
  {http://adsabs.harvard.edu/abs/2011MNRAS.412.1926W} {412, 1926}

\bibitem[\protect\citeauthoryear{{Zackrisson}, {Rydberg}, {Schaerer},
  {{\"O}stlin}  \& {Tuli}}{{Zackrisson} et~al.}{2011}]{zack11}
{Zackrisson} E.,  {Rydberg} C.-E.,  {Schaerer} D.,  {{\"O}stlin} G.,   {Tuli}
  M.,  2011, \mn@doi [\apj] {10.1088/0004-637X/740/1/13}, \href
  {http://adsabs.harvard.edu/abs/2011ApJ...740...13Z} {740, 13}

\makeatother
\end{thebibliography}


\bsp	
\label{lastpage}
\end{document}